%% file: main.tex
\documentclass[runningheads]{llncs}
\usepackage{subcaption}
\usepackage{graphicx}
\usepackage{hyperref}

\usepackage{mathtools}
\usepackage{bbding} 

\usepackage{prettyref}
\newcommand{\rref}[2][]{\prettyref{#2}}
\newrefformat{sec}{Section\,\ref{#1}}
\newrefformat{app}{Appendix\,\ref{#1}}
\newrefformat{def}{Def.\,\ref{#1}}
\newrefformat{thm}{Theorem\,\ref{#1}}
\newrefformat{prop}{Proposition\,\ref{#1}}
\newrefformat{lem}{Lemma\,\ref{#1}}
\newrefformat{cor}{Corollary\,\ref{#1}}
\newrefformat{ex}{Example\,\ref{#1}}
\newrefformat{tab}{Table\,\ref{#1}}
\newrefformat{fig}{Fig.\,\ref{#1}}
\newrefformat{rem}{Remark\,\ref{#1}}
\newrefformat{case}{Case\,\ref{#1}}

\newif\iflongversion\longversiontrue
\iflongversion
\newrefformat{app}{Appendix\,\ref{#1}}
\newenvironment{proofsketch}[1][TODO]{\proof[Proof Summary (\ifthenelse{\equal{#1}{TODO}}{TODO}{\rref{#1}})]}{\endproof}
\newcommand{\seeapp}[1]{(see #1)}
\newcommand{\derived}[1]{*#1}
\else
\newrefformat{app}{\cite{prebet2024uniform}}

\newcommand{\seeapp}[1]{#1}
\newcommand{\derived}[1]{}
\fi

\usepackage[prefixflatinterpret,prefixflatbracketmodalinterpret,prefixflatsetinterpret,sidenotecalculus,longseqcontext,shortmquant,mquantifiertype,mconnectiveformal,irunistore]{logic}
\usepackage[prefixflatinterpret,prefixflatbracketmodalinterpret,differentialdL,precisenames,fixformat]{dL}
\usepackage{math}
\newcommand{\I}{\interpretation[state=\nu, const=I]}
\newcommand{\Itil}{\interpretation[state=\tilde{\nu}, const=I]}
\newcommand{\It}{\interpretation[state=\omega, const=I]}
\newcommand{\J}{\interpretation[state=\tilde{\nu}, const=J]}
\newcommand{\Jb}{\interpretation[state=\tilde{\omega}, const=J]}
\newcommand{\Ittil}{\interpretation[state=\tilde{\omega}, const=I]}
\newcommand{\Itb}{\interpretation[state=\mu, const=I]}
\newcommand{\Itbi}[1]{\interpretation[state=\mu_{#1}, const=I]}
\newcommand{\Iadj}{\iadjointSubst{\sigma}{\omega}{\I}}
\newcommand{\Iadjb}{\iadjointSubst{\sigma}{\omega}{\Itb}}
\newcommand{\Idot}{\interpretation[const=I,state=]}

\newcommand{\sol}{\varphi}
\newcommand{\Isol}[1][t]{\interpretation[state=\sol(#1), const=I]}
\newcommand{\Imut}{\interpretation[state=\mu(t), const=I]}

\usepackage{dLmacros}

\definecolor{vgreen}{rgb}{.1,.5,0}
\definecolor{vdarkgreen}{rgb}{.06,.3,0}
\definecolor{vred}{rgb}{.7,0,0}
\definecolor{vblue}{rgb}{.1,.15,.62}
\definecolor{vgray}{rgb}{.35,.35,.35}
\definecolor{darkishgray}{rgb}{.35,.35,.35}
\definecolor{vvblue}{rgb}{.14,.21,.868}

\renewcommand*{\irrulename}[1]{\text{\textcolor{darkishgray}{\upshape(#1)}}}
\renewcommand{\axkey}[1]{\textcolor{vvblue}{#1}}
\renewcommand{\axeffect}[1]{\textcolor{vred}{#1}}

\begin{document}
\include{ruledefs}

\title{Uniform Substitution for\\ Differential Refinement Logic\thanks{%
Funding has been provided by an Alexander von Humboldt Professorship and the pilot program Core Informatics (KiKIT) of the Helmholtz Association (HGF).}}
%
%
\author{Enguerrand Prebet\Envelope\orcidID{0009-0008-0160-5219} \and
Andr\'e Platzer\orcidID{0000-0001-7238-5710}}
\authorrunning{E. Prebet, A. Platzer}
%
\institute{
  Karlsruhe Institute of Technology, Karlsruhe, Germany
  \email{\{enguerrand.prebet,platzer\}@kit.edu}
}
\maketitle              
\begin{abstract}
This paper introduces a uniform substitution calculus for differential refinement logic \dRL.
The logic \dRL extends the differential dynamic logic \dL such that one can simultaneously reason about properties of and relations between hybrid systems.
Refinements are useful e.g.\ for simplifying proofs by relating a concrete hybrid system to an abstract one from which the property can be proved more easily.
Uniform substitution is the key to parsimonious prover microkernels.
It enables the verbatim use of single axiom formulas instead of axiom schemata with soundness-critical side conditions scattered across the proof calculus.
The uniform substitution rule can then be used to instantiate all axioms soundly.
Access to differential variables in \dRL enables more control over the notion of refinement, which is shown to be decidable on a fragment of hybrid programs.

\end{abstract}
\section{Introduction}

Hybrid systems modeled by joint discrete dynamics and continuous dynamics are important and subtle systems in need of sound proofs \cite{DBLP:books/sp/Platzer18} on account of their important applications \cite{DBLP:journals/sttt/JeanninGKSGMP17,DBLP:journals/ijrr/MitschGVP17,DBLP:journals/tcad/KabraMP22,DBLP:journals/tase/PereiraBGA23,DBLP:conf/hybrid/StaunerMF97}.
Since such systems are important to get right, hybrid systems verification techniques themselves should be sound.
Uniform substitution \cite{DBLP:conf/cade/Platzer15,DBLP:journals/jar/Platzer17,DBLP:conf/cade/Platzer18,DBLP:conf/cade/Platzer19}, originally phrased by Church for first-order logic \cite[\S35,40]{Church_1956}, has been identified as the key technique reducing the soundness-critical core to a prover microkernel and is behind the \KeYmaeraX prover \cite{DBLP:conf/cade/FultonMQVP15}.

This paper designs a corresponding uniform substitution proof calculus for differential refinement logic (\dRL) \cite{DBLP:conf/lics/LoosP16}.
The logic \dRL is unique in its capabilities of proving simultaneous hybrid systems properties and hybrid systems refinement relations.
This ability of \dRL has been shown to be beneficial for establishing refinement relations of system implementations to verification abstractions and for relating time-triggered implementation models to event-triggered verification models \cite{Loos16}.
The latter relation overcomes a stark divide in embedded system design principles while combining ease of verification with ease of implementation in ways that neither design paradigm alone supports.
But such proving power only helps practical system verification if the theoretical proof calculi are implemented in a sound way and, in fact, \dRL has not yet been implemented at all.
Such an implementation is significantly simplified and significantly easier to get sound by identifying a uniform substitution calculus, which has no axiom schemata with their usual side conditions (and the algorithms implementing them) but merely a finite list of concrete \dRL formulas as axioms.
Reasoning directly with these concrete formulas also makes the proofs easier as the conditions are checked only when uniform substitution is used.
This means that a direct consequence of the axioms could have more admissible substitution instances than the axioms themselves, whereas with schemata, the side conditions would pile up and not generalize as well.
Other beneficial side effects include the fact that \dRL now acquires a Hilbert-style proof calculus that is significantly more flexible and also more modular than \dRL's previous sequent calculus.

Challenges include the fact that uniform substitution calculi for hybrid systems give a differential-form semantics to differentials and differential symbols \cite{DBLP:journals/jar/Platzer17}, which is critical to obtain logic-based decision procedures for differential equation invariants \cite{DBLP:journals/jacm/PlatzerT20}, but also renders some sequent calculus proof rules of \dRL unsound due to the resulting finer-grained view on differential equations.
The flip side is that this finer view distinguishes widely different classes of differential equations better, thereby making it easier to tell apart different differential equations that merely coincide on the overall reachable set while having different temporal behavior.
This difference is exploited here to obtain a decidability result for refinement for a fragment of hybrid systems.
Other challenges to overcome are the unexpected definition of free variables of refinements, which are required for soundness.
The core of the resulting calculus has been implemented in \KeYmaeraX\footnote{\url{https://github.com/LS-Lab/KeYmaeraX-release/tree/dRL}}, extending the prover microkernel in 4 hours of work with about 300 lines of code, mostly spent on writing down all the new axioms.

\section{Related Work}

Hybrid programs in \dRL form a Kleene algebra with tests \cite{DBLP:journals/toplas/Kozen97}.
Program equivalence for Kleene algebra with tests is known to be decidable for abstract atomic programs.
Refinement $\asprg \refines \bsprg$ can be recovered and defined as $\pchoice{\asprg}{\bsprg} \prgeq \bsprg$, but that duplicates reasoning about $\bsprg$.
Certain classes of hypotheses can be added to the theory, e.g.\ Hoare-like triples $\ptest{p};\asprg;\ptest{\lnot q} \prgeq {\ptest{\lfalse}}$, without breaking the decidability \cite{DBLP:conf/fossacs/DoumaneKPP19}.
This however does not extend when limited commutativity is allowed, which arises even in the discrete fragment: $(\umod{x}{2};\umod{y}{3}) \prgeq (\umod{y}{3};\umod{x}{2})$ but $(\umod{x}{2};\umod{x}{3}) \neq (\umod{x}{3};\umod{x}{2})$.
KAT with only discrete assignments has been studied as Schematic KAT \cite{10.5555/867190}.
\dRL can derive the axioms of Schematic KAT, but also allows reasoning with continuous dynamics and differential equations.

The Event-B method \cite{DBLP:books/daglib/0024570} is a formalism for reasoning about discrete models where the primary mechanism is refinement to check the conformance between abstract models and more detailed ones.
Multiple different formalisms have been proposed. 
Hybrid Event-B \cite{DBLP:conf/asm/AbrialSZ12,DBLP:journals/fac/BanachZSH14,DBLP:journals/scp/BanachBQVZ15} is an extension with tool support \cite{DBLP:books/crc/p/ButlerAB16} for hybrid systems with events corresponding to discrete and continuous evolutions.
These continuous steps are however abstracted by the invariants they are assumed to satisfy.
Event-B can also be extended with theories \cite{DBLP:conf/birthday/ButlerM13}.
By adding some axioms about differential equations, it allows refinement reasoning with some continuous dynamics \cite{DBLP:phd/hal/Dupont21,DBLP:conf/asm/AfendiLM20}.
In contrast, \dRL captures the continuous dynamics directly and proves the invariants as a consequence of the continuous dynamics.

Uniform substitution was proposed by Alonzo Church for first-order logic to capture axioms instead of axiom schemata \cite[\S35,40]{Church_1956}.
Modern uniform substitution originated for \dL to support hybrid systems theorem proving in simple ways \cite{DBLP:journals/jar/Platzer17}, extended to hybrid games in differential game logic \dGL \cite{DBLP:conf/cade/Platzer18}, and 
to communicating parallel programs \dLCHP \cite{DBLP:conf/cade/BriegerMP23}.
This work is complementing the approach by adding refinement reasoning in a uniform substitution calculus for hybrid systems.
Developing uniform substitution calculi are key to the design of small soundness-critical prover microkernels such as \KeYmaeraX \cite{DBLP:conf/cade/FultonMQVP15}.

\section{Differential Refinement Logic \dRL}\label{sec:dRL}

Differential refinement logic \dRL \cite{DBLP:conf/lics/LoosP16} extends the differential dynamic logic \dL for hybrid systems \cite{DBLP:journals/jar/Platzer08} with a first-class refinement operator $\refines$ on hybrid systems.
This section presents \emph{differential-form} \dRL, which prepares \dRL for the features needed for \dL's uniform substitution axiomatization, most notably the inclusion of differential terms alongside function symbols, predicate symbols, and program constant symbols, but also the requisite inclusion of differential variable symbols.
Differential terms $\der{\astrm}$ are the fundamental logical device with which to enable sound \cite{DBLP:journals/jar/Platzer17} and complete \cite{DBLP:conf/lics/PlatzerT18,DBLP:journals/jacm/PlatzerT20} reasoning about differential equations.

\subsection{Syntax}
This section defines the syntax of the differential refinement logic \dRL.
The set of all variables is $\allvars$.
To each variable $x \in \allvars$ is associated a \emph{differential symbol} $\D{x}$ which is also in $\allvars$.
Its purpose is to use $\D{x}$ to refer to the time-derivative of variable $x$ during a differential equation, but also to cleverly relay that information to surrounding formulas in a sound way \cite{DBLP:journals/jar/Platzer17}.
It is this (crucial) presence of differential symbols, that gives differential-form \dRL a refined notion of refinement, especially of differential equations, compared to its sequent calculus predecessor \cite{DBLP:conf/lics/LoosP16}.

\begin{definition}[Terms]
	\emph{Terms} are defined by the grammar below where $x \in \allvars$ is a variable, $\oustrm$ is a function symbol of arity $n$ and $\astrm,\bstrm,\istrm{1},\dots,\istrm{n}$ are terms:
$$\begin{array}{c c c}
	\astrm, \bstrm & \Coloneqq & x \OR \oustrm[\istrm{1},\dots,\istrm{n}] \OR \astrm + \bstrm \OR \astrm \cdot \bstrm \OR \der{\astrm}
\end{array}$$
\end{definition}
Terms have the usual arithmetic operations and function symbols.
They also have differentials of terms $\der{\astrm}$ which describe how the value of $\astrm$ changes locally depending on the values of the differential symbols associated to the variables of $\astrm$.

\begin{definition}[Formulas]
	\emph{Formulas} are defined by the grammar below where $\astrm,\bstrm,\istrm{1},\dots,\istrm{n}$ are terms, $\osfml$ is a predicate symbol of arity $n,\asfml,\bsfml$ are formulas and $\asprg, \bsprg$ are hybrid programs (\rref{def:HP}):
$$\begin{array}{c c c}
    \asfml, \bsfml & \Coloneqq & \astrm \leq \bstrm \OR \ousfml[\istrm{1},\dots,\istrm{n}] \OR \lnot \asfml \OR \asfml \land \bsfml \OR \lforall{x}{\asfml} \OR \dbox{\asprg}{\asfml} \OR \asprg \refines \bsprg
\end{array}$$
\end{definition}

In addition to the operators of first-order logic of real arithmetic, formulas also contain the \dL modality $\dbox{\asprg}{\asfml}$ which expresses that the formula $\asfml$ holds after all possible runs of the hybrid program $\asprg$.
\dRL extends \dL with the refinement operator $\asprg \refines \bsprg$ which expresses that $\asprg$ refines $\bsprg$ as $\bsprg$ has more behaviors than $\asprg$: it is true in a state $\iget[state]{\I}$ if all states reachable by hybrid program $\asprg$ from $\iget[state]{\I}$ can be reached by hybrid program $\bsprg$.
The program equivalence $\asprg \prgeq \bsprg$ is shorthand for $\asprg \refines \bsprg \land \bsprg \refines \asprg$.
This will be made explicit by axiom (\irref{leqantisym}) in \rref{sec:dRL-calculus}.

Note the fundamental difference between \dRL modal formula \(\dbox{\asprg}{\asfml}\), which expresses that all runs of hybrid program $\asprg$ satisfy \dRL formula $\asfml$, compared to the \dRL refinement formula \(\asprg \refines \bsprg\), which expresses that all runs of hybrid program $\asprg$ are also runs of hybrid program $\bsprg$.
Both \dRL formulas refer to the runs of a hybrid program $\asprg$, but only the former states a property of the (final) states reached, while only the latter relates the overall transition behavior of hybrid program $\asprg$ to that of another program.
Just like \(\dbox{\asprg}{\asfml}\), formula \(\asprg \refines \bsprg\) is a \dRL formula and not just a judgment, so it can be true in some states and false in others.
This makes it possible to easily express conditional refinement as \(\asfml \limply \asprg \refines \bsprg\) meaning that if $\asfml$ is true initially, then $\asprg$ refines $\bsprg$.
The logic \dRL is closed under all operators.
For example the \dRL formula \(\dbox{\asprg}{\bsprg\refines\csprg}\) expresses that after all runs of $\asprg$ it is the case that all runs of $\bsprg$ are also runs of $\csprg$.
Just like in an ordinary implication,  \(\asfml \limply \asprg \refines \bsprg\) says nothing about what happens when the initial state does not satisfy $\asfml$.
Just like ordinary dynamic logic modalities, \(\dbox{\asprg}{\bsprg\refines\csprg}\)  says nothing about what happens before program $\asprg$ ran.
Indeed, this extended capabilities that \dRL is closed under all operators will add to its expressibility and the eloquence of its uniform substitution proof calculus.

\begin{definition}[Hybrid Programs]\label{def:HP}
	\emph{Hybrid programs} are defined by the grammar below where $x$ is a variable, $\astrm$ is a term, $\ausprg$ is a program constant, $\bsfml$ is a differential-free formula and $\asprg,\bsprg$ are hybrid programs:
$$\begin{array}{c c c}
	\asprg, \bsprg & \Coloneqq & \ausprg \OR \ptest{\bsfml} \OR \pupdate{\pumod{x}{\astrm}} \OR \prandom{x} \OR \pode{\D{x}=\astrm}{\bsfml} \OR \pchoice{\asprg}{\bsprg} \OR \asprg;\bsprg \OR \prepeat{\asprg}
\end{array}$$
\end{definition}
The \emph{test} $\ptest{\bsfml}$ behaves like a skip if the formula $\bsfml$ is true in the current state and blocks the system otherwise.
The \emph{assignment} $\pupdate{\pumod{x}{\astrm}}$ instantaneously updates the value of the variable $x$ to the value of the term $\astrm$.
The \emph{nondeterministic assignment} $\prandom{x}$ updates the value of the variable $x$ to an arbitrary value.
The \emph{differential equation} $\pode{\D{x}=\astrm}{\bsfml}$ behaves like a continuous evolution where both the differential equation $\D{x}=\astrm$ and the evolution domain $\bsfml$ holds.
The \emph{nondeterministic choice} $\pchoice{\asprg}{\bsprg}$ can behave like either $\asprg$ or $\bsprg$.
The \emph{sequence} $\asprg;\bsprg$ behaves like $\asprg$ followed by $\bsprg$.
The \emph{nondeterministic repetition} $\prepeat{\asprg}$ behaves like $\asprg$ repeated an arbitrary natural number of times.
\begin{example}[Modelling safe breaking]\label{ex:carbrake}
	Let us consider a car that needs to stop before a wall at distance $m$. It starts from a safe position and can accelerate with acceleration $A$ if some safety condition $\text{safe}_T(x)$ is true or brake with braking force $B$. The controller is run at most every $T$ seconds. Proving its safety can be achieved by proving the following \dRL formula:
	\[\begin{array}{c}
	A \geq 0 \land B > 0 \land x + v^2/2B \leq m \limply [car_{T}] x \leq m
	\vspace*{0.3em}\\
	car_{T} \Coloneqq (\pchoice{\pupdate{\pumod{a}{-B}}}{\ptest{\text{safe}_T(x)};\pupdate{\pumod{a}{A}}});\pupdate{\pumod{t_0}{t}};\pode{\D{x} = v, \D{v} = a, \D{t} = 1}{t - t_0\leq T}
	\end{array}
	\]
	Such system, called \emph{time-triggered}, can be refined to a \emph{event-triggered} system where the controller is sure to run before a critical event, leaving the domain $E(x)$, occurs. Event-triggered systems are easier to verify but less realistic. With \dRL and the axiom (\irref{boxleq}) below, the time-triggered system can be proved safe by proving the safety of the event-triggered system and the refinement between the two systems:
	\[A \geq 0 \land B \geq 0 \land x + v^2/2B \leq m \limply car_{T} \refines car_{E} \land [car_{E}] x \leq m\]
	\[car_{E} \Coloneqq (\pchoice{\pupdate{\pumod{a}{-B}}}{\ptest{\text{safe}_E(x)};\pupdate{\pumod{a}{A}}});\pupdate{\pumod{t_0}{t}};\pode{\D{x} = v, \D{v} = a, \D{t} = 1}{E(x)}\]
\end{example}
\subsection{Semantics}
A state $\iget[state]{\I}$ is a mapping $\allvars\rightarrow\R$.
The state $\modif{\iget[state]{\I}}{x}{r}$ agrees with the state $\iget[state]{\I}$ except for the variable $x$ whose value is $r\in\R$.
State $\iget[state]{\It}$ is a \emph{$U$-variation} of $\iget[state]{\I}$ if $\iget[state]{\It}$ and $\iget[state]{\I}$ are equal on the complement $\scomplement{U}$ of that set of variables $U$.
For instance, $\modif{\iget[state]{\I}}{x}{r}$ is an $\set{x}$-variation of $\iget[state]{\I}$.
The set of all states is $\State$.
The interpretation of a function symbol of arity $n$ in \emph{interpretation} $\iget[const]{\I}$ is a smooth function $\getInterp{\I}{f}: \R^n\rightarrow\R$.
\begin{definition}[Term semantics]
	The \emph{semantics of a term} $\astrm$ in interpretation $\iget[const]{\I}$ and state $\iget[state]{\I}$ is its value $\ivaluation{\I}{\astrm}\in\R$ and is defined as follows:
	\begin{enumerate}
		\item $\ivaluation{\I}{x} = \getValue{\I}{x}$
		\item $\ivaluation{\I}{\oustrm[\istrm{1},\dots,\istrm{n}]} = \getInterp{\I}{\oustrm}(\ivaluation{\I}{\istrm{1}},\dots,\ivaluation{\I}{\istrm{n}})$
		\item $\ivaluation{\I}{\astrm + \bstrm} = \ivaluation{\I}{\astrm} + \ivaluation{\I}{\bstrm}$
		\item $\ivaluation{\I}{\astrm \cdot \bstrm} = \ivaluation{\I}{\astrm} \cdot \ivaluation{\I}{\bstrm}$
		\item $\ivaluation{\I}{\der{\astrm}} = \sum\limits_{x\in\allvars} \getValue{\I}{\D{x}} \Dp[x]{\ivaluation{\Idot}{\theta}}(\iget[state]{\I}) = \sum\limits_{x\in\allvars}\getValue{\I}{\D{x}}\Dp[x]{\ivaluation{\I}{\astrm}}$
	\end{enumerate}
\end{definition}
The partial derivative $\Dp[x]{\ivaluation{\I}{\astrm}}$ corresponds to the derivative of the one-dimensional function $X \mapsto \ivaluation{\imodif[state]{\I}{x}{X}}{\astrm}$ at $X=\getValue{\I}{x}$.
Since $\ivaluation{\I}{\astrm}$ denotes a smooth function, the derivative always exists.

Since hybrid programs appear in formulas and vice versa, the interpretation of hybrid programs and formulas is defined by simultaneous induction.
The interpretation of a predicate symbol of arity $n$ in interpretation $\iget[const]{\I}$ is an $n$-ary relation $\getInterp{\I}{\osfml} \subseteq \R^n$.
The interpretation of a program constant symbol $\ausprg$ in interpretation $\iget[const]{\I}$ is a state-transition relation $\getInterp{\I}{\ausprg} \subseteq \State \times \State$ where $\iaccessible[\ausprg]{\I}{\It}$ iff the program constant $\ausprg$ can reach the state $\iget[state]{\It}$ starting from the state $\iget[state]{\I}$.

\begin{definition}[\dRL semantics]
    The \emph{semantics of a formula} $\asfml$ for an interpretation $\iget[const]{\I}$ is the subset $\imodel{\I}{\asfml} \subseteq \State$ of states in which $\asfml$ is true and defined as:
    \begin{enumerate}
        \item $\imodels{\I}{\astrm \leq \bstrm}$ iff $\ivaluation{\I}{\astrm}\leq \ivaluation{\I}{\bstrm}$
        \item $\imodels{\I}{\ousfml[\istrm{1},\dots,\istrm{n}]}$ iff $(\ivaluation{\I}{\istrm{1}},\dots,\ivaluation{\I}{\istrm{n}})\in \getInterp{\I}{\osfml}$
        \item $\imodels{\I}{\lnot \asfml}$ iff $\inonmodels{\I}{\asfml}$
        \item $\imodels{\I}{\asfml \land \bsfml}$ iff $\imodels{\I}{\asfml}$ and $\imodels{\I}{\bsfml}$
        \item $\imodels{\I}{\lforall{x}{\asfml}}$ iff $\imodels{\imodif[state]{\I}{x}{r}}{\asfml}$ for all $r\in\R$
        \item $\imodels{\I}{\dbox{\asprg}{\asfml}}$ iff $\imodels{\It}{\asfml}$ for all $\iaccessible[\asprg]{\I}{\It}$
        \item $\imodels{\I}{\asprg \refines \bsprg}$ iff $\iaccessible[\bsprg]{\I}{\It}$ for all $\iaccessible[\asprg]{\I}{\It}$
    \end{enumerate}
\end{definition}
A formula $\asfml$ is \emph{valid in }$\iget[const]{\I}$ if $\imodel{\I}{\asfml} = \State$. A formula $\asfml$ is \emph{valid} if it is valid in all interpretations.
\begin{definition}[Transition semantics of programs]
	The \emph{semantics of a hybrid program} $\asprg$ for an interpretation $\iget[const]{\I}$ is the transition relation $\iaccess[\asprg]{\I} \subseteq \State \times \State$ and is defined as follows:
	\begin{enumerate}
		\item $\iaccess[\ausprg]{\I} = \getInterp{\I}{\ausprg}$
		\item $\iaccess[\ptest\bsfml]{\I} = \set{ (\iget[state]{\I}, \iget[state]{\I}) \with \imodels{\I}{\bsfml}}$

		\item $\iaccess[\prandom{x}]{\I} = \set{(\iget[state]{\I}, \modif{\iget[state]{\I}}{x}{r}) \with \text{for all }r\in\R}$
		
		\item $\iaccess[\pupdate{\pumod{x}{\astrm}}]{\I} = \set{(\iget[state]{\I}, \modif{\iget[state]{\I}}{x}{r}) \with r = \ivaluation{\I}{\astrm}}$
		
		\item $\iaccess[\pode{\D{x}=\astrm}{\bsfml}]{\I} = \set{(\iget[state]{\I}, \iget[state]{\It}) \with \sol(0) \text{ is a }\set{\D{x}}\text{-variation of }\iget[state]{\I}\text{ and } \iget[state]{\It} = \sol(r) \\\text{for some function } \sol : [0, r ] \rightarrow \State \text{ where } \sol(t) \text{ is a }\set{x,\D{x}}\text{-variation of }\iget[state]{\I}, \\\text{satisfies } \imodels{\Isol}{\D{x} = \astrm \land \bsfml} \text{ and } \Ddiff[t]{\ivaluation{\Isol}{x}}=\ivaluation{\Isol}{\astrm} \text{ for all } t\in[0,r]}$
		
		\item $\iaccess[\pchoice{\asprg}{\bsprg}]{\I} = \iaccess[\asprg]{\I}\cup\iaccess[\bsprg]{\I}$
		
		\item $\iaccess[\asprg;\bsprg]{\I} = \iaccess[\asprg]{\I}\circ\iaccess[\bsprg]{\I} = \set{(\iget[state]{\I},\iget[state]{\It}) : \iaccessible[\asprg]{\I}{\Itb}, \iaccessible[\bsprg]{\Itb}{\It} \text{ for some } \iget[state]{\Itb}\in\State}$
		
		\item $\iaccess[\prepeat{\asprg}]{\I} = \bigcup_{n\in\N}\iaccess[{\prepeat[n]{\asprg}}]{\I}$
	\end{enumerate}
\end{definition}
Most importantly, \(\asprg \refines \bsprg\) is true in a state $\iget[state]{\I}$ iff all states $\iget[state]{\It}$ reachable from $\iget[state]{\I}$ by running program $\asprg$ are also reachable by running $\bsprg$ from $\iget[state]{\I}$.

The transition for a differential equation $\pode{\D{x}=\astrm}{\bsfml}$ synchronizes the differential symbol $\D{x}$ with the current time-derivative of $x$, i.e.\ $\astrm$, and then evolves the system continuously along the solution $\sol$ of the differential equation $\D{x}=\astrm$ within the domain $\bsfml$.
Differential equations are the only hybrid programs that intrinsically relate variables with their associated differential symbol.

As differential equations effectively \emph{change} the value of differential symbols, this is taken into account in the semantics of refinements.
The differential equations $\D{x} = 1$ and $\D{x} = 2$ are \emph{not} equivalent: although both can reach the same values for $x$, their respective end states will always have a different value for $\D{x}$.
This behavior differs from the original semantics of \dRL \cite{DBLP:conf/lics/LoosP16}.
Intuitively, this notion of refinement corresponds to assuming that differential equations evolve with a global time $\D{t}=1$.
Other extensions of \dL like \dLCHP \cite{DBLP:conf/cade/BriegerMP23} already assume the presence of such global time.
This property allows to express refinements of differential equations as a \dL formula as shown in the axiom (\irref{ode}) below.

\subsection{Static Semantics}
Uniform substitution relies on the notions of free and bound variables to prevent any unsound substitution attempts.
Static semantics gives a definition for free and bound variables of terms, formulas and hybrid programs based on their (dynamic) semantics, which can be defined as in \dL \cite{DBLP:journals/jar/Platzer17}:
\begin{definition}[Static semantics] \label{def:static-semantics}
	The \emph{static semantics} defines the free variables $\freevarsdef{\astrm}$, $\freevarsdef{\asfml}$ and $\freevarsdef{\asprg}$, which are the variables whose values the expression depends on, and the bound variables $\boundvarsdef{\asprg}$, which are the variables whose values may change during the execution of $\asprg$.
They are defined formally as follows:
	\begin{align*}
		\freevarsdef{\astrm} = \set{x\in\allvars \with & \mexists{\iget[const]{\I},\iget[state]{\I},\iget[state]{\Itil}}{\text{a }\set{x}\text{-variation of }\iget[state]{\I}\text{ such that }\ivaluation{\I}{\astrm}\neq \ivaluation{\Itil}{\astrm}} }\\
		\freevarsdef{\asfml} = \set{x\in\allvars \with & \mexists{\iget[const]{\I},\iget[state]{\I},\iget[state]{\Itil}}{\text{a }\set{x}\text{-variation of }\iget[state]{\I}\text{ such that }\imodels[\astrm]{\I}{\asfml}\not\ni \iget[state]{\Itil}} }\\
		\freevarsdef{\asprg} = \set{x\in\allvars \with & \mexists{\iget[const]{\I},\iget[state]{\I},\iget[state]{\It},\iget[state]{\Itil}}{\text{a }\set{x}\text{-variation of }\iget[state]{\I}\text{ such that }\iaccessible[\asprg]{\I}{\It}} \\
		& \text{and }\mforall{\,\iget[state]{\Ittil}}{\set{x}\text{-variation of }\iget[state]{\It}\text{ such that }\iinaccessible[\asprg]{\Itil}{\Ittil}}}\\
		\boundvarsdef{\asprg} = \set{x\in\allvars \with & \mexists{\iget[const]{\I},\iget[state]{\I},\iget[state]{\It}}{\text{such that }\iaccessible[\asprg]{\I}{\It}\text{ and }\getValue{\I}{x}\neq\getValue{\It}{x}}}
	\end{align*}
\end{definition}
Free and bounds variables are the only information needed about the logic to ensure that the result of uniform substitution is only defined when sound.
The coincidence lemmas \cite{DBLP:journals/jar/Platzer17} show that the truth-values of formulas only depend on their free variables and the interpretation of the symbols appearing in them (similarly for terms and hybrid programs).
The set of function, predicate, and program symbols appearing in a formula, term or hybrid program is denoted $\intsigns{\cdot}$.

\begin{lemma}[Coincidence for terms \cite{DBLP:journals/jar/Platzer17}] \label{lem:coincidence-term}
	The set $\freevarsdef{\astrm}$ is the smallest set with the coincidence property for $\astrm$: If $\iget[state]{\I} = \iget[state]{\J}$ on $V\supseteq\freevarsdef{\astrm}$ and $\iget[const]{\I} = \iget[const]{\J}$ on $\intsigns{\astrm}$, then $\ivaluation{\I}{\astrm} = \ivaluation{\J}{\astrm}$.
\end{lemma}

\begin{lemma}[Coincidence for formulas \cite{DBLP:journals/jar/Platzer17}] \label{lem:coincidence}
	The set $\freevarsdef{\asfml}$ is the smallest set with the coincidence property for $\asfml$: If $\iget[state]{\I} = \iget[state]{\J}$ on $V\supseteq\freevarsdef{\asfml}$ and $\iget[const]{\I} = \iget[const]{\J}$ on $\intsigns{\asfml}$, then $\imodels{\I}{\asfml}$ iff $\imodels{\J}{\asfml}$.
\end{lemma}

\begin{lemma}[Coincidence for hybrid programs \cite{DBLP:journals/jar/Platzer17}] \label{lem:coincidence-HP}
	The set $\freevarsdef{\asprg}$ is the smallest set with the coincidence property for $\asprg$: If $\iget[state]{\I} = \iget[state]{\J}$ on $V\supseteq\freevarsdef{\asprg}$ and $\iget[const]{\I} = \iget[const]{\J}$ on $\intsigns{\asprg}$, then $\iaccessible[\asprg]{\I}{\It}$ implies $\iaccessible[\asprg]{\J}{\Jb}$ for some $\iget[state]{\Jb}$ with $\iget[state]{\It} = \iget[state]{\Jb}$ on $V$.
\end{lemma}
The proof \cite{DBLP:journals/jar/Platzer17} requires a mutual induction on the structure of the formula and hybrid program to show that $\imodel{\I}{\asfml} = \imodel{\J}{\asfml}$ and $\iaccess[\asprg]{\I} = \iaccess[\asprg]{\J}$ which extends to the refinement case.
The rest is done by induction on the set of variables $S$ where the states $\iget[state]{\I}$ and $\iget[state]{\J}$ can differ.

\begin{lemma}[Bound effect \cite{DBLP:journals/jar/Platzer17}]\label{lem:boundeffect}
	The set $\boundvarsdef{\asprg}$ is the smallest set with the bound effect property for $\asprg$: If $\iaccessible[\asprg]{\I}{\It}$, then $\iget[state]{\I} = \iget[state]{\It}$ on $\scomplement{\boundvarsdef{\asprg}}$.
\end{lemma}

These sets are the smallest sets with the coincidence property, which means that all conservative extensions of these sets can also be used soundly.
We define $\freevars{\astrm},\freevars{\asfml},\freevars{\asprg}$ and $\boundvars{\asprg}$ as such overapproximations that can be computed syntactically.
Computing the free variables for a formula $\dbox{\asprg}{\asfml}$ requires the \emph{must-bound variables} of the hybrid program $\asprg$, written $\mustboundvars{\alpha}$.
They represent the variables that will be written in all executions of $\asprg$.
These sets are given in \rref{app:static} and are constructed in a standard way \cite{DBLP:journals/jar/Platzer17}, except for the new refinement operator.

Since the behavior of hybrid program $\asprg$ and $\bsprg$ only depends on their respective free variables (\rref{lem:coincidence-HP}), it would be tempting to define \(\freevars{\asprg\refines\bsprg} = \freevars{\asprg}\cup\freevars{\bsprg}\) stating that the refinement depends on the variables for which either program depends on.
Somewhat surprisingly, this would be unsound for reasons that truly touch on the nature of refinement.
Take the refinement formula $\ptest{\ltrue}\refines \pupdate{\umod{x}{1}}$ and a state $\iget[state]{\I}$ with $\getValue{\I}{x} = 0$.
Then $\inonmodels{\I}{\ptest{\ltrue}\refines \pupdate{\umod{x}{1}}}$.
However if the initial value of $x$ is $1$, then the refinement holds: $\imodels{\imodif[state]{\I}{x}{1}}{\ptest{\ltrue}\refines \pupdate{\umod{x}{1}}}$, because the assignment \(\pupdate{\umod{x}{1}}\) has no effect.
In fact $\freevarsdef{\ptest{\ltrue}\refines \pupdate{\umod{x}{1}}} = \set{x}$ even though \(\freevarsdef{\ptest{\ltrue}}=\freevarsdef{\pupdate{\umod{x}{1}}}=\emptyset\).
To obtain a sound definition of $\freevars{\asprg\refines\bsprg}$, one needs to take into account the variables that may be written in one program, $\boundvars{\asprg}\cup\boundvars{\bsprg}$, but that can also remain unmodified (which makes them depend on their initial values), so not in $\mustboundvars{\asprg}\cap\mustboundvars{\bsprg}$.
Hence, the (syntactic) free variables of a refinement are defined as follows:
\[
\freevars{\asprg\refines\bsprg} = \freevars{\asprg}\cup\freevars{\bsprg}\cup((\boundvars{\asprg}\cup\boundvars{\bsprg})\setminus(\mustboundvars{\asprg}\cap\mustboundvars{\bsprg}))
\]
With this definition for refinements as the only but notable outlier to an otherwise standard definition of the syntatic computations for a static semantics  \cite{DBLP:journals/jar/Platzer17}, the static semantics $\freevars{\asfml}$ etc.\ can be proved to be sound overapproximations of the static semantics $\freevarsdef{\asfml}$ from \rref{def:static-semantics} and thereby enjoy the coincidence lemmas \ref{lem:coincidence-term}--\ref{lem:coincidence-HP} and the bound effect lemma \ref{lem:boundeffect}, respectively.
\begin{lemma}[Soundness of static semantics]\label{lem:static}
	For all terms $\astrm$, formulas $\asfml$ and hybrid programs $\asprg$:
	$$
		\freevars{\astrm}\supseteq\freevarsdef{\astrm} \quad
		\freevars{\asfml}\supseteq\freevarsdef{\asfml} \quad
		\freevars{\asprg}\supseteq\freevarsdef{\asprg} \quad
		\boundvars{\asprg}\supseteq\boundvarsdef{\asprg}
	$$
\end{lemma}
The proof of $\freevars{\cdot}\supseteq\freevarsdef{\cdot}$ for formulas and hybrid programs is the only case affected by the addition of refinement operators compared to prior proofs \cite[Lem.\,17]{DBLP:journals/jar/Platzer17}.
It is proved by induction on the structure of the formulas and hybrid programs.
For hybrid programs, the property shown for $\freevars{\asprg}$ is stronger than the coincidence property from \rref{lem:coincidence-HP}, enforcing $\iget[state]{\It} = \iget[state]{\Jb}$ on $V\cup\mustboundvars{\asprg}$ rather than $V$.

 For the case of the refinement operator $\asprg\refines\bsprg$, the main insight is visible when proving that $\imodels{\J}{\asprg\refines\bsprg}$ implies $\imodels{\I}{\asprg\refines\bsprg}$ with $\iget[state]{\I} = \iget[state]{\J}$ on $V$ and $\iget[const]{\I} = \iget[const]{\J}$ on $\intsigns{\asprg\refines\bsprg}$.
For any $\iaccessible[\asprg]{\I}{\It}$, we have $\iaccessible[\asprg]{\J}{\Jb}$, $\iaccessible[\bsprg]{\J}{\Jb}$ and $\iaccessible[\bsprg]{\I}{\Itb}$ for some states $\iget[state]{\Jb},\iget[state]{\Itb}$ by repeated use of the induction hypothesis and the definition of refinement.
Both the induction hypothesis and \rref{lem:boundeffect} give us information on $\iget[state]{\Jb}$ and $\iget[state]{\Itb}$. As $V \supseteq \freevars{\asprg\refines\bsprg}$, the definition of $\freevars{\asprg\refines\bsprg}$ is crucial for ensuring that this knowledge is enough to fully determine $\iget[state]{\Jb}$ and $\iget[state]{\Itb}$ from $\iget[state]{\I},\iget[state]{\It}$ and $\iget[state]{\J}$, and then that $\iget[state]{\It} = \iget[state]{\Itb}$.

\section{Uniform Substitution}\label{sec:usubst}

A \emph{uniform substitution} $\sigma$ is a mapping from terms of the form $\oustrm[\usarg]$ to terms $\applyusubst{\sigma}{\oustrm[\usarg]}$, from formulas of the form $\ousfml[\usarg]$ to formulas $\applyusubst{\sigma}{\ousfml[\usarg]}$, and from program constants $\ausprg$ to hybrid programs $\applyusubst{\sigma}{\ausprg}$.
The reserved 0-ary function symbol $\usarg$ marks the position where the argument, e.g.\ $\astrm$ in $\ousfml[\astrm]$, will be substituted in the resulting expression.
Soundness of such substitutions requires that the substitution does not introduce new free variables in a context where they are bound \cite{Church_1956}.

\begin{figure}[tbhp]
	\begin{align*}
		\usubstapp{U}{\sigma}{(x)} &= \applyusubst{\sigma}{x} & \text{for }x\in\allvars\\
		\usubstapp{U}{\sigma}{(\oustrm[\astrm])} &= \usubstapp{\emptyset}{\usubstlist{\usubstmod{\usarg}{\usubstapp{U}{\sigma}{\astrm}}}}{\applyusubst{\sigma}{\oustrm[\usarg]}} &\text{if }\freevars{\applyusubst{\sigma}{\oustrm[\usarg]}}\cap U = \emptyset\\
		\usubstapp{U}{\sigma}{(\astrm + \bstrm)} &= \usubstapp{U}{\sigma}{\astrm} + \usubstapp{U}{\sigma}{\bstrm} &\\
		\usubstapp{U}{\sigma}{(\astrm \cdot \bstrm)} &= \usubstapp{U}{\sigma}{\astrm} \cdot \usubstapp{U}{\sigma}{\bstrm} &\\
		\usubstapp{U}{\sigma}{(\der{\astrm})} &= \der{\usubstapp{\allvars}{\sigma}{\astrm}} &\\
		\hline
		\usubstapp{U}{\sigma}{(\ousfml[\astrm])} &= \usubstapp{\emptyset}{\usubstlist{\usubstmod{\usarg}{\usubstapp{U}{\sigma}{\astrm}}}}{\applyusubst{\sigma}{\ousfml[\usarg]}} &\text{if }\freevars{\applyusubst{\sigma}{\ousfml[\usarg]}}\cap U = \emptyset\\
		\usubstapp{U}{\sigma}{(\lnot \asfml)} &= \lnot \usubstapp{U}{\sigma}{\asfml} &\\
		\usubstapp{U}{\sigma}{(\asfml \land \bsfml)} &= \usubstapp{U}{\sigma}{\asfml} \land \usubstapp{U}{\sigma}{\bsfml} &\\
		\usubstapp{U}{\sigma}{(\lforall{x}{\asfml})} &= \lforall{x}{\usubstapp{U\cup\set{x}}{\sigma}{\asfml}} &\\
		\usubstapp{U}{\sigma}{(\dbox{\asprg}{\asfml})} &= \dbox{\usubstappp{V}{U}{\sigma}{\asprg}}{\usubstapp{V}{\sigma}{\asfml}} &\\
		\usubstapp{U}{\sigma}{(\asprg \refines \bsprg)} &= \usubstappp{V}{U}{\sigma}{\asprg} \refines \usubstappp{W}{U}{\sigma}{\bsprg} &\\
		\hline
		\usubstappp{U\cup\boundvars{\applyusubst{\sigma}{\ausprg}}}{U}{\sigma}{(\ausprg)} &= \applyusubst{\sigma}{\ausprg} &\\
		\usubstappp{U}{U}{\sigma}{(\ptest{\asfml})} &= \ptest{\usubstapp{U}{\sigma}{\asfml}} &\\
		\usubstappp{U\cup\set{x}}{U}{\sigma}{(\pupdate{\pumod{x}{\astrm}})} &= \pupdate{\pumod{x}{\usubstapp{U}{\sigma}{\astrm}}} &\\
		\usubstappp{U\cup\set{x}}{U}{\sigma}{(\prandom{x})} &= \prandom{x} &\\
		\usubstappp{U\cup\set{x,\D{x}}}{U}{\sigma}{(\pode{\D{x}=\astrm}{\asfml})} &= \pode{\D{x}=\usubstapp{U\cup\set{x,\D{x}}}{\sigma}{\astrm}}{\usubstapp{U\cup\set{x,\D{x}}}{\sigma}{\asfml}} &\\
		\usubstappp{V\cup W}{U}{\sigma}{(\pchoice{\asprg}{\bsprg})} &= \pchoice{\usubstappp{V}{U}{\sigma}{\asprg}}{\usubstappp{W}{U}{\sigma}{\bsprg}} &\\
		\usubstappp{W}{U}{\sigma}{(\asprg;\bsprg)} &= \usubstappp{V}{U}{\sigma}{\asprg};\usubstappp{W}{V}{\sigma}{\bsprg} &\\
		\usubstappp{V}{U}{\sigma}{(\prepeat{\asprg})} &= \prepeat{(\usubstappp{V}{V}{\sigma}{\asprg})} & \text{where }\usubstappp{V}{U}{\sigma}{\asprg} \text{ is defined}
	\end{align*}
	\caption{Recursive application of uniform substitution with input taboos $U\subseteq\allvars$}\label{fig:usubst}
\end{figure}

\rref{fig:usubst} defines the result \(\usubstapp{U}{\sigma}{\asfml}\) of applying a uniform substitution $\sigma$ with taboo set $U\subseteq\allvars$ to a formula $\asfml$ (or term $\astrm$, or hybrid programs $\asprg$ respectively) \cite{DBLP:conf/cade/Platzer19}.
For hybrid programs $\asprg$, the substitution result \(\usubstappp{V}{U}{\sigma}{\asprg}\) for input taboo $U\subseteq\allvars$ also outputs a taboo set $V\subseteq\allvars$, written in subscript notation, that will be tabooed after program $\asprg$.
Taboos $U,V$ are sets of variables that cannot be substituted in free during the application of the substitution, because they have been bound within the context and, thus, potentially changed their meaning compared to the original substitution $\sigma$.
The difference is that the input $U$ is already taboo when the substitution $\sigma$ is applied to $\asprg$ while $V$ is the new output taboo after $\asprg$.
Finally, \(\applyusubst{\sigma}{\asfml}\) is short for \(\usubstapp{\emptyset}{\sigma}{\asfml}\) started without initial taboos.
The key advantage to working with uniform substitution applications with taboo passing is that they enable an efficient one-pass substitution \cite{DBLP:conf/cade/Platzer19} compared to the classical Church-style uniform substitution application mechanism that checks admissibility at every binding operator along the way \cite{DBLP:journals/jar/Platzer17}.
One-pass uniform substitution postpones admissibility checks till the actual substitutions of function and predicate symbols according to explicit taboos carried around.

Despite the surprising definition of the free variables of a refinement, defining uniform substitution for the refinement case is standard, the input taboo $U$ is given to both programs except that their output taboos $V,W$ are discarded: 
\[\usubstapp{U}{\sigma}{(\asprg\refines\bsprg)} = \usubstappp{V}{U}{\sigma}{\asprg}\refines\usubstappp{W}{U}{\sigma}{\bsprg}\]
The reason is two-fold: 
\begin{enumerate}
	\item Unlike quantifiers and modalities, refinements do not subsequently bind any variables.
	\item The free variables of a refinement introduced by a substitution can only be introduced free in the programs, and thus checking these against the input taboo set $U$ is sufficient.
\end{enumerate}
This last statement is a consequence of $\boundvars{\usubstappp{}{}{\sigma}{\asprg}} \subseteq \boundvars{\asprg}$ and $\mustboundvars{\usubstappp{}{}{\sigma}{\asprg}} \supseteq \mustboundvars{\asprg}$, which is proved by a direct induction.

\subsection{Uniform Substitutions and Adjoint Interpretations}

The proof of the soundness of uniform substitution follows the same structure as the proof of the uniform substitution lemma for \dGL \cite{DBLP:conf/cade/Platzer19} but adapted to hybrid programs instead of hybrid games and generalized to the presence of refinements.
The output taboo $V$ of a uniform substitution \(\usubstappp{V}{U}{\sigma}{\asprg}\) will include the original taboo set $U$ and all variables bound in the program $\asprg$.
\begin{lemma}[Taboo set computation \cite{DBLP:conf/cade/Platzer19}]\label{lem:taboo}
	If $\usubstappp{V}{U}{\sigma}{\asprg}$ is defined, then $V \supseteq U \cup \boundvarsdef{\usubstappp{V}{U}{\sigma}{\asprg}}$.
\end{lemma}
Whereas uniform substitutions are syntactic transformations on expressions, their semantic counterparts are semantic transformations on interpretations.
The two are related by Lemmas~\ref{lem:usubst-term} and \ref{lem:usubst-fmlprg}.
Let $\iget[const]{\imodif[const]{\I}{\usarg}{d}}$ denote the interpretation that agrees with interpretation $\iget[const]{\I}$ except for the constant function symbol $\usarg$ which is interpreted as the constant $d\in\R$.
\begin{definition}[Adjoint interpretation]
	For an interpretation $\iget[const]{\I}$ and a state $\iget[state]{\It}$, the \emph{adjoint interpretation} $\iget[const]{\Iadj}$ modifies the interpretation of each function symbol $\oustrm\in\sigma$, predicate symbol $\osfml\in\sigma$ and program constant $\ausprg\in\sigma$ as follows:
	\begin{align*}
		\getInterp{\Iadj}{\oustrm} &: \R \to \R; d \mapsto \ivaluation{\imodif[const]{\It}{\usarg}{d}}{\applysubst{\sigma}{\oustrm[\usarg]}}\\
		\getInterp{\Iadj}{\osfml} &= \set{d\in\R \with \imodels{\imodif[const]{\It}{\usarg}{d}}{\applysubst{\sigma}{\ousfml[\usarg]}}}\\
		\getInterp{\Iadj}{\ausprg} &= \iaccess[\applysubst{\sigma}{\ausprg}]{\It}
	\end{align*}
\end{definition}

\begin{lemma}[Uniform substitution for terms \cite{DBLP:conf/cade/Platzer19}]\label{lem:usubst-term}
	The uniform substitution $\sigma$ for taboo $U \subseteq \allvars$ and its adjoint interpretation $\iget[const]{\Iadj}$ for $\iget[const]{\I},\iget[state]{\It}$ have the same semantics on $U$-variations $\iget[state]{\Iadj}$ of $\iget[state]{\It}$ for all \emph{terms} $\astrm$:
	\begin{align*}
		\ivaluation{\I}{\usubstapp{U}{\sigma}{\astrm}} &= \ivaluation{\Iadj}{\astrm}
	\end{align*}
\end{lemma}

\begin{lemma}[Uniform substitution for formulas, programs]\label{lem:usubst-fmlprg}
	Uniform substitution $\sigma$ for taboo $U \subseteq \allvars$ and its adjoint interpretation $\iget[const]{\Iadj}$ for $\iget[const]{\I},\iget[state]{\It}$ have the same semantics on $U$-variations $\iget[state]{\Iadj}$ of $\iget[state]{\It}$ for all \emph{formulas} $\asfml$ and \emph{hybrid programs} $\asprg$:
	\begin{align*}
		\text{for all }U\text{-variations }\iget[state]{\I}\text{ of }\iget[state]{\It}&:\imodels{\I}{\usubstapp{U}{\sigma}{\asfml}} \text{ iff } \imodels{\Iadj}{\asfml} \\
		\text{for all states }\iget[state]{\Itb}\text{ and all }U\text{-variations }\iget[state]{\I}\text{ of }\iget[state]{\It}&:\iaccessible[\usubstappp{V}{U}{\sigma}{\asprg}]{\I}{\Itb} \text{ iff } \iaccessible[\asprg]{\Iadj}{\Itb}
	\end{align*}
\end{lemma}
The proof is done by simultaneous induction on the structure of $\sigma$, $\asprg$ and $\asfml$ for all $U,\iget[state]{\I},\iget[state]{\It}$ and $\iget[state]{\Itb}$ \seeapp{\rref{app:sec34}}.
The use of $U$-variations is critical when the induction hypothesis needs to be used in a state other than $\iget[state]{\I}$, e.g.\ for quantifiers and modalities.
Without considering the extension of the refinement operator, this result was previously proved in a weaker form ($U = \emptyset$) for \dL \cite{DBLP:journals/jar/Platzer17} or for more complex semantics like hybrid games \cite{DBLP:conf/cade/Platzer19}.

\subsection{Soundness of Uniform Substitution}

\rref{lem:usubst-fmlprg} is essentially all that is required to ensure the sound application of uniform substitution.
First, uniform substitution can be used to have a sound instantiation of the axioms, using the uniform substitution rule (\irref{US}).
A proof rule is \emph{sound} if the validity of the premises implies the validity of the conclusion.

\begin{theorem}[Soundness of uniform substitution \cite{DBLP:conf/cade/Platzer19}]
	The proof rule (\irref{US}) is sound.
	\[\cinferenceRuleQuoteDef{US}\]
\end{theorem}
Uniform substitution can also be used on rules or whole inferences, as long as they are \emph{locally sound}, i.e.\ the conclusion is valid in any interpretation where the premises are valid.
Locally sound inferences are also sound.

\begin{theorem}[Soundness of uniform substitution for rules \cite{DBLP:conf/cade/Platzer19}]
	All locally sound inferences remain locally sound when substituted with a uniform substitution $\sigma$ with taboo set $\allvars$.

	$\linfer{\isfml{1} & \dots & \isfml{n}}{\bsfml}$ locally sound implies $\linfer{\usubstapp{\allvars}{\sigma}{\isfml{1}} & \dots & \usubstapp{\allvars}{\sigma}{\isfml{n}}}{\usubstapp{\allvars}{\sigma}{\bsfml}}$ locally sound.
\end{theorem}

\section{Proof Calculus} \label{sec:dRL-calculus}

Most notably, uniform substitution makes it possible to use concrete \dRL formulas as axioms instead of axiom schemata that accept infinitely many formulas as axioms.
Axioms are finite syntactic objects, and are thus easy to implement, while axiom schemata are ultimately algorithms accepting certain formulas as input while rejecting others \cite{DBLP:journals/jar/Platzer17}.
Figure~\ref{fig:axioms} lists the axioms of \dRL.
\dRL also satisfies the axioms of KAT \cite{DBLP:journals/toplas/Kozen97}, Schematic KAT \cite{10.5555/867190} and the axioms of \dL \seeapp{\rref{app:axioms}}.
Some axioms use the reverse implication $\asfml \lylpmi \bsfml$ instead of $\bsfml \limply \asfml$ for emphasis.

In the axiom (\irref{boxleq}), $\bar{x}$ stands for the (finite) vector of all relevant variables (alternative treatments \cite{DBLP:journals/jar/Platzer17,DBLP:conf/cade/Platzer19} of $p(\bar{x})$ use quantifier symbols or additional program constants instead, but are not necessary for this paper).
This characteristic axiom of \dRL expresses that if formula $\ausfml$ holds after all runs of hybrid program $\busprg$, then it also holds after any refinement $\ausprg$.
Thus, as long as a proof of the refinement is given, it is possible to replace hybrid programs inside modalities.
In general, axioms are meant to be applied to the axiom key (marked \textcolor{vvblue}{blue}).

Refinement is transitive (\irref{leqtrans}), allowing the introduction of intermediate refinements $\cusprg$ similar to the role that cuts play in first-order logic.

Axioms (\irref{choicel}) and (\irref{choicer}) decompose the choice operator using logical connectives.
As the choice $\pchoice{\ausprg}{\busprg}$ can behave like either subprograms, whenever it refines a program $\cusprg$, both $\ausprg$ and $\busprg$ must refine $\cusprg$.
Axiom (\irref{choicer}) is not an equivalence though. \(\ausprg \refines \busprg \lor \ausprg \refines \cusprg\) says that for each initial state, one of the two refinement holds.
However, when $\ausprg$ is nondeterministic, and so can have multiple end states for one initial state, it may not be the case despite the left-hand side being true.

Axiom (\irref{sequence}) helps proving a refinement between two sequences of programs ($\ausprg;\busprg \refines \cusprg;\dusprg$) by proving the refinement of the first programs ($\ausprg \refines \cusprg$) and the refinement of the second programs, but only after all executions of $\ausprg$ ($\dbox{\ausprg}{\busprg \refines \dusprg}$).
Axioms (\irref{testdet}) and (\irref{assigndet}) are particular cases of the axiom (\irref{sequence}) where the implication can be strengthened to an equivalence.
As such, the implication from right to left is not required for both axioms \seeapp{\rref{app:derived}}.
\begin{figure}[t!b!h!]
\begin{minipage}{0.48\textwidth}
	\begin{calculus}
		\cinferenceRule[leqtrans|$\refines_{t}$]{Transitivity of $\refines$}
		{\linferenceRule[lpmi]
		  {\axkey{\ausprg \refines \busprg}}
		  {\ausprg \refines \cusprg \land \cusprg \refines \busprg}
		}{}
		\cinferenceRule[leqantisym|$\prgeq$]{Antisymmetry of $\refines$}
		{\linferenceRule[equiv]
		  {\ausprg \refines \busprg \land \busprg \refines \ausprg}
		  {\axkey{\ausprg \prgeq \busprg}}
		}{}
		\cinferenceRule[boxleq|{[$\refines$]}]{Box refinement}
		{\linferenceRule[impl]
		  {\ausprg \refines \busprg}
		  {(\axkey{\dbox{\ausprg}\ausfml}\lylpmi\dbox{\busprg}{\ausfml})}
		}{}
		\cinferenceRule[test|?]{test axiom}
		{\linferenceRule[equiv]
		{(\osfml \limply \osfmlb)}
		{(\axkey{\ptest{\osfml} \refines  {\ptest{\osfmlb}}})}
		}{}
		
		\cinferenceRule[update|${:}{=}$]{Update axiom}
		{\linferenceRule[eq]
			{\prandom{x};\ptest{x = \oustrm}}
			{\pupdate{\pumod{x}{\oustrm}}}
		}{}
		
		\cinferenceRule[testdet|?$_{\text{det}}$]{deterministic test}
		{\linferenceRule[equiv]
		{\dbox{\ptest{\osfml}}{\ausprg \leq \busprg}}
		{\axkey{\ptest{\osfml};\ausprg \leq \ptest{\osfml};\busprg}}
		}{}
		\cinferenceRule[stutter|stutter]{stuttering}
		{\linferenceRule[eq]
		{{\ptest{\ltrue}}}
		{\axkey{\pupdate{\pumod{x}{x}}}}
		}{}
	\end{calculus}		
\end{minipage}%
\hfill%
\begin{minipage}{0.52\textwidth}
	\begin{calculus}
		\cinferenceRule[choicel|$\cup_l$]{choice left axiom}
		{\linferenceRule[equiv]
		{\ausprg \refines \cusprg \land \busprg \refines \cusprg}
		{\axkey{\pchoice{\ausprg}{\busprg} \refines \cusprg}}
		}{}
		
		\cinferenceRule[choicer|$\cup_r$]{choice right axiom}
		{\linferenceRule[lpmi]
		{\axkey{\ausprg \refines \pchoice{\busprg}{\cusprg}}}
		{\ausprg \refines \busprg \lor \ausprg \refines \cusprg}
		}{}
		
		\cinferenceRule[sequence|;]{sequence axiom}
		{\linferenceRule[lpmi]
		{\axkey{\ausprg;\busprg \refines \cusprg;\dusprg}}
		{\ausprg \refines \cusprg \land \dbox{\ausprg}{\busprg \refines \dusprg}}
		}{}
		
		\cinferenceRule[loopl|$\text{loop}_l$]{loop left axiom}
		{\linferenceRule[lpmi]
		{\axkey{\prepeat{\ausprg};\busprg \refines \busprg}}
		{\dbox{\prepeat{\ausprg}}{\ausprg;\busprg \refines \busprg}}
		}{}
		
		\cinferenceRule[loopr|$\text{loop}_r$]{loop right axiom}
		{\linferenceRule[lpmi]
		{\axkey{\ausprg;\prepeat{\busprg} \refines \ausprg}}
		{\ausprg;\busprg \refines \ausprg}
		}{}

		\cinferenceRule[unloop|$\text{unloop}$]{unloop axiom}
		{\linferenceRule[lpmi]
			{\axkey{\prepeat{\ausprg}\refines\prepeat{\busprg}}}
			{\dbox{\prepeat{\ausprg}}(\ausprg\refines\busprg)}
		}{}

		\cinferenceRule[assigndet|$:=_{\text{det}}$]{deterministic assignment}
		{\linferenceRule[equiv]
		{\dbox{\pupdate{\pumod{x}{\oustrm}}}{\ausprg \leq \busprg}}
		{\axkey{\pupdate{\pumod{x}{\oustrm}};\ausprg \leq \pupdate{\pumod{x}{\oustrm}};\busprg}}
		}{}	

	\end{calculus}		
\end{minipage}
\begin{calculus}

		\cinferenceRule[randtestmerge|${:}*_{\text{merge}}$]{Merging randoms with test}
		{\linferenceRule[eq]
			{\prandom{x};\ptest{\lexists y \ousfml[y]}}
			{\axkey{\prandom{x};\ptest{\ousfml[x]};\prandom{x}}}
		}{}

\cinferenceRule[randtestassignmerge|${:}{=}*_{\text{merge}}$]{Merging random and assignement with test}
{\linferenceRule[eq]
	{\prandom{x};\ptest{\lexists y (\ousfml[y] \land x = \oustrm[y])}}
	{\axkey{\prandom{x};\ptest{\ousfml[x]};\pupdate{\umod{x}{\oustrm[x]}}}}
}{}
	
\cinferenceRule[ode|ODE]{ODE axiom}
{\linferenceRule[equiv]
  {\dbox{\pode{\D{x}=\oustrm[x]}{\ousfml[x]}}{(\D{x}=\oustrmb[x] \land \ousfmlb[x])}}
  {\axkey{\pode{\D{x}=\oustrm[x]}{\ousfml[x]}\refines\pode{\D{x}=\oustrmb[x]}{\ousfmlb[x]}}}
}{}

\cinferenceRule[DWref|$\text{DW}_{\prgeq}$]{DW axiom}
{\linferenceRule[eq]
  {{\ptest{\ousfml[x]};\pode{\D{x}=\oustrm[x]}{\ousfml[x]};\ptest{\ousfml[x]}}}
  {\axkey{\pode{\D{x}=\oustrm[x]}{\ousfml[x]}}}
}{}

\cinferenceRule[DEref|$\text{DE}_{\prgeq}$]{DE axiom}
{\linferenceRule[eq]
  {\pode{\D{x}=\oustrm[x]}{\ousfml[x]};\pupdate{\umod{\D{x}}{\oustrm[x]}}}
  {\axkey{\pode{\D{x}=\oustrm[x]}{\ousfml[x]}}}
}{}

\cinferenceRule[DX|DX]{DX axiom}
{\linferenceRule[leq]
  {\axkey{\pode{\D{x}=\oustrm[x]}{\ousfml[x]}}}
  {\pupdate{\umod{\D{x}}{\oustrm[x]}};\ptest{\ousfml[x]}}
}{}

\cinferenceRule[ODEidemp|ODE$_{\text{idemp}}$]{ODE idempotent axiom}
{\linferenceRule[eq]
  {\pode{\D{x}=\oustrm[x]}{\ousfml[x]}}
  {\axkey{\pode{\D{x}=\oustrm[x]}{\ousfml[x]};\pode{\D{x}=\oustrm[x]}{\ousfml[x]}}}
}{}

\end{calculus}

\caption{Axioms of \dRL}\label{fig:axioms}
\end{figure}

Axioms (\irref{loopl}), (\irref{loopr}) and (\irref{unloop}) are used to prove refinements of loops. The first two state that if adding a program before or after only leads to less executions, then adding an unbounded number of executions, i.e.\ a loop, will also lead to less executions. The axiom (\irref{unloop}) is useful for comparing two loops, as it allows to reduce the problem to comparing the loop bodies. Both axioms (\irref{loopl}) and (\irref{unloop}) need a box modality when proving the refinement of the loop body, as the refinement must be proved after any number of iterations of $a$.

The axiom (\irref{ode}) describes how to prove refinements between differential equations.
A refinement $\pode{\D{x}=\oustrm[x]}{\ousfml[x]} \refines \pode{\D{x}=\oustrmb[x]}{\ousfmlb[x]}$ is true iff throughout the execution of the former ODE, it always satisfies the latter differential equation and evolution domain.
Along with the axioms (\irref{DWref}) and (\irref{DEref}), these axioms subsume differential cut (\irref{DCax}), differential weakening (\irref{diffweakenax}) and differential effect (\irref{DEax}) from \dL \seeapp{\rref{app:axioms}}.
The equivalence in the axiom (\irref{ode}) effectively means that refinements of differential equations can \emph{always} be reduced to standard \dL formulas, which is essential to our decidability result.

The axiom (\irref{DX}) states that a differential equation always has a solution for the interval $[0,0]$.
In that case, the execution succeeds only if the domain holds, and the correct value $\oustrm[x]$ is assigned to the differential variable $\D{x}$.
The axiom (\irref{ODEidemp}) states that following the same differential equation twice in a row is equivalent to following it only once, because the concatenation of solutions of the same differential equation is still a solution of the same differential equation.

Compared to the original sequent calculus for \dRL \cite{DBLP:conf/lics/LoosP16}, the proof rule schemata matching infinitely many instances are now replaced by a \emph{finite} number of axioms that are concrete \dRL formulas rather than standing for infinitely many instances.
The infinitely many possible instances can then be recovered soundly using the uniform substitution rule (\irref{US}).
Because of this two-step mechanism, reasoning with the axioms can be done without considering the possible instantiations.
Take for instance the sound equivalence ${\pupdate{\umod{x}{\oustrm}};\prandom{x}} \prgeq {\prandom{x}}$.
The proof can be done by transitivity (\irref{leqtrans}) with $\prandom{x};\ptest{x = \oustrm};\prandom{x}$ as intermediate step \seeapp{\rref{app:derived}}.
But the same proof cannot be done by replacing $\oustrm$ by any term $\astrm$: the intermediate program is not always equivalent to the other two (e.g.\ for $\astrm = x+1$).
On the other hand, by proving the equivalence for $\oustrm$ and then using rule (\irref{US}), the equivalence can be proved for all terms $\astrm$.

The \dRL axioms are also more modular than its cast-in-stone sequent calculus rules.
For instance, with rule (\irref{Gax}) and axiom (\irref{Kax}), any implication $\asfml \limply \bsfml$, e.g.\ (\irref{choicer}), can be used to prove $\dbox{\ausprg}{\asfml} \limply \dbox{\ausprg}{\bsfml}$.
This would not fit the shape of the corresponding sequent rule, which requires $\bsfml$ at the top level.
The lack of differential symbols in the original sequent calculus \cite{DBLP:conf/lics/LoosP16} changes the soundness of some rules: the match direction field rule (MDF) would allow rescaling the right-hand side of a differential equation, which is unsound here as it would change the resulting differential symbols.
Conversely, only the reverse implication of the axiom (\irref{ode}) would be sound in the original calculus, again for lack of differential symbols.
The \dRL axioms are proved sound \seeapp{\rref{app:proofaxioms}}:

\begin{theorem}[Soundness of \dRL axioms]\label{thm:soundness}
	All axioms of \dRL are sound.
\end{theorem}

\section{Decidability of Refinement for a Fragment of \dRL}\label{sec:decidable}

This section identifies a subset of hybrid programs for which the refinement problem is decidable.
It is focused on concrete programs, i.e.\ programs without function symbols, predicate symbols or program constants.
They have the following high-level structure: $\prepeat{(ctrl;plant)}$ where a discrete, loop-free program $ctrl$, modelling a controller that sets some parameters $\bar{u}$, then a continuous program $plant$ that describes the dynamics of the variables $\bar{y}$ according to the choice of the parameters $\bar{u}$.
These steps are then repeated nondeterministically.
The continuous variables $\bar{y}$ (and by extension $\D{\bar{y}}$) are expected to be distinct from the discrete variables $\bar{u}$ and also contain a global clock $t$ which follows the differential equation $\D{t}=1$.
The presence of the clock $t$ is not needed for comparing the differential equations, but to distinguish between discrete executions and hybrid executions.

For two such programs, $\prepeat{(ctrl_a;plant_a)}$ and $\prepeat{(ctrl_b;plant_b)}$, a canonical proof of the refinement has the following shape (omitting uses of \irref{MPax} for brevity):
\[\linfer[unloop]{
	\linfer[Gax]{
		\linfer[sequence]{
			\linfer{\dots}{ctrl_a \refines ctrl_b}
			& \linfer{\dots}{\dbox{ctrl_a}(plant_a \refines plant_b)}}
		{ctrl_a;plant_a\refines ctrl_b;plant_b}}
	{\dbox{\prepeat{(ctrl_a;plant_a)}}(ctrl_a;plant_a\refines ctrl_b;plant_b)}}
{\prepeat{(ctrl_a;plant_a)} \refines \prepeat{(ctrl_b;plant_b)}}\]

This means that proving the refinement of the whole programs is reduced to proving the refinement of the controllers, $ctrl_a \refines ctrl_b$ and the refinement of the plants after all $ctrl_a$ executions, $\dbox{ctrl_a}(plant_a \refines plant_b)$.
With our restrictions on the controllers, the first refinement is always decidable.
\begin{lemma}\label{lem:decidiscret}
	For concrete, discrete and loop-free controllers $ctrl_a$ and $ctrl_b$, the validity of $ctrl_a \refines ctrl_b$ is decidable by \dRL proof.
\end{lemma}
Given a controller $ctrl_a$, it is possible to synthesize a first-order formula $\asfml_a(x,x^+)$ that characterizes the behavior of $ctrl_a$, where $x$ (resp.\ $x^+$) corresponds to the variables after (resp.\ before) the controller \cite{DBLP:journals/fmsd/MitschP16}.
Using the \dRL axioms, $ctrl_a \refines ctrl_b$ is provable from $\asfml_a(x,x^+) \limply \asfml_b(x,x^+)$. 
The validity of the latter is decidable as it is first-order real arithmetic \cite{TarskiMcKinsey+1951}.
The full proof is in \rref{app:sec6}.

The second refinement, $\dbox{ctrl_a}(plant_a \refines plant_b)$, is more complex.
Let us write the two plants as $plant_a \mequiv \pode{\D{\bar{y}}=p(\bar{y},\bar{u})}{Q}$ and $plant_b \mequiv \pode{\D{\bar{y}}=q(\bar{y},\bar{u})}{R}$ for some polynomials $p(\bar{y},\bar{u}), q(\bar{y},\bar{u})$ and formulas $Q,R$.
The axiom \irref{ode} entails that we must prove $\dbox{ctrl_a}\dbox{plant_a}(p(\bar{y},\bar{u}) = q(\bar{y},\bar{u}) \land R)$, which no longer contains any refinement.
For the decidability result (\rref{thm:deciderefine}) to hold, we require that the validity of this formula is decidable.

There are two cases which always ensure this.
First, if the differential equation $plant_a$ admits a solution expressible in \dRL (e.g.\ a polynomial), then using standard \dL reasoning, the formula can be reduced to a first-order formula and thus its validity can be decided.
The differential equation from \rref{ex:carbrake}, $\D{x} = v,\D{v} = a$, is such a case.

The second case is when domain $R$ is algebraic, i.e.\ of the form \(\landfold_i\lorfold_j p_{ij}(x) = 0\) for some polynomial $p_{ij}$
and $Q$, the domain of $plant_a$, is a semialgebraic set \cite{DBLP:journals/jacm/PlatzerT20}.

The remaining question is now to show that the approach presented above is complete, meaning it always succeeds when the refinement holds.
The only additional constraint we require is that the controller $ctrl_b$ is idempotent.
\begin{definition}[Idempotent controller]
	A controller $ctrl$ is \emph{idempotent} if it satisfies $ctrl;ctrl \prgeq ctrl$.
\end{definition}
An idempotent controller cannot reach more states by executing multiple times without any continuous dynamics happening.
Pure reactive controllers, i.e.\ controllers for which the parameters' values only depend on the values of the continuous variables, are always idempotent.
This is the case for the controllers in \rref{ex:carbrake}: $\pchoice{\pupdate{\pumod{x}{-B}}}{\ptest{\text{safe}_T(x)};\pupdate{\pumod{x}{A}}}$.
On the other hand, counting the number of times the controller has been executed would not be idempotent.

\begin{lemma}\label{lem:refineidemp}
This derived rule is invertible, if $ctrl_b$ is idempotent.
\[\linfer{ctrl_a;plant_a\refines ctrl_b;plant_b}{\prepeat{(ctrl_a;plant_a)} \refines \prepeat{(ctrl_b;plant_b)}}\]
\end{lemma}
The derivation of the rule is given in the canonical proof.
The converse, that the conclusion implies the premise, is more involved \seeapp{\rref{app:sec6}}.
Proving $ctrl_a;plant_a \refines \prepeat{(ctrl_b;plant_b)}$ from $\prepeat{(ctrl_a;plant_a)} \refines \prepeat{(ctrl_b;plant_b)}$ is done by unfolding the loop on the left.
To get rid of the loop on the right, we use the fact that $ctrl_b$ is idempotent.
It means that if the global time is not modified, then we can assume without loss of generality that the controller (and thus also the plant) is executed only once.
The case when the global time is modified additionally considers the value of the derivative to ensure that there is an execution of the right program that does not require looping.

With the above lemma, we can now state the decidability result. 

\begin{theorem}[Decidability of refinement for idempotent controllers]\label{thm:deciderefine}
	For concrete hybrid programs $ctrl_a;plant_a$ and $ctrl_b;plant_b$ discrete loop-free $ctrl_a, ctrl_b$ and with $plant_a \mequiv \pode{\D{\bar{y}}=p(\bar{y},\bar{u})}{Q}$ and $plant_b \mequiv \pode{\D{\bar{y}}=q(\bar{y},\bar{u})}{R}$, if $ctrl_b$ is idempotent, and the validity of
	$\dbox{ctrl_a}\dbox{plant_a}{(p(\bar{y},\bar{u}) = q(\bar{y},\bar{u}) \land R)}$ is decidable,
	then the validity of $\prepeat{(ctrl_a;plant_a)} \refines \prepeat{(ctrl_b;plant_b)}$ is also decidable.
\end{theorem}

In particular, the theorem applies to the event-triggered model and the time-triggered model templates used to show how to prove that the latter refines the former \cite{DBLP:conf/lics/LoosP16}. Indeed, their controller template is loop-free and idempotent and the differential equation are assumed to be solvable. \rref{thm:deciderefine} strengthens their result by showing the completeness of the approach.

\section{Conclusion}
This paper introduced a uniform substitution proof calculus for differential refinement logic \dRL.
This yields a parsimonious prover microkernel for hybrid systems verification that simultaneously works for properties of and relations between hybrid systems.
The handling of refinement relations between hybrid systems is subtle even only in its static semantics, which makes the correctness proofs of this paper particularly interesting.
The uniform substitution is one-pass \cite{DBLP:conf/cade/Platzer19} giving it respectable performance advantages compared to Church-style uniform substitutions.
While the joint presence of differential equations reasoning and refinement reasoning causes challenges, a resulting benefit besides soundness is that a finer notion of differential equation refinement is obtained with logical decidability properties on a fragment of hybrid systems refinements.

Future work involves improving the implementation of the uniform substitution calculus in \KeYmaeraX.
Although the prover microkernel was straightforward following the uniform substitution process and list of \dRL's uniform substitution axioms, the prover would benefit from quality of life features, e.g.\ using the axioms to rewrite on subprograms, and an implementation of the refinement decision algorithm for the decidable fragment.
Another axis of research is to combine refinements with hybrid games, with a proper semantics and adapt the new axioms of \dRL to games, some of which would not be sound as is.

\renewcommand{\doi}[1]{doi: \href{https://doi.org/#1}{\nolinkurl{#1}}}
\bibliographystyle{splncs04}
\bibliography{platzer,dRL}

\clearpage
\appendix
\section{Additional \dRL Axioms}\label{app:axioms}
\vspace*{-0.1cm}
\begin{figure}[t!bhp]
	\begin{subfigure}{\textwidth}
	\begin{minipage}{0.6\textwidth}
	\begin{calculus}
		\cinferenceRuleQuoteDef{testbax}
		\cinferenceRuleQuoteDef{assignbax}
		\derived{\cinferenceRule[assignbeqax|$\dibox{:=}_=$]{assignment equational axiom}
		{\linferenceRule[equiv]
		{\lforall x (x = \aconst \limply p(x))}
		{\axkey{\dbox{\pupdate{\pumod{x}{\aconst}}}p(x)}}
		}{}}
		\cinferenceRuleQuoteDef{randomb}
		\cinferenceRuleQuoteDef{choicebax}
		*\cinferenceRuleQuoteDef{iteratebax}
		\cinferenceRuleQuoteDef{Kax}
		\cinferenceRuleQuoteDef{Ieqax}
			
	\end{calculus}
	\end{minipage}
	\begin{minipage}{0.4\textwidth}
	\begin{calculus}
		\cinferenceRuleQuoteDef{composebax}
		\cinferenceRuleQuoteDef{Vax}
		\cinferenceRuleQuoteDef{Gax}
		\cinferenceRuleQuoteDef{genaax}
		\cinferenceRuleQuoteDef{MPax}
	\end{calculus}
	\end{minipage}
	\begin{calculus}
		\iflongversion
		\cinferenceRuleQuoteDef{Dconst}
		\cinferenceRuleQuoteDef{Dvar}
		\cinferenceRuleQuoteDef{Dplusax}
		\cinferenceRuleQuoteDef{Dminusax}
		\cinferenceRuleQuoteDef{Dtimesax}
		\else
		\fi
		*\cinferenceRuleQuoteDef{DEax}
		*\cinferenceRuleQuoteDef{diffweakenax}
		\cinferenceRuleQuoteDef{DIax}
		*\cinferenceRuleQuoteDef{DCax}
		\cinferenceRuleQuoteDef{DGax}
		\cinferenceRuleQuoteDef{DSax}
	\end{calculus}
	\caption{Axioms of \dL}\label{fig:dLaxioms}
	\end{subfigure}
\iflongversion
\caption{Additional axioms of \dRL}
\end{figure}
\begin{figure}
	\ContinuedFloat
\fi

	\begin{subfigure}{\textwidth}
	\begin{minipage}{0.6\textwidth}
	\begin{calculus}
		\cinferenceRule[leqrefl|$\refines_{\text{refl}}$]{Reflexivity of $\refines$}
		{\linferenceRule[leq]
			{\ausprg}
			{\ausprg}
		}{}
		\derived{\cinferenceRule[cupidemp|$\cup_{\text{idemp}}$]{Idempotence of $\cup$}
		{\linferenceRule[eq]
		  {\ausprg}
		  {\axkey{\pchoice{\ausprg}{\ausprg}}}
		}{}}
		\derived{\cinferenceRule[cupassoc|$\cup_{\text{assoc}}$]{Associativity of $\cup$}
		{\linferenceRule[eq]
		  {\pchoice{\ausprg}{(\pchoice{\busprg}{\cusprg})}}
		  {\pchoice{(\pchoice{\ausprg}{\busprg})}{\cusprg}}
		}{}}
		\cinferenceRule[seqidl|$;_{\text{id-l}}$]{Left neutral of $;$}
		{\linferenceRule[eq]
		  {\ausprg}
		  {\axkey{\ptest{\ltrue};\ausprg}}
		}{}
		\cinferenceRule[seqdistl|dist-l]{Left distributivity of $;$}
		{\linferenceRule[eq]
		  {\ausprg;(\pchoice{\busprg}{\cusprg})}
		  {\pchoice{(\ausprg;\busprg)}{(\ausprg;\cusprg)}}
		}{}
		\cinferenceRule[seqannih-l|annih-l]{$;$ left annihilator}
		{\linferenceRule[eq]
		  {\ptest{\lfalse}}
		  {\axkey{\ptest{\lfalse};\ausprg}}
		}{}
		\derived{\cinferenceRule[testand|$?_{\text{and}}$]{Test and}
		{\linferenceRule[eq]
			{\ptest{\osfml \land \osfmlb}}
			{\axkey{\ptest{\osfml};\ptest{\osfmlb}}}
		}{}}
	
		\cinferenceRule[unfold-l|unfold-l]{Left unfolding of $\prepeat{\ausprg}$}
		{\linferenceRule[eq]
		  {\prepeat{\ausprg}}
		  {\pchoice{\ptest{\ltrue}}{(\ausprg;\prepeat{\ausprg})}}
		}{}
	\end{calculus}
	\end{minipage}
	\begin{minipage}{0.4\textwidth}
	\begin{calculus}
		\cinferenceRule[cupid|$\cup_{\text{id}}$]{Neutral of $\cup$}
		{\linferenceRule[eq]
		  {\ausprg}
		  {\axkey{\pchoice{\ausprg}{\ptest{\lfalse}}}}
		}{}
		\derived{\cinferenceRule[cupcomm|$\cup_{\text{comm}}$]{Commutativity of $\cup$}
		{\linferenceRule[eq]
		  {\pchoice{\ausprg}{\busprg}}
		  {\pchoice{\busprg}{\ausprg}}
		}{}}
		\cinferenceRule[seqassoc|$;_{\text{assoc}}$]{Associativity of $;$}
		{\linferenceRule[eq]
		  {(\ausprg;\busprg);\cusprg}
		  {\ausprg;(\busprg;\cusprg)}
		}{}
		\cinferenceRule[seqidr|$;_{\text{id-r}}$]{Right neutral of $;$}
		{\linferenceRule[eq]
		  {\ausprg}
		  {\axkey{\ausprg;\ptest{\ltrue}}}
		}{}
		\derived{\cinferenceRule[seqdistr|dist-r]{Right distributivity of $;$}
		{\linferenceRule[eq]
		  {(\pchoice{\ausprg}{\busprg});\cusprg}
		  {\pchoice{(\ausprg;\cusprg)}{(\busprg;\cusprg)}}
		}{}}
		\cinferenceRule[seqannih-r|annih-r]{$;$ right annihilator}
		{\linferenceRule[eq]
		  {\ptest{\lfalse}}
		  {\axkey{\ausprg;\ptest{\lfalse}}}
		}{}
		\derived{\cinferenceRule[testor|$?_{\text{or}}$]{Test or}
		{\linferenceRule[eq]
			{\ptest{\osfml \lor \osfmlb}}
			{\axkey{\pchoice{\ptest{\osfml}}{\ptest{\osfmlb}}}}
		}{}}
		\cinferenceRule[unfold-r|unfold-r]{Right unfolding of $\prepeat{\ausprg}$}
		{\linferenceRule[eq]
		  {\prepeat{\ausprg}}
		  {\pchoice{\ptest{\ltrue}}{(\prepeat{\ausprg};\ausprg)}}
		}{}
	\end{calculus}
	\end{minipage}
	\caption{Axioms of KAT}\label{fig:KATaxioms}
	\end{subfigure}

	\begin{subfigure}{\textwidth}
	\begin{minipage}{0.6\textwidth}
	\begin{calculus}
		\cinferenceRule[randcomm|${:}{}*_{\text{comm}}$]{Commutativity of randoms}
		{\linferenceRule[eq]
		  {\prandom{y};\prandom{x}}
		  {\prandom{x};\prandom{y}}
		}{}
		\derived{\cinferenceRule[assignsub|$:=_{\text{sub}}$]{Substitution of assignments}
		{\linferenceRule[eq]
		{\pumod{x}{\oustrm};\pumod{y}{\oustrmb[\oustrm]}}
		{\pumod{x}{\oustrm};\pumod{y}{\oustrmb[x]}}
		}{}}
		\derived{\cinferenceRule[assigntest|$:=_{\text{test}}$]{Commutativity of assignment and test}
		{\linferenceRule[eq]
		{\ptest{\ousfml[\oustrm]};\pumod{x}{\oustrm}}
		{\pumod{x}{\oustrm};\ptest{\ousfml[x]}}
		}{}}
		\derived{\cinferenceRule[assignmerge|$:=_{\text{merge}}$]{Merging assignments}
		{\linferenceRule[eq]
		{\pumod{x}{\oustrmb[\oustrm]}}
		{\axkey{\pumod{x}{\oustrm};\pumod{x}{\oustrmb[x]}}}
		}{}}
	\end{calculus}		
	\end{minipage}
	\begin{minipage}{0.4\textwidth}
	\begin{calculus}
		\derived{\cinferenceRule[assigncomm|$:=_{\text{comm}}$]{Commutativity of assignments}
		{\linferenceRule[eq]
		{\pumod{y}{\oustrmb[\oustrm]};\pumod{x}{\oustrm}}
		{\pumod{x}{\oustrm};\pumod{y}{\oustrmb[x]}}
		}{}}
	\derived{\cinferenceRule[assignrandcomm|${:}{=}*_{\text{comm}}$]{Commutativity of assignement and random}
	{\linferenceRule[eq]
	{\prandom{y};\pupdate{\pumod{x}{\oustrm}}}
	{\pupdate{\pumod{x}{\oustrm}};\prandom{y}}
	}{}}
	\cinferenceRule[randtest|${:}*_{\text{test}}$]{Commutativity of random and test}
	{\linferenceRule[eq]
	{\ptest{\osfml};\prandom{x}}
	{\prandom{x};\ptest{\osfml}}
	}{}
	\end{calculus}		
	\end{minipage}
	\caption{Axioms of SKAT with nondeterministic assignment}\label{fig:SKATaxioms}
\end{subfigure}
\caption{Additional axioms of \dRL}
\end{figure}
\iflongversion
\else
\noindent
Axioms of \dL also include the differential axioms, e.g.\ $\D{(x)} = \D{x}$ \cite{DBLP:journals/jar/Platzer17}, to reason on terms, which are omitted as it is not the main focus of this paper.
\fi
\noindent
Axioms preceded by a star can be derived from other axioms \seeapp{\rref{app:derived}}.
\iflongversion
\newpage
\section{Syntactic Static Semantics}\label{app:static}
	The (syntactically) \emph{bound variables} of a hybrid program $\asprg$ are defined as follows:
	\begin{align*}
		\boundvars{\ausprg} &= \allvars\\
		\boundvars{\ptest{\asfml}} &= \emptyset\\
		\boundvars{\pupdate{\pumod{x}{\astrm}}} &= \set{x} \\
		\boundvars{\prandom{x}} &= \set{x} \\
		\boundvars{\pode{\D{x}=\astrm}{\asfml}} &= \set{x,\D{x}} \\
		\boundvars{\pchoice{\asprg}{\bsprg}} &= \boundvars{\asprg} \cup \boundvars{\bsprg}\\
		\boundvars{\asprg;\bsprg} &= \boundvars{\asprg} \cup \boundvars{\bsprg} \\
		\boundvars{\prepeat{\asprg}} &= \boundvars{\asprg}
	\end{align*}
	The definition of free variables for formulas requires the definition of \emph{must-bound variables} for a hybrid program $\asprg$ which are written on all runs of $\asprg$:
	\begin{align*}
		\mustboundvars{\ausprg} &= \emptyset\\
		\mustboundvars{\ptest{\asfml}} &= \emptyset\\
		\mustboundvars{\pupdate{\pumod{x}{\astrm}}} &= \set{x} \\
		\mustboundvars{\prandom{x}} &= \set{x} \\
		\mustboundvars{\pode{\D{x}=\astrm}{\asfml}} &= \set{x,\D{x}}\\
		\mustboundvars{\pchoice{\asprg}{\bsprg}} &= \mustboundvars{\asprg} \cap \mustboundvars{\bsprg}\\
		\mustboundvars{\asprg;\bsprg} &= \mustboundvars{\asprg} \cup \mustboundvars{\bsprg} \\
		\mustboundvars{\prepeat{\asprg}} &= \emptyset
	\end{align*}
	The (syntactically) \emph{free variables for terms} are defined as follows where \(\D{U} \mdefeq \{\D{x} \with x\in U\}\) are the differential variables corresponding to a set $U$ of variables:
	\begin{align*}
		\freevars{x} &= \set{x}\\
		\freevars{\oustrm[\istrm{1},\dots,\istrm{n}]} &= \freevars{\istrm{1}} \cup \dots \cup \freevars{\istrm{n}}\\
		\freevars{\astrm + \bstrm} &= \freevars{\astrm} \cup \freevars{\bstrm}\\
		\freevars{\astrm \cdot \bstrm} &= \freevars{\astrm} \cup \freevars{\bstrm}\\
		\freevars{\D{\astrm}} &= \freevars{\astrm} \cup \D{\freevars{\astrm}}
	\end{align*}
	The (syntactically) \emph{free variables for formulas} are defined as follows:
	\begin{align*}		
		\freevars{\astrm \leq \bstrm} &= \freevars{\astrm} \cup \freevars{\bstrm}\\
		\freevars{\ousfml[\istrm{1},\dots,\istrm{n}]} &= \freevars{\istrm{1}} \cup \dots \cup \freevars{\istrm{n}}\\
		\freevars{\lnot \asfml} &= \freevars{\asfml}\\
		\freevars{\asfml \land \bsfml} &= \freevars{\asfml} \cup \freevars{\bsfml}\\
		\freevars{\lforall{x}{\asfml}} &= \freevars{\asfml} \setminus \set{x}\\
		\freevars{\dbox{\asprg}{\asfml}} &= \freevars{\asprg} \cup (\freevars{\asfml} \setminus \mustboundvars{\asprg})\\
		\freevars{\asprg \refines \bsprg} &= \freevars{\asprg}\cup\freevars{\bsprg}\cup((\boundvars{\asprg}\cup\boundvars{\bsprg})\setminus(\mustboundvars{\asprg}\cap\mustboundvars{\bsprg}))
	\end{align*}
	The (syntactically) \emph{free variables for hybrid programs} are defined as follows:
	\begin{align*}
		\freevars{\ausprg} &= \allvars\\
		\freevars{\ptest{\asfml}} &= \freevars{\asfml}\\
		\freevars{\pupdate{\pumod{x}{\astrm}}} &= \freevars{\astrm} \\
		\freevars{\prandom{x}} &= \set{x} \\
		\freevars{\pode{\D{x}=\astrm}{\asfml}} &= \set{x} \cup \freevars{\astrm} \cup \freevars{\asfml} \\
		\freevars{\pchoice{\asprg}{\bsprg}} &= \freevars{\asprg} \cup \freevars{\bsprg} \\
		\freevars{\ausprg;\busprg} &= \freevars{\asprg} \cup (\freevars{\bsprg} \setminus \mustboundvars{\asprg})\\
		\freevars{\prepeat{\asprg}} &= \freevars{\asprg}
	\end{align*}

\section{Proofs for \rref{sec:dRL} and \rref{sec:usubst}}\label{app:sec34}
\begin{proof}[\rref{lem:static}]
	We focus on the proof of $\freevars{\cdot}\supseteq\freevarsdef{\cdot}$ for formulas and hybrid programs, because it is the only case affected by the addition of refinement operators compared to prior proofs \cite[Lem.\,17]{DBLP:journals/jar/Platzer17}.
	We reason by induction on the structure of the formulas and hybrid programs.
	For hybrid programs, the property shown for $\freevars{\asprg}$ is stronger than the coincidence property, it shows that if $\iget[state]{\I} = \iget[state]{\J}$ on $V\supseteq\freevarsdef{\asprg}$ and $\iget[const]{\I} = \iget[const]{\J}$ on $\intsigns{\asprg}$, then $\iaccessible[\asprg]{\I}{\It}$ implies $\iaccessible[\asprg]{\J}{\Jb}$ for some $\iget[state]{\Jb}$ with $\iget[state]{\It} = \iget[state]{\Jb}$ on $V\cup\mustboundvars{\asprg}$.

	We show the case for the refinement operator $\asprg\refines\bsprg$.
	Take $\iget[state]{\I},\iget[state]{\J},\iget[const]{\I},\iget[const]{\J}$ such that $\iget[state]{\I} = \iget[state]{\J}$ on $V\mdefeq\freevars{\asprg\refines\bsprg}$ and $\iget[const]{\I} = \iget[const]{\J}$ on $\intsigns{\asprg\refines\bsprg}$.

	We prove that $\imodels{\J}{\asprg\refines\bsprg}$ implies $\imodels{\I}{\asprg\refines\bsprg}$.
	For $\iaccessible[\asprg]{\I}{\It}$, by induction hypothesis, we have $\iaccessible[\asprg]{\J}{\Jb}$ for some $\iget[state]{\Jb}$ with $\iget[state]{\It} = \iget[state]{\Jb}$ on $V\cup\mustboundvars{\asprg}$.
	By definition of the refinement, $\iaccessible[\bsprg]{\J}{\Jb}$ follows from $\imodels{\J}{\asprg\refines\bsprg}$ and $\iaccessible[\asprg]{\J}{\Jb}$.
	Similarly, we have $\iaccessible[\bsprg]{\I}{\Itb}$ for some state $\iget[state]{\Itb}$ with $\iget[state]{\Itb} = \iget[state]{\Jb}$ on $V\cup\mustboundvars{\bsprg}$. Thus, $\iget[state]{\Itb} = \iget[state]{\It}$ on $V\cup(\mustboundvars{\asprg}\cap\mustboundvars{\bsprg})$.
	Additionally, by \rref{lem:boundeffect}, we also have $\iget[state]{\Itb} = \iget[state]{\I} = \iget[state]{\It}$ on $\scomplement{\boundvars{\asprg}}$. This means that in fact $\iget[state]{\Itb} = \iget[state]{\It}$ and thus $\iaccessible[\bsprg]{\I}{\It}$.

	By a symmetric argument, we have $\imodels{\I}{\asprg\refines\bsprg}$ iff $\imodels{\J}{\asprg\refines\bsprg}$.
	This means that $\freevars{\asprg\refines\bsprg}$ satisfies the coincidence property for $\asprg\refines\bsprg$ and thus $\freevars{\asprg\refines\bsprg} \supseteq \freevarsdef{\asprg\refines\bsprg}$ by \rref{lem:coincidence}.
\qed
\end{proof}
\begin{proof}[\rref{lem:usubst-fmlprg}]
	We prove the lemma by simultaneous induction on the structure of $\sigma$, $\asprg$ and $\asfml$ for all $U,\iget[state]{\I},\iget[state]{\It}$ and $\iget[state]{\Itb}$.
	Take a $U$-variation $\iget[state]{\I}$ of $\iget[state]{\It}$.

	For formulas, we need to prove:
	\[\text{for all }U\text{-variations }\iget[state]{\I}\text{ of }\iget[state]{\It}:\imodels{\I}{\usubstapp{U}{\sigma}{\asfml}} \text{ iff } \imodels{\Iadj}{\asfml}
	\]
	\begin{enumerate}
		\item For $\asfml \mequiv \astrm \leq \bstrm$, we have $\imodels{\I}{\usubstapp{U}{\sigma}{(\astrm \leq \bstrm)}}$ iff $\ivaluation{\I}{\usubstapp{U}{\sigma}{\astrm}} \leq \ivaluation{\I}{\usubstapp{U}{\sigma}{\bstrm}}$, by \rref{lem:usubst-term} iff $\ivaluation{\Iadj}{\astrm} \leq \ivaluation{\Iadj}{\bstrm}$ iff $\imodels{\Iadj}{\astrm \leq \bstrm}$.

		\item \label{case:pred} For $\asfml \mequiv \ousfml[\astrm]$, where $\osfml$ is not substituted by anything else, we have $\imodels{\I}{\usubstapp{U}{\sigma}{\ousfml[\astrm]}} = \imodel{\I}{\ousfml[\usubstapp{U}{\sigma}{\astrm}]}$ iff $\ivaluation{\I}{\usubstapp{U}{\sigma}{\astrm}}\in \getInterp{\I}{\osfml}$, by \rref{lem:usubst-term} iff $(\ivaluation{\Iadj}{\astrm})\in \getInterp{\I}{\osfml} = \getInterp{\Iadj}{\osfml}$ iff $\imodels{\Iadj}{\ousfml[\astrm]}$.

		\item For $\asfml \mequiv \ousfml[\astrm]$, we write $d \mdefeq \ivaluation{\I}{\usubstapp{U}{\sigma}{\astrm}} = \ivaluation{\Iadj}{\astrm}$ by \rref{lem:usubst-term}.
		We have $\imodels{\I}{\usubstapp{U}{\sigma}{\ousfml[\astrm]}} = \imodel{\I}{\usubstapp{\emptyset}{\usubstlist{\usubstmod{\usarg}{\usubstapp{U}{\sigma}{\astrm}}}}{\applyusubst{\sigma}{\ousfml[\usarg]}}}$, by IH iff $\imodels{\imodif[const]{\I}{\usarg}{d}}{\applyusubst{\sigma}{\ousfml[\usarg]}}$, by \rref{lem:coincidence} iff
		$\imodels{\imodif[const]{\It}{\usarg}{d}}{\applyusubst{\sigma}{\ousfml[\usarg]}}$, by definition iff $d=(\ivaluation{\Iadj}{\astrm})\in \getInterp{\Iadj}{\osfml}$ iff $\imodels{\Iadj}{\ousfml[\astrm]}$.
		The IH for $\usubstlist{\usubstmod{\usarg}{\usubstapp{U}{\sigma}{\astrm}}}$ is used for a potentially bigger $\applyusubst{\sigma}{\ousfml[\usarg]}$ but simpler substitution that does not substitute a predicate so is covered by \rref{case:pred}.
		\rref{lem:coincidence} is applicable as $\iget[state]{\I}$ is a $U$-variation of $\iget[state]{\It}$ and $\freevarsdef{\applyusubst{\sigma}{\ousfml[\usarg]}}\cap U = \emptyset$, so $\iget[state]{\I} = \iget[state]{\It}$ on $\freevars{\applyusubst{\sigma}{\ousfml[\usarg]}}$.
		
		\item For $\asfml \mequiv \lnot \bsfml$, we have $\imodels{\I}{\usubstapp{U}{\sigma}{\lnot \bsfml}} =\imodel{\I}{\lnot \usubstapp{U}{\sigma}{\bsfml}}$ iff $\inonmodels{\I}{\usubstapp{U}{\sigma}{\bsfml}}$, by IH iff $\inonmodels{\Iadj}{\bsfml}$ iff $\imodels{\Iadj}{\lnot \bsfml}$.

		\item For $\asfml \mequiv \bsfml \land \csfml$, we have $\imodels{\I}{\usubstapp{U}{\sigma}{\bsfml \land \csfml}} = \imodel{\I}{\usubstapp{U}{\sigma}{\bsfml} \land \usubstapp{U}{\sigma}{\csfml}} = \imodel{\I}{\usubstapp{U}{\sigma}{\bsfml}} \cap \imodel{\I}{\usubstapp{U}{\sigma}{\csfml}}$, by IH iff $\imodels{\Iadj}{\bsfml} \cap \imodel{\Iadj}{\csfml}$ iff $\imodels{\Iadj}{\bsfml \land \csfml}$.

		\item For $\asfml \mequiv \lforall{x}{\bsfml}$, we have $\imodels{\I}{\usubstapp{U}{\sigma}{(\lforall{x}{\bsfml})}} = \imodel{\I}{\lforall{x}{\usubstapp{U\cup\set{x}}{\sigma}{\bsfml}}}$ iff for all $d$, $\imodels{\imodif[state]{\I}{x}{d}}{\usubstapp{U\cup\set{x}}{\sigma}{\bsfml}}$, by IH iff for all $d$, $\imodels{\imodif[state]{\Iadj}{x}{d}}{\bsfml}$ iff $\imodels{\Iadj}{\lforall{x}{\bsfml}}$.
		The IH is applicable as $\modif{\iget[state]{\I}}{x}{d}$ is a $(U\cup\set{x})$-variation of $\iget[state]{\It}$.

		\item \label{case:box} For $\asfml \mequiv \dbox{\asprg}{\bsfml}$, we have $\imodels{\I}{\usubstapp{U}{\sigma}{(\dbox{\asprg}{\bsfml})}} = \imodel{\I}{\dbox{\usubstappp{V}{U}{\sigma}{\asprg}}{\usubstapp{V}{\sigma}{\bsfml}}}$ iff for all states $\iget[state]{\Itb}$, $\iaccessible[\usubstappp{V}{U}{\sigma}{\asprg}]{\I}{\Itb}$ implies $\imodels{\Itb}{\usubstapp{V}{\sigma}{\bsfml}}$.
		Since $\iget[state]{\I}$ is a $U$-variation of $\iget[state]{\It}$, by IH, $\iaccessible[\usubstappp{V}{U}{\sigma}{\asprg}]{\I}{\Itb}$ iff $\iaccessible[\asprg]{\Iadj}{\Itb}$.
		Take a state $\iget[state]{\Itb}$ with $\iaccessible[\usubstappp{V}{U}{\sigma}{\asprg}]{\I}{\Itb}$.
		By \rref{lem:boundeffect}, $\iget[state]{\Itb}$ is a $\boundvarsdef{\asprg}$-variation of $\iget[state]{\I}$, and thus a $(U\cup\boundvarsdef{\asprg})$-variation of $\iget[state]{\It}$.
		By \rref{lem:taboo}, it means that $\iget[state]{\Itb}$ is a $V$-variation of $\iget[state]{\It}$ because $\usubstappp{V}{U}{\sigma}{\asprg}$ is defined so $V\supseteq U\cup\boundvarsdef{\asprg}$.
		Thus, we can apply the induction hypothesis: $\imodels{\Itb}{\usubstapp{V}{\sigma}{\bsfml}}$ iff $\imodels{\Iadjb}{\bsfml}$.
		Finally, we can conclude $\imodels{\I}{\dbox{\usubstappp{V}{U}{\sigma}{\asprg}}{\usubstapp{V}{\sigma}{\bsfml}}}$ iff $\imodels{\Iadj}{\dbox{\asprg}{\bsfml}}$.

		\item For $\asfml \mequiv \asprg \refines \bsprg$, we have by definition  that for all states $\iget[state]{\Itb}$, $\iaccessible[\asprg]{\Iadj}{\Itb}$ implies $\iaccessible[\bsprg]{\Iadj}{\Itb}$.
		Since $\iget[state]{\I}$ is a $U$-variation of $\iget[state]{\It}$, by induction hypothesis for both $\asprg$ and $\bsprg$, this is equivalent to having, for all states $\iget[state]{\Itb}$, $\iaccessible[\usubstappp{V}{U}{\sigma}{\asprg}]{\I}{\Itb}$ implies $\iaccessible[\usubstappp{W}{U}{\sigma}{\bsprg}]{\I}{\Itb}$.
		This is, by definition, equivalent to $\imodels{\I}{\usubstappp{V}{U}{\sigma}{\asprg}\refines\usubstappp{W}{U}{\sigma}{\bsprg}} = \imodel{\I}{\usubstapp{U}{\sigma}{(\asprg\refines\bsprg)}}$.

	\end{enumerate}

	For programs, we need to prove:
	\[\text{for all states }\iget[state]{\Itb}\text{ and all }U\text{-variations }\iget[state]{\I}\text{ of }\iget[state]{\It}:\iaccessible[\usubstappp{V}{U}{\sigma}{\asprg}]{\I}{\Itb} \text{ iff } \iaccessible[\asprg]{\Iadj}{\Itb}\]
	\begin{enumerate} 
		\setcounter{enumi}{7}
		\item For $\asprg \mequiv \ausprg$, we have $\iaccess[\usubstappp{V}{U}{\sigma}{\ausprg}]{\I} = \iaccess[\applysubst{\sigma}{\ausprg}]{\I} = \iaccess[\ausprg]{\Iadj}$.
		
		\item For $\asprg \mequiv \ptest{\asfml}$, we have $\iaccessible[\usubstappp{U}{U}{\sigma}{\ptest{\asfml}}]{\I}{\I}$ iff $\imodels{\I}{\usubstapp{U}{\sigma}{\asfml}}$, by IH iff $\imodels{\Iadj}{\asfml}$ iff $\iaccessible[\ptest{\asfml}]{\Iadj}{\I}$.

		\item For $\asprg \mequiv \pupdate{\pumod{x}{\astrm}}$, we have $\iaccessible[\usubstappp{U\cup\set{x}}{U}{\sigma}{(\pupdate{\pumod{x}{\astrm}})}]{\I}{\Itb} = \iaccess[\pupdate{\pumod{x}{\usubstapp{U}{\sigma}{\astrm}}}]{\I}$ iff $\iget[state]{\Itb} = \modif{\iget[state]{\I}}{x}{d}$ with $d = \ivaluation{\I}{\usubstapp{U}{\sigma}{\astrm}}$.
		By \rref{lem:usubst-term}, this is equivalent to $\iget[state]{\Itb} = \modif{\iget[state]{\Iadj}}{x}{d}$ with $d = \ivaluation{\Iadj}{\astrm}$, i.e.\ $\iaccessible[\pupdate{\pumod{x}{\astrm}}]{\Iadj}{\Itb}$.
		
		\item For $\asprg \mequiv \prandom{x}$, we have $\iaccessible[\usubstappp{U\cup\set{x}}{U}{\sigma}{\prandom{x}}]{\I}{\Itb} = \iaccess[\prandom{x}]{\I}$ iff $\iget[state]{\Itb} = \modif{\iget[state]{\I}}{x}{d}$ for some $d$ iff $\iaccessible[\prandom{x}]{\Iadj}{\Itb}$.

		\item For $\asprg \mequiv \pode{\D{x}=\astrm}{\asfml}$, we have $\iaccessible[\usubstappp{U\cup\set{x,\D{x}}}{U}{\sigma}{(\pode{\D{x}=\astrm}{\asfml})}]{\I}{\Itb} = \break \iaccess[\pode{\D{x}=\usubstapp{U\cup\set{x,\D{x}}}{\sigma}{\astrm}}{\usubstapp{U\cup\set{x,\D{x}}}{\sigma}{\asfml}}]{\I}$ iff $\imodels{\Isol}{\D{x} = \usubstapp{U\cup\set{x,\D{x}}}{\sigma}{\astrm} \land \usubstapp{U\cup\set{x,\D{x}}}{\sigma}{\asfml}}$ and $\Ddiff[t]{\ivaluation{\Isol}{x}}=\ivaluation{\Isol}{\usubstapp{U\cup\set{x,\D{x}}}{\sigma}{\astrm}}$ for all $t\in[0,r]$ for some function $\sol : [0, r ] \rightarrow \State$ where $\sol(t)$  is a $\set{x,\D{x}}$-variation of $\iget[state]{\I}$, $\sol(0)$ is a $\set{\D{x}}$-variation of $\iget[state]{\I}$, and $\sol(r) = \iget[state]{\Itb}$.
		$\iget[state]{\Isol}$ is a $(U\cup\set{x,\D{x}})$-variation of $\iget[state]{\It}$, so by \rref{lem:usubst-term}, $\ivaluation{\Isol}{\usubstapp{U\cup\set{x,\D{x}}}{\sigma}{\astrm}}= \ivaluation{\iadjointSubst{\sigma}{\omega}{\Isol}}{\astrm}$ and by IH $\imodels{\Isol}{\D{x} = \usubstapp{U\cup\set{x,\D{x}}}{\sigma}{\astrm} \land \usubstapp{U\cup\set{x,\D{x}}}{\sigma}{\asfml}}$ is equivalent to $\imodels{\iadjointSubst{\sigma}{\omega}{\Isol}}{\D{x} = \astrm \land \asfml}$.
		Thus, we have $\iaccessible[\usubstappp{U\cup\set{x,\D{x}}}{U}{\sigma}{(\pode{\D{x}=\astrm}{\asfml})}]{\I}{\Itb}$ iff $\iaccessible[\pode{\D{x}=\astrm}{\asfml}]{\Iadj}{\Itb}$.

		\item For $\asprg \mequiv \pchoice{\bsprg}{\csprg}$, we have $\iaccessible[\usubstappp{V\cup W}{U}{\sigma}{(\pchoice{\bsprg}{\csprg})}]{\I}{\Itb} = \iaccess[\pchoice{\usubstappp{V}{U}{\sigma}{\bsprg}}{\usubstappp{W}{U}{\sigma}{\csprg}}]{\I}$ iff \break $\iaccessible[\usubstappp{V}{U}{\sigma}{\bsprg}]{\I}{\Itb} \cup \iaccess[\usubstappp{W}{U}{\sigma}{\csprg}]{\I}$, by IH iff $\iaccessible[\bsprg]{\Iadj}{\Itb} \cup \iaccess[\csprg]{\Iadj}$ iff $\iaccessible[\pchoice{\bsprg}{\csprg}]{\Iadj}{\Itb}$.

		\item \label{case:seq} For $\asprg \mequiv \bsprg;\csprg$, we have $\iaccessible[\usubstappp{W}{U}{\sigma}{(\bsprg;\csprg)}]{\I}{\Itb}$ iff $\iaccessible[\usubstappp{V}{U}{\sigma}{\bsprg}]{\I}{\Itbi{1}}$ and \break $\iaccessible[\usubstappp{W}{V}{\sigma}{\csprg}]{\Itbi{1}}{\Itb}$ for some $\iget[state]{\Itbi{1}}$.
		By IH, $\iaccessible[\usubstappp{V}{U}{\sigma}{\bsprg}]{\I}{\Itbi{1}}$ iff $\iaccessible[\bsprg]{\Iadj}{\Itbi{1}}$.
		Similarly to \rref{case:box}, by Lemmas~\ref{lem:boundeffect} and~\ref{lem:taboo}, we have that $\iget[state]{\Itbi{1}}$ is a $V$-variation of $\iget[state]{\It}$, so by IH, $\iaccessible[\usubstappp{W}{V}{\sigma}{\csprg}]{\Itbi{1}}{\Itb}$ iff $\iaccessible[\csprg]{\Iadj}{\Itb}$.
		Thus, $\iaccessible[\usubstappp{W}{U}{\sigma}{(\bsprg;\csprg)}]{\I}{\Itb}$ iff $\iaccessible[\bsprg;\csprg]{\Iadj}{\Itb}$.
		
		\item For $\asprg \mequiv \prepeat{\bsprg}$, we have $\iaccessible[\usubstappp{V}{U}{\sigma}{(\prepeat{\bsprg})}]{\I}{\Itb} = \iaccess[\prepeat{(\usubstappp{V}{V}{\sigma}{\bsprg})}]{\I}$ iff $\iaccessible[(\usubstappp{V}{V}{\sigma}{\bsprg})^{n}]{\I}{\Itb}$ for some $n$.
		By IH, we have that $\iaccessible[\usubstappp{V}{V}{\sigma}{\bsprg}]{\I}{\Itbi{1}}$ iff $\iaccessible[\bsprg]{\Iadj}{\Itbi{1}}$ for all states $\iget[state]{\Itbi{1}}$.
		By repeated application of \rref{case:seq}, we obtain by induction that $\iaccessible[(\usubstappp{V}{V}{\sigma}{\bsprg})^{n}]{\I}{\Itb}$ iff $\iaccessible[\bsprg^{n}]{\Iadj}{\Itb}$. Thus $\iaccessible[\prepeat{\bsprg}]{\Iadj}{\Itb}$.
		The other direction is similar.
	\end{enumerate}
\end{proof}

\section{Proof of Axioms' Soundness}\label{app:proofaxioms}

In this section, we prove the soundness of the axioms of \dRL. This means that for each axiom $\asfml$, we show that for all interpretations $\iget[const]{\I}$, $\imodel{\I}{\asfml} = \State$.

The soundness of \dL axioms (\rref{fig:dLaxioms}) was proved in \cite{DBLP:journals/jar/Platzer17}.
The proof of soundness of most \dRL axioms is similar to the corresponding rules in the original presentation of \dRL \cite{DBLP:conf/lics/LoosP16}.
In particular, all KAT axioms in \rref{fig:KATaxioms} are instances of the KAT axioms from \cite{DBLP:conf/lics/LoosP16}, so their soundness follow from the original proof \cite{Loos16}.
Here is a proof for the remaining axioms from \rref{fig:axioms} and \rref{fig:SKATaxioms}.
\begin{itemize}
	\item[(\irref{leqtrans})]
Take $\imodels{\I}{\ausprg \refines \busprg \land \busprg \refines \cusprg}$ and $\iaccessible[\ausprg]{\I}{\It}$.
As $\imodels{\I}{\ausprg \refines \busprg}$, we have $\iaccessible[\busprg]{\I}{\It}$.
Then, as $\imodels{\I}{\busprg \refines \cusprg}$, we have $\iaccessible[\cusprg]{\I}{\It}$.
Thus, $\imodels{\I}{\ausprg \refines \cusprg}$.

\item[(\irref{boxleq})]
Take $\imodels{\I}{\ausprg \refines \busprg}$ and $\imodels{\I}{\dbox{\busprg}{\ausfml}}$.
We need to show that $\imodels{\I}{\dbox{\ausprg}{\ausfml}}$.
Let us consider a state $\iget[state]{\It}$ such that $\iaccessible[\ausprg]{\I}{\It}$.
By refinement, i.e.\ \(\imodels{\I}{\ausprg\refines\busprg}\), this implies $\iaccessible[\busprg]{\I}{\It}$.
This then implies $\imodels{\It}{\ausfml}$, as $\imodels{\I}{\dbox{\busprg}{\ausfml}}$.
Thus, we can conclude that $\imodels{\I}{\dbox{\ausprg}{\ausfml}}$.

\item[(\irref{test})]
By definition, $\iaccessible[\ptest{\osfml}]{\I}{\It}$ is equivalent to $\iget[state]{\It} = \iget[state]{\I}$ and $\imodels{\I}{\osfml}$.
Thus, $\imodels{\I}{\ptest{\osfml} \refines  {\ptest{\osfmlb}}}$ holds iff
$\iaccessible[\ptest{\osfml}]{\I}{\I}$ implies $\iaccessible[\ptest{\osfmlb}]{\I}{\I}$, which by definition is equivalent to $\imodels{\I}{\osfml}$ implies $\imodels{\I}{\osfmlb}$ and so $\imodels{\I}{\osfml \limply \osfmlb}$.

\item[(\irref{choicel})]
Take $\imodels{\I}{\pchoice{\ausprg}{\busprg} \refines \cusprg}$ and $\iaccessible[\ausprg]{\I}{\It}$.
Then, by the semantics of the choice, \(\iaccessible[\pchoice{\ausprg}{\busprg}]{\I}{\It} = \iaccess[\ausprg]{\I}\cup\iaccess[\busprg]{\I}\) and by refinement, $\iaccessible[\cusprg]{\I}{\It}$.
Similarly, if $\iaccessible[\busprg]{\I}{\It}$, we have $\iaccessible[\cusprg]{\I}{\It}$.
Thus, \(\imodels{\I}{\pchoice{\ausprg}{\busprg} \refines \cusprg \limply \ausprg \refines \cusprg \land \busprg \limply \cusprg}\).

On the other hand, take \(\imodels{\I}{\ausprg \refines \cusprg \land \busprg \refines \cusprg}\).
If \(\iaccessible[\pchoice{\ausprg}{\busprg}]{\I}{\It} = \iaccess[\ausprg]{\I}\cup\iaccess[\busprg]{\I}\), then $\iaccessible[\ausprg]{\I}{\It}$ or $\iaccessible[\busprg]{\I}{\It}$.
In both cases, by refinement, we have $\iaccessible[\cusprg]{\I}{\It}$.
Thus \(\imodels{\I}{\ausprg \refines \cusprg \land \busprg \refines \cusprg \limply \pchoice{\ausprg}{\busprg} \refines \cusprg}\).

\item[(\irref{choicer})]
Take $\imodels{\I}{\ausprg \refines \cusprg \lor \busprg \refines \cusprg}$.
Then $\imodels{\I}{\ausprg \refines \busprg}$ or $\imodels{\I}{\ausprg \refines \cusprg}$.
If $\iaccessible[\ausprg]{\I}{\It}$, then depending on the case above, $\iaccessible[\busprg]{\I}{\It}$ or $\iaccessible[\cusprg]{\I}{\It}$.
In either case, we have $\iaccessible[\pchoice{\busprg}{\cusprg}]{\I}{\It}$.
Thus \(\imodels{\I}{\ausprg \refines \cusprg \lor \ausprg \refines \busprg \limply \ausprg \refines \pchoice{\busprg}{\cusprg}}\).

\item[(\irref{sequence})]
Take $\imodels{\I}{\ausprg \refines \cusprg}$ and $\imodels{\I}{\dbox{\ausprg}{\busprg \refines \dusprg}}$.
If $\iaccessible[\ausprg;\cusprg]{\I}{\It}$, then there exists $\iget[state]{\Itb}$ such that $\iaccessible[\ausprg]{\I}{\Itb}$ and $\iaccessible[\cusprg]{\Itb}{\It}$.
Additionally, by semantics of the box operator, $\imodels{\Itb}{\busprg \refines \dusprg}$.
Thus, by refinement, we have both $\iaccessible[\cusprg]{\I}{\Itb}$ and $\iaccessible[\dusprg]{\Itb}{\It}$.
We can conclude $\iaccessible[\busprg;\dusprg]{\I}{\It}$, and thus \(\imodels{\I}{\ausprg;\cusprg \refines \busprg;\dusprg}\).

\item[(\irref{loopl})]
Assume that $\imodels{\I}{\dbox{\prepeat{\ausprg}}{\ausprg;\busprg \refines \ausprg}}$.
We show by induction on $n$ that $\imodels{\I}{\ausprg^n;\busprg \refines \busprg}$.

For $n = 0$, $\ausprg^0 = \ptest{\ltrue}$ by definition and $\ptest{\ltrue};\busprg \refines \busprg$ by (\irref{seqidl}).

If $\imodels{\I}{\ausprg^n;\busprg \refines \busprg}$, we show that $\imodels{\I}{\ausprg^{n+1};\busprg \refines \busprg}$ by (\irref{leqtrans}), that is\break $\imodels{\I}{\ausprg^{n+1};\busprg \refines \ausprg^n;\busprg \land \ausprg^n;\busprg \refines \busprg}$.
The second part holds by induction hypothesis.
For the first part, by (\irref{sequence}), we must show $\imodels{\I}{\ausprg^n \refines \ausprg^n \land \dbox{\ausprg^n}{\ausprg;\busprg \refines \busprg}}$.
By semantics of the loop operator, $\iaccess[\prepeat{\ausprg}]{\I} \supseteq \iaccess[\ausprg^n]{\I}$ for all $n$, so $\ausprg^n \refines \prepeat{\ausprg}$ holds.
Thus, we can conclude the proof by (\irref{leqrefl}) for the left part, and by (\irref{boxleq}) for the right part, as $\imodels{\I}{\dbox{\prepeat{\ausprg}}{\ausprg;\busprg \refines \busprg}}$.

\item[(\irref{loopr})]
Assume that $\imodels{\I}{\ausprg;\busprg \refines \ausprg}$.
We show by induction on $n$ that $\imodels{\I}{\ausprg;\busprg^n \refines \ausprg}$.

For $n = 0$, $\busprg^0 = \ptest{\ltrue}$ by definition and $\ausprg;\ptest{\ltrue} \refines \ausprg$ by (\irref{seqidr}).

If $\imodels{\I}{\ausprg;\busprg^n \refines \ausprg}$, we show that $\imodels{\I}{\ausprg;\busprg^{n+1} \refines \ausprg}$ by (\irref{leqtrans}), that is \break $\imodels{\I}{\ausprg;\busprg^{n+1} \refines \ausprg;\busprg^n \land \ausprg;\busprg^n \refines \ausprg}$.
The second part holds by induction hypothesis.
For the first part, by (\irref{sequence}), we must show $\imodels{\I}{\ausprg;\busprg \refines \ausprg \land \dbox{\ausprg;\busprg}{\busprg^n \refines \busprg^n}}$.
We can conclude the proof by assumption for the left part, and by (\irref{Gax}) and (\irref{leqrefl}) for the right part.

\item[(\irref{unloop})]
Assume that $\imodels{\I}{\dbox{\prepeat{\ausprg}}{\ausprg \refines \busprg}}$.
We show by induction on $n$ that $\imodels{\I}{\ausprg^n \refines \busprg^n}$.

For $n = 0$, $\ausprg^0 = \ptest{\ltrue} = \busprg^0$ by definition, so it holds by (\irref{leqrefl}).

If $\imodels{\I}{\ausprg^n \refines \busprg^n}$, then by (\irref{sequence}), we must show $\imodels{\I}{\ausprg^n \refines \busprg^n \land \dbox{\ausprg^n}{\ausprg \refines \busprg}}$.
As in the proof of (\irref{loopl}), we have $\ausprg^n \refines \prepeat{\ausprg}$.
Thus, the first part holds by induction hypothesis, and the second part holds by (\irref{boxleq}) and by assumption.

\item[(\irref{update})]
The main insight is that as $\ivaluation{\I}{\oustrm} = \getInterp{\I}{\oustrm} = \ivaluation{\imodif[state]{\I}{x}{r}}{\oustrm}$ for any $r$, $\imodels{\imodif[state]{\I}{x}{r}}{x = \oustrm}$ iff $r = \ivaluation{\I}{\oustrm}$.
As $\iaccessible[\pupdate{\umod{x}{\oustrm}}]{\I}{\It}$ iff $\iget[state]{\It} = \imodif[state]{\I}{x}{r}$ for $r = \ivaluation{\I}{\oustrm}$, and $\iaccessible[\prandom{x};\ptest{x = \oustrm}]{\I}{\It}$ iff $\iget[state]{\It} = \imodif[state]{\I}{x}{r}$ for some $r$ with $\imodels{\It}{x = \oustrm}$, the two programs are equivalent.

\item[(\irref{testdet})]
Take $\imodels{\I}{\ptest{\osfml};\ausprg\refines  {\ptest{\osfml}};\busprg}$.
For all $\iaccessible[\ptest{\osfml}]{\I}{\It}$ and $\iaccessible[\ausprg]{\It}{\Itb}$, we need to show $\iaccessible[\busprg]{\It}{\Itb}$.
As, $\iaccessible[\ptest{\osfml};\ausprg]{\I}{\Itb}$, by refinement, we have $\iaccessible[\ptest{\osfml};\busprg]{\I}{\Itb}$, which means that there exists $\iget[state]{\Ittil}$ such that $\iaccessible[\ptest{\osfml}]{\I}{\Ittil}$ and $\iaccessible[\busprg]{\Ittil}{\Itb}$.

By definition of $\iaccess[\ptest{\osfml}]{\I}$, we have $\iget[state]{\It} = \iget[state]{\I} = \iget[state]{\Ittil}$, so $\iaccessible[\busprg]{\It}{\Itb}$.
We can conclude that $\imodels{\I}{\dbox{\ptest{\osfml}}{\ausprg \refines \busprg}}$.

The other direction is derivable from (\irref{sequence}) \seeapp{\rref{app:derived}}.

\item[(\irref{assigndet})]
Take $\imodels{\I}{\pupdate{\pumod{x}{\oustrm}};\ausprg\refines \pupdate{\pumod{x}{\oustrm}};\busprg}$.
For all $\iaccessible[\pupdate{\pumod{x}{\oustrm}}]{\I}{\It}$ and $\iaccessible[\ausprg]{\It}{\Itb}$, we need to show $\iaccessible[\busprg]{\It}{\Itb}$.
As, $\iaccessible[\pupdate{\pumod{x}{\oustrm}};\ausprg]{\I}{\Itb}$, by refinement, we have $\iaccessible[\pupdate{\pumod{x}{\oustrm}};\busprg]{\I}{\Itb}$, which means that there exists $\iget[state]{\Ittil}$ such that $\iaccessible[\pupdate{\pumod{x}{\oustrm}}]{\I}{\Ittil}$ and $\iaccessible[\busprg]{\Ittil}{\Itb}$.

By definition of $\iaccess[\pupdate{\pumod{x}{\oustrm}}]{\I}$, we have $\iget[state]{\It} = \modif{\iget[state]{\I}}{x}{r} = \iget[state]{\Ittil}$ for $r = \getInterp{\I}{\oustrm}$, so $\iaccessible[\busprg]{\It}{\Itb}$.
We can conclude that $\imodels{\I}{\dbox{\pupdate{\pumod{x}{\oustrm}}}{\ausprg \refines \busprg}}$.

The other direction is derivable from (\irref{sequence}) \seeapp{\rref{app:derived}}.

\item[(\irref{stutter})]
$\iaccessible[\pupdate{\umod{x}{x}}]{\I}{\It}$ iff $\iget[state]{\It} = \modif{\iget[state]{\I}}{x}{r}$ for $r = \ivaluation{\I}{x}$.
It means that $\iget[state]{\It} = \iget[state]{\I}$.
So $\iaccessible[\pupdate{\umod{x}{x}}]{\I}{\It}$ iff $\iget[state]{\It} = \iget[state]{\I}$.

On the other hand, $\iaccessible[\ptest{\ltrue}]{\I}{\I}$ iff $\imodels{\I}{\ltrue}$ which is always true.
So $\iaccess[\ptest{\ltrue}]{\I} = \set{(\iget[state]{\I},\iget[state]{\I})} = 
\iaccess[\pupdate{\umod{x}{x}}]{\I}$, and the programs are thus equivalent.

\item[(\irref{randtestmerge})]
For all states $\iget[state]{\I}, \iget[state]{\It}$, we have $\iaccessible[\prandom{x};\ptest{\ousfml[x]};\prandom{x}]{\I}{\It}$ iff there exists $r_1, r_2,\iget[state]{\Itb}$ with $\iget[state]{\Itb} = \modif{\iget[state]{\I}}{x}{r_1}$, and
$\imodels{\Itb}{\osfml}$, and $\iget[state]{\It} = \modif{\iget[state]{\Itb}}{x}{r_2}$.

This also means that $\iget[state]{\It} = \modif{\iget[state]{\I}}{x}{r_2}$ and $\iget[state]{\Itb} = \modif{\iget[state]{\It}}{x}{r_1}$.
Thus it is equivalent to having some $r_2$ such that $\iget[state]{\It} = \modif{\iget[state]{\I}}{x}{r_2}$ and some $r_1$ such that $\imodels{\imodif[state]{\It}{x}{r_1}}{\osfml}$, and so $\iaccessible[\prandom{x};\ptest{\lexists y \ousfml[y]}]{\I}{\It}$.

\item[(\irref{randtestassignmerge})]
For all states $\iget[state]{\I}, \iget[state]{\It}$, we have $\iaccessible[\prandom{x};\ptest{\ousfml[x]};\pupdate{\umod{x}{\oustrm[x]}}]{\I}{\It}$ iff there exists $r_1, r_2,\iget[state]{\Itb}$ with $\iget[state]{\Itb} = \modif{\iget[state]{\I}}{x}{r_1}$, and
$\imodels{\Itb}{\osfml}$, and $\iget[state]{\It} = \modif{\iget[state]{\Itb}}{x}{r_2}$ with $r_2 = \ivaluation{\Itb}{\oustrm[x]}$.

This also means that $\iget[state]{\It} = \modif{\iget[state]{\I}}{x}{r_2}$ and $\iget[state]{\Itb} = \modif{\iget[state]{\It}}{x}{r_1}$.
Thus it is equivalent to having some $r_2$ such that $\iget[state]{\It} = \modif{\iget[state]{\I}}{x}{r_2}$ and some $r_1$ such that both $\imodels{\imodif[state]{\It}{x}{r_1}}{\osfml}$ and $r_2 = \ivaluation{\imodif[state]{\It}{x}{r_1}}{\oustrm[x]}$.
The latter can be rewritten as $\imodels{\imodif[state]{\It}{x}{r_1}}{r_2 = \oustrm[x]}$ so we can conclude that it is equivalent to $\iaccessible[\prandom{x};\ptest{\lexists y (\ousfml[y] \land y = \oustrm[y])}]{\I}{\It}$.

\item[(\irref{DWref})]
$\ptest{\ousfml[x]};\pode{\D{x}=\oustrm[x]}{\ousfml[x]};\ptest{\ousfml[x]} \refines \pode{\D{x}=\oustrm[x]}{\ousfml[x]}$ is a consequence of (\irref{seqidl}), (\irref{seqidr}), (\irref{test}) and (\irref{sequence}).

For the other direction, take $\iaccessible[\pode{\D{x}=\oustrm[x]}{\ousfml[x]}]{\I}{\It}$. Then there is $\sol: [0, r] \rightarrow \State$ with $\sol(0)$ being a $\set{\D{x}}$-variation of $\iget[state]{\I}$ and $\iget[state]{\It} = \sol(r)$ where $\sol(t)$ is a $\set{x,\D{x}}$-variation of $\iget[state]{\I}$, and satisfies $\imodels{\Isol}{\D{x} = \oustrm[x] \land \ousfml[x]}$.
In particular, $\sol(0)$ and $\sol(r) = \imodels{\It}{\ousfml[x]}$.
As $\sol(0) = \iget[state]{\I}$ on $\freevars{\ousfml[x]} = \set{x}$, then $\imodels{\I}{\ousfml[x]}$ by \rref{lem:coincidence}.
Thus, both $(\iget[state]{\I},\iget[state]{\I})$ and $\iaccessible[\ptest{\ousfml[x]}]{\It}{\It}$.
This implies \break $\iaccessible[\ptest{\ousfml[x]};\pode{\D{x}=\oustrm[x]}{\ousfml[x]};\ptest{\ousfml[x]}]{\I}{\It}$.
As it holds for all end states $\iget[state]{\It}$, we can conclude that $\imodels{\I}{\pode{\D{x}=\oustrm[x]}{\ousfml[x]} \refines {\ptest{\ousfml[x]};\pode{\D{x}=\oustrm[x]}{\ousfml[x]};\ptest{\ousfml[x]}}}$.

\item[(\irref{DEref})]
Similarly to above, $\iaccessible[\pode{\D{x}=\oustrm[x]}{\ousfml[x]}]{\I}{\It}$ entails $\imodels{\It}{\D{x} = \oustrm[x]}$. Thus $\modif{\iget[state]{\It}}{\D{x}}{r} = \iget[state]{\It}$ for $r = \ivaluation{\It}{\oustrm[x]}$.
So $(\iget[state]{\I}, \modif{\iget[state]{\It}}{\D{x}}{r}) = \iaccessible[\pode{\D{x}=\oustrm[x]}{\ousfml[x]};\pupdate{\umod{\D{x}}{\ousfml[x]}}]{\I}{\It}$ iff \break$\iaccessible[\pode{\D{x}=\oustrm[x]}{\ousfml[x]}]{\I}{\It}$.
As it holds for all end states $\iget[state]{\It}$, we can conclude that $\imodels{\I}{\pode{\D{x}=\oustrm[x]}{\ousfml[x]} \refines {\pode{\D{x}=\oustrm[x]}{\ousfml[x]};\pupdate{\umod{\D{x}}{\ousfml[x]}}}}$.

\item[(\irref{ode})]
The forward implication is derivable from (\irref{DWref}), (\irref{DEref}), (\irref{boxleq}).

For the converse implication, take $\imodels{\I}{\dbox{\pode{\D{x}=\oustrm[x]}{\ousfml[x]}}{(\D{x}=\oustrmb[x] \land \ousfmlb[x])}}$ and take $\iget[state]{\It}$ with $\iaccessible[\pode{\D{x}=\oustrm[x]}{\ousfml[x]}]{\I}{\It}$.
Then there is a function $\sol: [0, r] \rightarrow \State$ with $\sol(0)$ being a $\set{\D{x}}$-variation of $\iget[state]{\I}$ and $\iget[state]{\It} = \sol(r)$ where $\sol(t)$ is a $\set{x,\D{x}}$-variation of $\iget[state]{\I}$, and satisfies $\imodels{\Isol}{\D{x} = \oustrm[x] \land \ousfml[x]}$ and $\Ddiff[t]{\ivaluation{\Isol}{x}}=\ivaluation{\Isol}{\astrm}$ for all $t\in[0,r]$.
This also means that $\iaccessible[\pode{\D{x}=\oustrm[x]}{\ousfml[x]}]{\I}{\Isol}$ for all $t\in[0,r]$.
By the semantics of the box operator, \break\(\imodels{\I}{\dbox{\pode{\D{x}=\oustrm[x]}{\ousfml[x]}}{(\D{x}=\oustrmb[x] \land \ousfmlb[x])}}\) implies $\imodels{\Isol}{\D{x}=\oustrmb[x] \land \ousfmlb[x]}$ for all $t\in[0,r]$.
Thus $\iaccessible[\pode{\D{x}=\oustrmb[x]}{\ousfmlb[x]}]{\I}{\It}$.
As it holds for all end states $\iget[state]{\It}$, we can conclude that $\imodels{\I}{\pode{\D{x}=\oustrm[x]}{\ousfml[x]} \refines \pode{\D{x}=\oustrmb[x]}{\ousfmlb[x]}}$.

\item[(\irref{DX})]
Take $\iaccessible[\pupdate{\umod{\D{x}}{\oustrm[x]}};\ptest{\ousfml[x]}]{\I}{\It}$. We have that $\iget[state]{\It} = \modif{\iget[state]{\I}}{\D{x}}{r}$ and $\imodels{\It}{\ousfml[x]}$ for $r = \ivaluation{\It}{\oustrm[x]}$.
In particular, this also means \(\ivaluation{\It}{\D{x}} = \ivaluation{\It}{\oustrm[x]}\) and thus $\imodels{\It}{\D{x} = \oustrm[x]}$.
We define $\sol : \set{0} \rightarrow \State$ with $\sol(0) = \iget[state]{\It}$. The function $\sol$ satisfies that $\sol(0)$ is a $\set{\D{x}}$-variation of $\iget[state]{\I}$, $\iget[state]{\Isol}$ is a $\set{x,\D{x}}$-variation of $\iget[state]{\I}$ for $t\in[0,0]$,
$\imodels{\Isol}{\D{x} = \oustrm[x] \land \ousfml[x]}$
for all $t\in[0,0]$.
Thus, $(\iget[state]{\I},\iget[state]{\It}) = \iaccessible[\pode{\D{x}=\oustrm[x]}{\ousfml[x]}]{\I}{\Isol[0]}$.

\item[(\irref{ODEidemp})]
$\pode{\D{x}=\oustrm[x]}{\ousfml[x]} \refines \pode{\D{x}=\oustrm[x]}{\ousfml[x]};\pode{\D{x}=\oustrm[x]}{\ousfml[x]}$ derives from (\irref{DX}), (\irref{DWref}) and (\irref{DEref}).

For the other direction, take any states $\iget[state]{\I},\iget[state]{\Itb},\iget[state]{\It}$ with $\iaccessible[\pode{\D{x}=\oustrm[x]}{\ousfml[x]}]{\I}{\Itb}$ and $\iaccessible[\pode{\D{x}=\oustrm[x]}{\ousfml[x]}]{\Itb}{\It}$.
There are two functions $\sol_1 : [0,r_1]$ and $\sol_2 : [0,r_2]$ with $\iget[state]{\I} = \sol_1(0)$ on $\scomplement{\set{\D{x}}}$, and $\iget[state]{\Itb} = \sol_1(r_1)$, and $\iget[state]{\Itb} = \sol_2(0)$ on $\scomplement{\set{\D{x}}}$, and $\iget[state]{\It} = \sol_2(r_2)$.
Since both $\sol_1(r_1)\in{\D{x} = \oustrm[x] \land \ousfml[x]}$ and 
$\sol_2(0)\in{\D{x} = \oustrm[x] \land \ousfml[x]}$, this means that $\sol_1(r_1) = \sol_2(0)$. By defining $\sol : [0, r_1+r_2]$ with $\sol(t) = \sol_1(t)$ on $[0,r_1]$ and $\sol(t) = \sol_2(t-r_1)$ on $[r_1,r_2]$, we have that $(\iget[state]{\I}, \sol(r_1+r_2)) = \iaccessible[\pode{\D{x}=\oustrm[x]}{\ousfml[x]}]{\I}{\It}$.

\item[(\irref{randcomm})]
If $\iaccessible[\prandom{x};\prandom{y}]{\I}{\It}$, then there is $\iget[state]{\Itb}, r_1, r_2$ such that $\iget[state]{\Itb} = \modif{\iget[state]{\I}}{x}{r_1}$ and $\iget[state]{\It} = \modif{\iget[state]{\Itb}}{y}{r_2}$.
But then, with $\iget[state]{\Itil} = \modif{\iget[state]{\I}}{y}{r_2}$, we have $\iget[state]{\It} = \modif{\iget[state]{\Itil}}{x}{r_1}$, so $\iaccessible[\prandom{y};\prandom{x}]{\I}{\It}$.
As this holds for all $\iget[state]{\It}$, we can conclude that $\imodels{\I}{\prandom{x};\prandom{y} \refines \prandom{y};\prandom{x}}$.
The equivalence holds by symmetry.

\item[(\irref{randtest})]
Take $\iaccessible[\prandom{x};\ptest{\osfml}]{\I}{\It}$.
Then $\iaccessible[\prandom{x}]{\I}{\It}$ and $\imodels{\It}{\osfml}$.
By definition, it means $() \in \getInterp{\It}{\osfml}$ where $()$ is the 0-ary tuple so, $\imodel{\It}{\osfml} = \State$.
In particular, $\imodels{\I}{\osfml}$, which implies $\iaccessible[\ptest{\osfml};\prandom{x}]{\I}{\It}$.

The other direction is similar.
\end{itemize}

\section{Derived Axioms}\label{app:derived}
In this section, we use the axioms of the logic to derive some additional axioms.
The proofs follow the same structure of introducing intermediate formulas/programs where each step is justified by some axioms.
For this purpose, we write the following for proving $\asfml \lylpmi \csfml$ via an intermediate formula $\bsfml$:
\begin{align*}
	\asfml &\lylpmi \bsfml \text{ by }(A)\\
	&\lbisubjunct \csfml \text{ by }(B)
\end{align*}
The empty space to the left of the implication/equivalence symbol corresponds to the right hand side of the previous line.
In the example above, this notation means that $\asfml \lylpmi \bsfml$ is derivable by the axiom $(A)$ and $\bsfml \lbisubjunct \csfml$ is derivable by the axiom $(B)$.
The derivation of $\asfml \lylpmi \csfml$ is then done by combining the two results using standard propositional logic reasoning.
The proof of (\irref{cupcomm}) below makes the left-hand side (in blue) explicit to show the structure of the proof.
The same notation is used for programs:
\begin{align*}
	\asprg &\prgeq \bsprg \text{ by }(C)\\
	&\refines \csprg \text{ by }(D)
\end{align*}
This means that $\asprg \prgeq \bsprg$ is derivable by the axiom $(C)$ and $\bsprg \refines \csprg$ is derivable by the axiom $(D)$.
The derivation of $\asprg \refines \csprg$ is then obtained via the transitivity of the refinement relation (\irref{leqtrans}).
The proof of (\irref{assigntest}) makes the left-hand side (in blue) explicit to show the structure of the proof.

\begin{itemize}
\item[(\irref{cupcomm})]
\begin{align*}
	\pchoice{\ausprg}{\busprg} \refines \pchoice{\busprg}{\ausprg} &\lbisubjunct (\ausprg \refines \pchoice{\busprg}{\ausprg})\land (\busprg \refines \pchoice{\busprg}{\ausprg}) &\text{by }(\irref{choicel})\\
	\axkey{(\ausprg \refines \pchoice{\busprg}{\ausprg})\land (\busprg \refines \pchoice{\busprg}{\ausprg})} &\lbisubjunct (\busprg \refines \pchoice{\busprg}{\ausprg})\land (\ausprg \refines \pchoice{\busprg}{\ausprg}) &\text{by FOL}\\
	\axkey{(\busprg \refines \pchoice{\busprg}{\ausprg})\land (\ausprg \refines \pchoice{\busprg}{\ausprg})} &\lbisubjunct \pchoice{\busprg}{\ausprg} \refines \pchoice{\busprg}{\ausprg} &\text{by }(\irref{choicel})\\
	\axkey{\pchoice{\busprg}{\ausprg} \refines \pchoice{\busprg}{\ausprg}} &\lbisubjunct \ltrue &\text{by }(\irref{leqrefl})
\end{align*}
The other case holds by symmetry.

\item[(\irref{cupidemp})]
\begin{align*}
	\pchoice{\ausprg}{\ausprg} = \ausprg &\lbisubjunct \pchoice{\ausprg}{\ausprg} \refines \ausprg \land \ausprg \refines \pchoice{\ausprg}{\ausprg} &\text{by }(\irref{leqantisym})\\
	&\lylpmi (\ausprg\refines \ausprg \land \ausprg \refines \ausprg) \land (\ausprg \refines \ausprg \lor \ausprg \refines \ausprg) &\text{by }(\irref{choicel}) \text{ and }(\irref{choicer})\\
	&\lbisubjunct \ltrue &\text{by }(\irref{leqrefl})
\end{align*}

\item[(\irref{cupassoc})]
We write $\asprg \eqdef \pchoice{\ausprg}{(\pchoice{\busprg}{\cusprg})}$.
\begin{align*}
	\pchoice{(\pchoice{\ausprg}{\busprg})}{\cusprg} \refines \asprg &\lbisubjunct (\ausprg \refines \asprg \land \busprg \refines \asprg) \land \cusprg \refines \asprg &\text{by }(\irref{choicel})\\
	&\lbisubjunct \ausprg \refines \asprg \land (\busprg \refines \asprg \land \cusprg \refines \asprg) &\text{by FOL}\\
	&\lbisubjunct \pchoice{\busprg}{\ausprg} \refines \pchoice{\busprg}{\ausprg} &\text{by }(\irref{choicel})\\
	&\lbisubjunct \ltrue &\text{by }(\irref{leqrefl})
\end{align*}

\item[(\irref{testand})]
\begin{align*}
	\ptest{\osfml \land \osfmlb} \refines  {\ptest{\osfml}};\ptest{\osfmlb}
	&\lbisubjunct {\ptest{\osfml \land \osfmlb}};\ptest{\ltrue} \refines  {\ptest{\osfml}};\ptest{\osfmlb} &\text{by } (\irref{seqidr})\\
	&\lylpmi {\ptest{\osfml \land \osfmlb}} \refines  {\ptest{\osfml}} \land \dbox{\ptest{\osfml \land \osfmlb}}\ptest{\ltrue} \refines  {\ptest{\osfmlb}} &\text{by } (\irref{sequence})\\
	&\lbisubjunct (\osfml \land \osfmlb \limply \osfml) \land \dbox{\ptest{\osfml \land \osfml}}(\ltrue \limply \osfmlb) &\text{by } (\irref{test})\\
	&\lbisubjunct (\osfml \land \osfmlb \limply \osfml) \land (\osfml \land \osfmlb \limply \ltrue \limply \osfmlb) &\text{by } (\irref{testbax})\\
	&\lbisubjunct \ltrue &\text{by FOL}\\
	\\
	\ptest{\osfml};\ptest{\osfmlb} \refines  {\ptest{\osfml \land \osfmlb}}
	&\lbisubjunct {\ptest{\osfml}};\ptest{\osfmlb} \refines  {\ptest{\ltrue}};\ptest{\osfml \land \osfmlb}  &\text{by } (\irref{seqidl})\\
	&\lylpmi  \ptest{\osfml} \refines  {\ptest{\ltrue}} \land \dbox{\ptest{\osfml}}\ptest{\osfmlb} \refines  {\ptest{(\osfml \land \osfmlb)}} &\text{by } (\irref{sequence})\\
	&\lbisubjunct (\osfml \limply \ltrue) \land \dbox{\ptest{\osfml \limply \osfmlb}}(\osfml \land \osfmlb) &\text{by } (\irref{test})\\
	&\lbisubjunct (\osfml \limply \ltrue) \land (\osfml \limply \osfmlb \limply \osfml \land \osfmlb) &\text{by } (\irref{testbax})\\
	&\lbisubjunct \ltrue &\text{by FOL}\\
\end{align*}

Then we can conclude by (\irref{leqantisym}).

\item[(\irref{testor})]
\begin{align*}
	\ptest{\osfml \lor \osfmlb} \refines \pchoice{\ptest{\osfml}}{\ptest{\osfmlb}}
	&\lylpmi (\ptest{\osfml \lor \osfmlb} \refines  {\ptest{\osfml}}) \lor (\ptest{\osfml \lor \osfmlb} \refines  {\ptest{\osfmlb}}) &\text{by } (\irref{choicer})\\
	&\lbisubjunct (\osfml \lor \osfmlb \limply \osfml) \lor (\osfml \lor \osfmlb \limply \osfmlb) &\text{by } (\irref{test})\\
	&\lbisubjunct \ltrue &\text{by FOL}\\
	\\
	\pchoice{\ptest{\osfml}}{\ptest{\osfmlb}} \refines  {\ptest{\osfml \lor \osfmlb}}
	&\lbisubjunct (\ptest{\osfml} \refines  {\ptest{\osfml \lor \osfmlb}}) \land (\ptest{\osfmlb} \refines  {\ptest{\osfml \lor \osfmlb}}) &\text{by } (\irref{choicel})\\
	&\lbisubjunct (\osfml \limply \osfml \lor \osfmlb) \land (\osfmlb \limply \osfml \lor \osfmlb) &\text{by } (\irref{test})\\
	&\lbisubjunct \ltrue &\text{by FOL}\\
\end{align*}
Then we can conclude by (\irref{leqantisym}).

\item[(\irref{seqdistr})] By (\irref{leqantisym}), we prove both directions.
\begin{align*}
	\pchoice{( \ausprg;\cusprg)}{( \busprg;\cusprg)} \refines ( \pchoice{\ausprg}{\busprg});\cusprg &\lbisubjunct (\ausprg;\cusprg \refines ( \pchoice{\ausprg}{\busprg});\cusprg) \land (\busprg;\cusprg \refines ( \pchoice{\ausprg}{\busprg});\cusprg) &\text{by } (\irref{choicel})\\
	&\lylpmi \ausprg \refines \pchoice{\ausprg}{\busprg} \land \dbox{\ausprg}\cusprg \refines \cusprg \land \busprg \refines \pchoice{\ausprg}{\busprg} \land \dbox{\busprg}\cusprg \refines \cusprg &\text{by } (\irref{sequence})\\
	&\lbisubjunct \ausprg \refines \pchoice{\ausprg}{\busprg} \land \busprg \refines \pchoice{\ausprg}{\busprg} &\text{by } (\irref{leqrefl})\\
	&\lbisubjunct \pchoice{\ausprg}{\busprg} \refines \pchoice{\ausprg}{\busprg} &\text{by } (\irref{choicel})\\
	&\lbisubjunct \ltrue &\text{by } (\irref{leqrefl})
\end{align*}

\begin{alignat*}{3}
	(\pchoice{\ausprg}{\busprg});\cusprg \refines &\,\pchoice{( \ausprg;\cusprg)}{( \busprg;\cusprg)}\\
	&\lbisubjunct ((\pchoice{\ausprg}{\busprg});\cusprg \refines \ausprg;\cusprg) \lor ((\pchoice{\ausprg}{\busprg});\cusprg \refines \busprg;\cusprg) &\text{by } (\irref{choicer})\\
	&\lylpmi (\pchoice{\ausprg}{\busprg} \refines \ausprg \land \dbox{\pchoice{\ausprg}{\busprg}}\cusprg \refines \cusprg) \lor (\pchoice{\ausprg}{\busprg} \refines \busprg \land \dbox{\pchoice{\ausprg}{\busprg}}\cusprg \refines \cusprg) &\text{by } (\irref{sequence})\\
	&\lbisubjunct (\pchoice{\ausprg}{\busprg} \refines \ausprg) \lor (\pchoice{\ausprg}{\busprg} \refines \busprg) &\text{by } (\irref{leqrefl})\\
	&\lbisubjunct \pchoice{\ausprg}{\busprg} \refines \pchoice{\ausprg}{\busprg} &\text{by } (\irref{choicel})\\
	&\lbisubjunct \ltrue &\text{by } (\irref{leqrefl})
\end{alignat*}

\end{itemize}

The reverse implication of (\irref{testdet}) and (\irref{assigndet}) is consequence of the general implication:
\begin{align*}
	\ausprg;\busprg \refines \ausprg;\cusprg &\lylpmi (\ausprg \refines \ausprg) \land \dbox{\ausprg}\busprg \refines \cusprg &\text{by } (\irref{sequence})\\
	&\lylpmi \dbox{\ausprg}\busprg \refines \cusprg &\text{by } (\irref{leqrefl})	
\end{align*}

As a consequence, we have the following derivation:
\[
\linfer[MPax]{
	\linfer{\lclose}{\dbox{\ausprg}\busprg \refines \cusprg \limply \ausprg;\busprg \refines \ausprg;\cusprg} &
	\linfer[Gax]{\busprg \refines \cusprg}{\dbox{\ausprg}\busprg \refines \cusprg}
}{\ausprg;\busprg \refines \ausprg;\cusprg}
\]
and symmetrically we can derive:
\[
\linfer{\ausprg \refines \busprg}
{\ausprg;\cusprg \refines \busprg;\cusprg}
\]

Associated with (\irref{leqtrans}), this rule means we can use axioms to rewrite inside sequences. This will be done implicitly in the following derivations.

\begin{itemize}
\item[(\irref{assignsub})]
\begin{align*}
	\pumod{x}{\oustrm};\pumod{y}{\oustrmb[\oustrm]} \refines \pumod{x}{\oustrm};\pumod{y}{\oustrmb[x]} &\lbisubjunct \dbox{\pumod{x}{\oustrm}}{\pumod{y}{\oustrmb[\oustrm]} \refines \pumod{y}{\oustrmb[x]}} &\text{by } (\irref{assigndet})\\
	&\lbisubjunct \pumod{y}{\oustrmb[\oustrm]} \refines \pumod{y}{\oustrmb[\oustrm]} &\text{by } (\irref{assignbax})\\
	&\lbisubjunct \ltrue &\text{by } (\irref{leqrefl})
\end{align*}
The other direction is similar.

In fact, by using the same reasoning, we can derive:
\[
	\cinferenceRule[assigntestaux|${:}{=}_{\text{test}_s}$]{Substitution in test}
	{\linferenceRule[eq]
	  {\pumod{x}{\oustrm};\ptest{\ousfml[\oustrm]}}
	  {\pumod{x}{\oustrm};\ptest{\ousfml[x]}}
	}{}
\]

\item[(\irref{assigntest})]
\begin{align*}
	\pumod{x}{\oustrm};\ptest{\ousfml[x]}
	&\prgeq \pumod{x}{\oustrm};\ptest{\ousfml[\oustrm]} &\text{by } (\irref{assigntestaux})\\
	\axkey{\pumod{x}{\oustrm};\ptest{\ousfml[\oustrm]}} &\prgeq \prandom{x};\ptest{(x = \oustrm \land \ousfml[\oustrm])} &\text{by } (\irref{testand})\\
	\axkey{\prandom{x};\ptest{(x = \oustrm \land \ousfml[\oustrm])}} &\prgeq \prandom{x};\ptest{(\ousfml[\oustrm]\land x = \oustrm)} &\text{by FOL}\\
	\axkey{\prandom{x};\ptest{(\ousfml[\oustrm]\land x = \oustrm)}}&\prgeq \prandom{x};\ptest{\ousfml[\oustrm]};\ptest{x = \oustrm} &\text{by } (\irref{testand})\\
	\axkey{\prandom{x};\ptest{\ousfml[\oustrm]};\ptest{x = \oustrm}}&\prgeq {\ptest{\ousfml[\oustrm]}};\prandom{x};\ptest{x = \oustrm} &\text{by } (\irref{randtest})\\
	\axkey{\ptest{\ousfml[\oustrm]};\prandom{x};\ptest{x = \oustrm}}&\prgeq {\ptest{\ousfml[\oustrm]}};\pumod{x}{\oustrm} &\text{by } (\irref{update})
\end{align*}

\item[(\irref{assignmerge})]
\begin{align*}
	\pumod{x}{\oustrm};\pumod{x}{\oustrmb[x]} &\prgeq \prandom{x};\ptest{x = \oustrm};\pumod{x}{\oustrmb[x]} &\text{by } (\irref{update})\\
	&\prgeq \prandom{x};\ptest{\lexists y (y = \oustrm \land x = \oustrmb[y])} &\text{by } (\irref{randtestassignmerge})\\
	&\prgeq \prandom{x};\ptest{x = \oustrmb[\oustrm]} &\text{by FOL}\\
	&\prgeq \pumod{x}{\oustrmb[\oustrm]} &\text{by } (\irref{update})
\end{align*}

\item[(\irref{assignrandcomm})]
\begin{align*}
	\prandom{y};\pupdate{\pumod{x}{\oustrm}} &\prgeq \prandom{y};\prandom{x};\ptest{x = \oustrm} &\text{by } (\irref{update})\\
	&\prgeq \prandom{x};\prandom{y};\ptest{x = \oustrm} &\text{by } (\irref{randcomm})\\
	&\prgeq \prandom{x};\ptest{x = \oustrm};\prandom{y} &\text{by } (\irref{randtest})\\
	&\prgeq \pupdate{\pumod{x}{\oustrm}};\prandom{y} &\text{by } (\irref{update})
\end{align*}

\item[(\irref{assigncomm})]
\begin{align*}
	\pupdate{\pumod{x}{\oustrm}};\pupdate{\pumod{y}{\oustrmb[x]}}
	&\prgeq \pupdate{\pumod{x}{\oustrm}};\pupdate{\pumod{y}{\oustrmb[\oustrm]}} &\text{by } (\irref{assignsub})\\
	&\prgeq \prandom{x};\ptest{x = \oustrm};\pupdate{\pumod{y}{\oustrmb[\oustrm]}} &\text{by } (\irref{update})\\
	&\prgeq \prandom{x};\pupdate{\pumod{y}{\oustrmb[\oustrm]}};\ptest{x = \oustrm} &\text{by } (\irref{assigntest})\\
	&\prgeq \pupdate{\pumod{y}{\oustrmb[\oustrm]}};\prandom{x};\ptest{x = \oustrm} &\text{by } (\irref{assignrandcomm})\\ 
	&\prgeq \pupdate{\pumod{y}{\oustrmb[\oustrm]}};\pupdate{\pumod{x}{\oustrm}} &\text{by } (\irref{update})
\end{align*}
\end{itemize}
We also derive the example from \rref{sec:dRL-calculus}: $\pupdate{\umod{x}{f}};\prandom{x} = \prandom{x}$

\begin{align*}
	\pupdate{\umod{x}{f}};\prandom{x} &\prgeq \prandom{x};\ptest{x = f};\prandom{x} &\text{by } (\irref{update})\\
	&\prgeq \prandom{x};\ptest{\exists y (y = f)} &\text{by } (\irref{randtestmerge})\\
	&\prgeq \prandom{x};\ptest{\ltrue} &\text{by FOL}\\
	&\prgeq \prandom{x} &\text{by } (\irref{seqidr})
\end{align*}

We can also obtain the following derivation:
\[
\linfer[MPax]{
	\linfer[Kax]{\lclose}{\dbox{\ausprg}{(\ausfml \limply \busfml)} \limply (\dbox{\ausprg}{\ausfml} \limply \dbox{\ausprg}{\busfml})}
	&
	\linfer[Gax]{\ausfml \limply \busfml}
	{\dbox{\ausprg}{(\ausfml \limply \busfml)}}
}
{\dbox{\ausprg}{\ausfml} \limply \dbox{\ausprg}{\busfml}}
\]
This derivation ensures that an implication, e.g.\ (\irref{testbax}), can be used inside a box operator.

\begin{itemize}
	\item[(\irref{assignbeqax})]
\begin{align*}
	\forall x(x = f \limply \ousfml[x]) &\lbisubjunct \dbox{\prandom{x}}{(x = f \limply \ousfml[x])} &\text{by } (\irref{randomb})\\
	&\lbisubjunct \dbox{\prandom{x}}{\dbox{\ptest{x = f}}{\ousfml[x]}} &\text{by } (\irref{testbax})\\
	&\lbisubjunct \dbox{\prandom{x};\ptest{x = f}}{\ousfml[x]} &\text{by } (\irref{composebax})\\
	&\lbisubjunct \dbox{\pupdate{\umod{x}{f}}}{\ousfml[x]} &\text{by } (\irref{boxleq} \text{ and } (\irref{update})
\end{align*}

\item[(\irref{iteratebax})]
\begin{align*}
	\ausfmlax \land \dbox{\ausprgax}{\dbox{\prepeat{\ausprgax}}{\ausfmlax}} &\lbisubjunct (\ltrue \limply \ausfmlax) \land \dbox{\ausprgax}{\dbox{\prepeat{\ausprgax}}{\ausfmlax}} &\text{by FOL}\\
	&\lbisubjunct \dbox{\ptest{\ltrue}}{\ausfmlax} \land \dbox{\ausprgax}{\dbox{\prepeat{\ausprgax}}{\ausfmlax}} &\text{by } (\irref{testbax})\\
	&\lbisubjunct \dbox{\ptest{\ltrue}}\ausfmlax \land \dbox{\ausprgax;\prepeat{\ausprgax}}{\ausfmlax} &\text{by } (\irref{composebax})\\
	&\lbisubjunct \dbox{\pchoice{\ptest{\ltrue}}{\ausprgax;\prepeat{\ausprgax}}}{\ausfmlax} &\text{by } (\irref{choicebax})\\
	&\lbisubjunct \dbox{\prepeat{\ausprgax}}{\ausfmlax} &\text{by } (\irref{boxleq}) \text{ and } (\irref{unfold-l})
\end{align*}

\item[(\irref{DEax})]
\begin{alignat*}{3}
	\dbox{\pevolvein{\D{x}=f(x)}{q(x)}}{&\dbox{\Dupdate{\Dumod{\D{x}}{\genDE{x}}}}{\ausfmlax}}\\
	& \lbisubjunct \dbox{\pevolvein{\D{x}=f(x)}{q(x)};\Dupdate{\Dumod{\D{x}}{\genDE{x}}}}{\ausfmlax} &\text{by } (\irref{composebax})\\
	&\lbisubjunct \dbox{\pevolvein{\D{x}=f(x)}{q(x)}}{\ausfmlax} &\text{by } (\irref{boxleq}) \text{ and } (\irref{DEref})
\end{alignat*}

\item[(\irref{diffweakenax})]
\begin{alignat*}{3}
	\dbox{\pevolvein{\D{x}=\genDE{x}}{q(x)}}{(q(x) &\limply p(x))}\\
	&\lbisubjunct \dbox{\pevolvein{\D{x}=\genDE{x}}{q(x)}}{\dbox{\ptest{q(x)}}{p(x)}} &\text{by } (\irref{testbax})\\
	&\lbisubjunct \dbox{\pevolvein{\D{x}=\genDE{x}}{q(x)};\ptest{q(x)}}{p(x)} &\text{by } (\irref{composebax})\\
	&\lbisubjunct \dbox{\pevolvein{\D{x}=\genDE{x}}{q(x)}}{p(x)} &\text{by } (\irref{boxleq}) \text{ and below}
\end{alignat*}
The last step requires a derivation of \(\pevolvein{\D{x}=\genDE{x}}{q(x)};\ptest{q(x)} \prgeq \pevolvein{\D{x}=\genDE{x}}{q(x)}\).
We derive both refinements:
\begin{align*}
	\pevolvein{\D{x}=\genDE{x}}{q(x)};\ptest{q(x)} &\refines \pevolvein{\D{x}=\genDE{x}}{\ltrue};\ptest{q(x)} &\text{by } (\irref{test})\\
	&\refines \pevolvein{\D{x}=\genDE{x}}{q(x)} &\text{by } (\irref{seqidr})
\end{align*}
and
\begin{align*}
	\pevolvein{\D{x}=\genDE{x}}{q(x)} &\prgeq {\ptest{q(x)}};\pevolvein{\D{x}=\genDE{x}}{q(x)};\ptest{q(x)} &\text{by } (\irref{DWref})\\
	&\refines {\ptest{\ltrue}};\pevolvein{\D{x}=\genDE{x}}{q(x)};\ptest{q(x)} &\text{by } (\irref{test})\\
	&\refines \pevolvein{\D{x}=\genDE{x}}{q(x)};\ptest{q(x)} &\text{by } (\irref{seqidl})
\end{align*}

\item[(\irref{DCax})]
First, we derive $(\pevolvein{\D{x}=\genDE{x}}{q(x)}) \prgeq (\pevolvein{\D{x}=\genDE{x}}{q(x)\land r(x)})$ from the assumption $\dbox{\pevolvein{\D{x}=\genDE{x}}{q(x)}}{r(x)}$ by deriving both refinements and (\irref{leqantisym}).
\begin{alignat*}{3}
	\pevolvein{\D{x}=\genDE{x}}{&q(x)} \refines \pevolvein{\D{x}=\genDE{x}}{(q(x)\land r(x))}\\
	&\lbisubjunct \dbox{\pevolvein{\D{x}=\genDE{x}}{q(x)}}{(\D{x}=\genDE{x}\land q(x)\land r(x))} &\text{by } (\irref{ode})\\
	&\lbisubjunct \dbox{\pevolvein{\D{x}=\genDE{x}}{q(x)}}{\dbox{\pupdate{\pumod{\D{x}}{\genDE{x}}}}{(\D{x}=\genDE{x}\land q(x)\land r(x))}} &\text{by } (\irref{DEax})\\
	&\lbisubjunct \dbox{\pevolvein{\D{x}=\genDE{x}}{q(x)}}{(\genDE{x}=\genDE{x}\land q(x)\land r(x))} &\text{by } (\irref{assignbax})\\
	&\lbisubjunct \dbox{\pevolvein{\D{x}=\genDE{x}}{q(x)}}{(q(x)\land r(x))} &\text{by FOL}\\
	&\lbisubjunct \dbox{\pevolvein{\D{x}=\genDE{x}}{q(x)}}{(q(x)\limply q(x)\land r(x))} &\text{by } (\irref{diffweakenax})\\
	&\lbisubjunct \dbox{\pevolvein{\D{x}=\genDE{x}}{q(x)}}{r(x)} &\text{by FOL}
\end{alignat*}
\begin{alignat*}{3}
	\pevolvein{\D{x}=\genDE{x}}{q(x) \land r(x)} &\refines \pevolvein{\D{x}=\genDE{x}}{q(x)}\\
	&\lylpmi \dbox{\pupdate{\pumod{\D{x}}{\genDE{x}}}}{\dbox{\ptest{q(x)\land r(x)}}{(\D{x}=\genDE{x}\land q(x))}} &\text{by } (\irref{DX})\\
	&\lylpmi \dbox{\ptest{q(x)\land r(x)}}{(\genDE{x}=\genDE{x}\land q(x))} &\text{by } (\irref{assignbax})\\
	&\lylpmi q(x)\land r(x) \limply \genDE{x}=\genDE{x}\land q(x) &\text{by } (\irref{testbax})\\
	&\lylpmi \ltrue &\text{by FOL}
\end{alignat*}

From that, the derivation of (\irref{DCax}) is:
\begin{alignat*}{3}
	(\dbox{\pevolvein{\D{x}=\genDE{x}}{q(x)}}{p(x)} &\lbisubjunct \dbox{\pevolvein{\D{x}=\genDE{x}}{q(x)\land r(x)}}{p(x)})\\
	&\lylpmi (\pevolvein{\D{x}=\genDE{x}}{q(x)}) \prgeq (\pevolvein{\D{x}=\genDE{x}}{q(x)\land r(x)}) &\text{by } (\irref{boxleq})\\
	&\lylpmi \dbox{\pevolvein{\D{x}=\genDE{x}}{q(x)}}{r(x)} &\text{by above}
\end{alignat*}
\end{itemize}
\section{Proofs for \rref{sec:decidable}}\label{app:sec6}

\begin{proof}[\rref{lem:decidiscret}]
	For any $x \notin \boundvars{ctrl_a}$, we have $ctrl_a \prgeq \pupdate{\umod{x}{x}};ctrl_a$ by (\irref{stutter}).
	Thus, without loss of generality, we assume $\boundvars{ctrl_a} = \boundvars{ctrl_b}$.
	Then, for all variables $x \in \boundvars{ctrl_a} = \boundvars{ctrl_b}$ and fresh variables $x^+$, we have $ctrl_a \refines ctrl_b$ iff $\dbox{\pupdate{\umod{x^+}{x}}}ctrl_a \refines ctrl_b$ by (\irref{assignbax}), iff $\pupdate{\umod{x^+}{x}};ctrl_a \refines \pupdate{\umod{x^+}{x}};ctrl_b$ by (\irref{assigndet}).
	In the following, we write $\pupdate{\umod{x^+}{x}}$ for the sequence of assignments where $x^+$ corresponds to all fresh variables introduced and $x$ corresponds to all bound variables.
	Axiom (\irref{sequence}) will be used without explicit mention in the schematic form
	\[\linfer{\asprg \refines \csprg & \bsprg \refines \dsprg}
	{\asprg;\bsprg \refines \csprg;\dsprg}\]

	We show that for any program $\asprg$ that is a sequence of tests and (possibly nondeterministic) assignments, $(\pupdate{\umod{x^+}{x}};\asprg) \prgeq (\pupdate{\umod{x^+}{x}};\prandom{x};\ptest{\asfml(x^+,x)})$ for some first-order formula $\asfml(x^+,x)$. 
	Using (\irref{seqassoc}) if required, the left side of a sequence is assumed to always be an (possibly nondeterministic) assignment or a test.
	We prove by induction on the structure of $\asprg$ that for all $x \cap \scomplement{\boundvars{\asprg}} \subseteq y \subseteq x$ and first-order formula $\asfml(x^+,y)$, the program $\pupdate{\umod{x^+}{x}};\prandom{y};\ptest{\asfml(x^+,y)};\asprg$ is equivalent to one of the form $\pupdate{\umod{x^+}{x}};\prandom{x};\ptest{\bsfml(x^+,x)}$.
	\begin{enumerate}
		\item For $\asprg \mequiv {\ptest{\bsfml(x)}}$, we have $\scomplement{\boundvars{\asprg}} = \emptyset$ so $y = x$.
		In that case, $(\pupdate{\umod{x^+}{x}};\prandom{x};\ptest{\asfml(x^+,x)};\ptest{\bsfml(x^+,x)}) \prgeq (\pupdate{\umod{x^+}{x}};\prandom{x};\ptest{\asfml(x^+,x) \land \bsfml(x^+,x)})$ by (\irref{testand}) and (\irref{sequence}).
		
		\item For $\asprg \mequiv \pupdate{\pumod{z}{f(x)}}$ when $z \notin y$, i.e.\ $x = y \cup \set{z}$, then 
		\begin{alignat*}{3}
			\pupdate{\umod{x^+}{x}};\prandom{y};&\ptest{\asfml(x^+,y)};\pupdate{\pumod{z}{f(y,z)}} \\
			&\prgeq \pupdate{\umod{y^+}{y}};\prandom{y};\pupdate{\umod{z^+}{z}};\ptest{\asfml(x^+,y)};\pupdate{\pumod{z}{f(y,z)}} &(\irref{assignrandcomm})\\
			&\prgeq \pupdate{\umod{y^+}{y}};\prandom{y};\pupdate{\umod{z^+}{z}};\pupdate{\pumod{z}{f(y,z)}};\ptest{\asfml(x^+,y)} &(\irref{assigntest})\\
			&\prgeq \pupdate{\umod{y^+}{y}};\prandom{y};\pupdate{\pumod{z^+}{z}};\pupdate{\pumod{z}{f(y,z^+)}};\ptest{\asfml(x^+,y)} &(\irref{assignsub})\\
			&\prgeq \pupdate{\umod{x^+}{x}};\prandom{y};\pupdate{\pumod{z}{f(y,z^+)}};\ptest{\asfml(x^+,y)} &(\irref{assignrandcomm})\\
			&\prgeq \pupdate{\umod{x^+}{x}};\prandom{y};\prandom{z};\ptest{z = f(y,z^+)};\ptest{\asfml(x^+,y)} & (\irref{update})\\
			&\prgeq \pupdate{\umod{x^+}{x}};\prandom{x};\ptest{z = f(y,z^+) \land \asfml(x^+,y)} & (\irref{testand})
		\end{alignat*}
		
		\item For $\asprg \mequiv \pupdate{\pumod{z}{f(x)}}$ when $y = y_2\cup\set{z}$, i.e.\ $x = y$, then
		\begin{alignat*}{3}
			\pupdate{\umod{x^+}{x}}&;\prandom{y};\ptest{\asfml(x^+,y)};\pupdate{\pumod{z}{f(y)}} \\
			&\prgeq \pupdate{\umod{x^+}{x}};\prandom{y_2};\prandom{z};\ptest{\asfml(x^+,y_2,z)};\pupdate{\pumod{z}{f(y_2,z)}} & (\irref{randcomm})\\
			&\prgeq \pupdate{\umod{x^+}{x}};\prandom{y_2};\prandom{z};\ptest{\lexists{z_2}{\asfml(x^+,y_2,z_2) \land z = f(y_2,z_2)}} &(\irref{randtestassignmerge})
		\end{alignat*}

		\item For $\asprg \mequiv \prandom{z}$ when $z \notin y$, i.e.\ $x = y\cup\set{z}$, then  we have by (\irref{randtest}) $$\pupdate{\umod{x^+}{x}};\prandom{y};\ptest{\asfml(x^+,y)};\prandom{z} \prgeq \pupdate{\umod{x^+}{x}};\prandom{y};\prandom{z};\ptest{\asfml(x^+,y)}$$
		
		\item For $\asprg \mequiv \prandom{z}$ when $y = y_2\cup\set{z}$, i.e.\ $y = x$, then 
		\begin{alignat*}{3}
			\pupdate{\umod{x^+}{x}};\prandom{y};\ptest{\asfml(x^+,y)}&;\prandom{z} \\
			&\prgeq \pupdate{\umod{x^+}{x}};\prandom{y_2};\prandom{z};\ptest{\asfml(x^+,y)};\prandom{z} &(\irref{randcomm})\\
			&\prgeq \pupdate{\umod{x^+}{x}};\prandom{y_2};\prandom{z}\ptest{\lexists{z}{\asfml(x^+,y)}} & (\irref{randtestmerge})
		\end{alignat*}
		
		\item For $\asprg \mequiv \csprg;\dsprg$, using (\irref{seqassoc}), $\pupdate{\umod{x^+}{x}};\prandom{y};\ptest{\asfml(x^+,y)};\csprg;\dsprg \prgeq (\pupdate{\umod{x^+}{x}};\prandom{y};\ptest{\asfml(x^+,y)};\csprg);\dsprg$. As we have that $\csprg$ is a test or an (possibly nondeterministic) assignment, we can apply a reasoning similar to the previous cases to $\csprg$. Then we can apply the induction hypothesis to $\dsprg$.
	\end{enumerate}

	This result can be extended to programs with choice.
	By (\irref{seqdistl}) and (\irref{seqdistr}), we can rewrite $\pupdate{\umod{x^+}{x}};\asprg$ as $\bigcup_i\pupdate{\umod{x^+}{x}};\asprg_i$ for some sequences of assignments and tests $\asprg_i$.
	By the proof above, this program is equivalent to $\bigcup_i\pupdate{\umod{x^+}{x}};\break\prandom{x};\asfml_i(x^+,x)$ for some formulas $\asfml_i(x^+,x)$.
	By (\irref{seqdistl}) and (\irref{testor}), we finally obtain that
	$\pupdate{\umod{x^+}{x}};\asprg \prgeq \pupdate{\umod{x^+}{x}};\prandom{x};\ptest{\bigvee_i\asfml_i(x^+,x)}$.

	As $ctrl_a$ and $ctrl_b$ are concrete, loop-free, discrete programs, they only contain (possibly nondeterministic) assignments, tests, sequences, and choices, so they correspond to the case above.
	Thus, we have that $ctrl_a \refines ctrl_b$ iff $\pupdate{\umod{x^+}{x}};\prandom{x};\ptest{\asfml(x^+,x)} \refines \pupdate{\umod{x^+}{x}};\prandom{x};\ptest{\bsfml(x^+,x)}$ for some formulas $\asfml(x^+,x)$ and $\bsfml(x^+,x)$.
	By (\irref{sequence}) and (\irref{Gax}), proving $\asfml(x^+,x) \limply \bsfml(x^+,x)$ entails $\pupdate{\umod{x^+}{x}};\prandom{x};\ptest{\asfml(x^+,x)} \refines \pupdate{\umod{x^+}{x}};\prandom{x};\ptest{\bsfml(x^+,x)}$, but by looking at the semantics, we can see that the reasoning is complete.
\end{proof}
\begin{proof}[\rref{lem:refineidemp}]
The derivation of the rule is given in the canonical proof.
We focus on the converse and use $\asprg$ (resp.\ $\bsprg$) for $ctrl_a;plant_a$ (resp.\ $ctrl_b;plant_b$). The proof is done in two steps.

First, we show that the following rule is derived.
 \[\linfer{\prepeat{\asprg} \refines \prepeat{\bsprg}}{\asprg\refines \prepeat{\bsprg}}\]
This can be done by transitivity:
\begin{align*}
	\asprg &\refines \asprg;\ptest{\ltrue} &(\irref{seqidr})\\
	&\refines \asprg;(\pchoice{\ptest{\ltrue}}{\asprg;\prepeat{\asprg}}) &(\irref{leqrefl})\text{ and }(\irref{choicer})\\
	&\refines \asprg;\prepeat{\asprg} &(\irref{unfold-l})\\
	&\refines \pchoice{\ptest{\ltrue}}{\asprg;\prepeat{\asprg}} &(\irref{leqrefl}) \text{ and } (\irref{choicer})\\
	&\refines \prepeat{\asprg} &(\irref{unfold-l})\\
	&\refines \prepeat{\bsprg} &(\text{hypothesis})
\end{align*}
Then, we have to prove why the loop can be removed from $\prepeat{\bsprg}$.
For all $\iaccessible[\asprg]{\I}{\It}$, $\iget[state]{\It} = \sol(r)$ for some function $\sol$, with a $\bar{u}$-variation $\iget[state]{\Itb}$ such that we have $\iaccessible[ctrl_a]{\I}{\Itb}$ and $\sol(0)$ a $\D{\bar{y}}$-variation of $\iget[state]{\Itb}$.
This means that $\iaccessible[\asprg]{\I}{\Isol}$ holds for all $t\in[0,r]$ and $\sol(t)$ is a $(\D{\bar{y}}\cup\bar{y})$-variation of $\iget[state]{\Itb}$.
As $\asprg \refines \prepeat{\bsprg}$, we have $\iaccessible[\prepeat{\bsprg}]{\I}{\Isol}$ for all $t\in[0,r]$.

Recall that the system has a global clock, which is written $t_0$ to prevent confusion.
As $t_0$ is only modified in the plant with the differential equation $\D{t_0} = 1$, we have $\ivaluation{\I}{t_0} = \ivaluation{\Itb}{t_0}$ and $\ivaluation{\Isol}{t_0} = \ivaluation{\Itb}{t_0} + t$.

As $\bar{y}$ is unchanged in $\sol(0)$ and contains the strictly increasing time $t_0$, we know that the values of $\bar{y}$ must remain constant throughout the execution of $\prepeat{\bsprg}$.
The only behavior for a differential equation with global time where the values of $\bar{y}$ remain constant is the one described in the axiom (\irref{DX}).
This can be rewritten as having $(\iget[state]{\I},\sol(0))\in\iaccess[\prepeat{(ctrl_b;\pupdate{\umod{\D{\bar{y}}}{q(\bar{y},\bar{u})}};\ptest{R})}]{\I}{\Isol}$ where $plant_b \mequiv (\pode{\D{\bar{y}}=q(\bar{y},\bar{u})}{R})$ for some polynomial $q$.

This program refines by $\prepeat{(ctrl_b;\pupdate{\umod{\D{\bar{y}}}{q(\bar{y},\bar{u})}})}$, by (\irref{seqidr}), (\irref{test}) and (\irref{unloop}).
By applying (\irref{unfold-r}) before using the refinement above, i.e.\ removing the test $\ptest{R}$, we obtain $(\iget[state]{\I},\sol(0))\in\iaccess[\pchoice{\ptest{\ltrue}}{\prepeat{(ctrl_b;\pupdate{\umod{\D{\bar{y}}}{q(\bar{y},\bar{u})}})};ctrl_b;\pupdate{\umod{\D{\bar{y}}}{q(\bar{y},\bar{u})}};\ptest{R}}]{\I}{\Isol}$.
Since $ctrl_b$ is idempotent and does not depend on $\D{\bar{y}}$, the first assignment to $\D{\bar{y}}$ is irrelevant in $ctrl_b;\pupdate{\umod{\D{\bar{y}}}{q(\bar{y},\bar{u})}};ctrl_b;\pupdate{\umod{\D{\bar{y}}}{q(\bar{y},\bar{u})}}$
The program $ctrl_b;\pupdate{\umod{\D{\bar{y}}}{q(\bar{y},\bar{u})}};\ptest{R}$ is actually idempotent. 
This means that $ctrl_b;\pupdate{\umod{\D{\bar{y}}}{q(\bar{y},\bar{u})}}$ is also idempotent.
Thus, by using (\irref{loopl}), we have that $(\prepeat{(ctrl_b;\pupdate{\umod{\D{\bar{y}}}{q(\bar{y},\bar{u})}})};ctrl_b;\pupdate{\umod{\D{\bar{y}}}{q(\bar{y},\bar{u})}}) \prgeq (ctrl_b;\pupdate{\umod{\D{\bar{y}}}{q(\bar{y},\bar{u})}})$.
So $(\iget[state]{\I},\sol(0))\in\iaccess[\pchoice{\ptest{\ltrue}}{ctrl_b;\pupdate{\umod{\D{\bar{y}}}{q(\bar{y},\bar{u})}};\ptest{R}}]{\I}{\Isol} \subseteq \iaccess[\pchoice{\ptest{\ltrue}}{\bsprg}]{\I}{\Isol}$.

Let us show that we can remove the left test.
As $ctrl_a$ does not depend on $\D{\bar{y}}$, take $\D{x}\in\D{\bar{y}}$ and $s \neq \getValue{\I}{\D{x}}$ we have $(\modif{\iget[state]{\I}}{\D{x}}{s}, \sol(0))\in\iaccess[\asprg]{\I}{\Isol}$, so $(\modif{\iget[state]{\I}}{\D{x}}{s}, \sol(0))\in\iaccess[\pchoice{\ptest{\ltrue}}{\bsprg}]{\I}{\Isol}$ holds.
$(\iget[state]{\I},\sol(0))\in\iaccess[\ptest{\ltrue}]{\I}{\Isol}$ only holds if $\iget[state]{\I} = \sol(0)$, meaning that as $\iget[state]{\I} \neq \modif{\iget[state]{\I}}{\D{x}}{s}$, we would have $(\modif{\iget[state]{\I}}{\D{x}}{s}, \sol(0))\in\iaccess[\bsprg]{\I}{\Isol}$.
The same reasoning as the one done for $\asprg$ can be done for $\bsprg$, which entails $(\iget[state]{\I}, \sol(0))\in\iaccess[\bsprg]{\I}{\Isol}$.

For $t > 0$, the global time $t_0$ is modified so at least one iteration must have occurred: $\iaccessible[\prepeat{\bsprg};\bsprg]{\I}{\Isol}$.
So for each $t\in(0,r]$, there is $\iget[state]{\Imut}$ such that we have $\iaccessible[\prepeat{\bsprg};ctrl_b]{\I}{\Imut}$ and $\iaccessible[\pode{\D{\bar{y}}=q(\bar{y},\bar{u})}{R}]{\Imut}{\Isol}$.
In particular, this entails $\getValue{\Isol}{\D{\bar{y}}} = q(\getValue{\Isol}{\bar{y}},\getValue{\Isol}{\bar{u}}) = q(\getValue{\Isol}{\bar{y}},\getValue{\Itb}{\bar{u}})$ for all $t\in[0,r]$.
The same equality can also be said about the plant of $\asprg$:
$\getValue{\Isol}{\D{\bar{y}}} = p(\getValue{\Isol}{\bar{y}},\getValue{\Itb}{\bar{u}})$ holds for all $t\in[0,r]$ where $plant_a \mequiv (\pode{\D{\bar{y}}=p(\bar{y},\bar{u})}{Q})$.
As the two polynomials $p(\cdot, \getValue{\Itb}{\bar{u}})$ and $q(\cdot, \getValue{\Itb}{\bar{u}})$ are equal at infinitely many points, they must be equal everywhere, meaning $\imodels{\Itb}{\dbox{plant_a}{p(\bar{y},\bar{u}) = q(\bar{y},\bar{u})}}$ holds.
Checking that $\imodels{\Isol}{R}$ follows from $\iaccessible[\prepeat{\bsprg}]{\I}{\Isol}$ for $t > 0$, and for $t = 0$, it follows from $(\iget[state]{\I},\sol(0))\in\iaccess[\bsprg]{\I}{\Isol}$.
In the end, we have $\imodels{\Itb}{\dbox{plant_a}{p(\bar{y},\bar{u}) = q(\bar{y},\bar{u}) \land R}}$.
By the axiom (\irref{ode}), we then have $\imodels{\Itb}{plant_a \refines plant_b}$, which implies $\iaccessible[plant_b]{\Itb}{\Isol}$.
As $ctrl_b$ does not affect differential variables and we have $(\iget[state]{\I},\sol(0))\in\iaccess[ctrl_b;plant_b]{\I}{\Isol}$, we have $\iaccessible[ctrl_b]{\I}{\Itb}$ and we can conclude $\iaccessible[\bsprg]{\I}{\Isol}$.
\qed
\end{proof}

\begin{proof}[\rref{thm:deciderefine}]
Following \rref{lem:refineidemp}, we are left to prove the decidability for one iteration of the loop: $ctrl_a;\pode{\D{\bar{y}}=p(\bar{y},\bar{u})}{Q} \refines ctrl_b;\pode{\D{\bar{y}}=q(\bar{y},\bar{u})}{R}$.
In practice, we do not use axiom (\irref{sequence}) immediately, as this would be incomplete if $ctrl_a$ have an execution which violates the domain constraint $Q$ and that $ctrl_b$ does not have.

Using (\irref{DWref}), we have $(\pode{\D{\bar{y}}=p(\bar{y},\bar{u})}{Q}) \prgeq (\ptest{Q};\pode{\D{\bar{y}}=p(\bar{y},\bar{u})}{Q})$, and similarly for $\pode{\D{\bar{y}}=q(\bar{y},\bar{u})}{R}$.
By using (\irref{sequence}), we are left with two formulas:
\begin{enumerate}
	\item $ctrl_a;\ptest{Q} \refines ctrl_b;\ptest{R}$
	\item $\dbox{ctrl_a;\ptest{Q}}{(\pode{\D{\bar{y}}=p(\bar{y},\bar{u})}{Q} \refines \pode{\D{\bar{y}}=q(\bar{y},\bar{u})}{R})}$
\end{enumerate}
By assumption and \rref{lem:decidiscret}, both of them are decidable.

The use of the axiom (\irref{sequence}) is complete for an argument similar to the one used in the proof of \rref{lem:refineidemp}.
If $ctrl_a;plant_a \refines ctrl_b;plant_b$, and $\iaccessible[ctrl_a]{\I}{\Itb}$, and $\iaccessible[plant_a]{\Itb}{\It}$ and $\ivaluation{\It}{t_0} = \ivaluation{\I}{t_0}$, then $\iaccessible[ctrl_a;\pupdate{\umod{\D{\bar{y}}}{p(\bar{y},\bar{u})}};\ptest{Q}]{\I}{\It}$ and $\iaccessible[ctrl_b;\pupdate{\umod{\D{\bar{y}}}{q(\bar{y},\bar{u})}};\ptest{R}]{\I}{\It}$.
As the continuous and discrete variables are distinct, this entails the first formula.
For the second one, we also need to consider the case where $\ivaluation{\It}{t_0} > \ivaluation{\I}{t_0}$, in which case the proof of \rref{lem:refineidemp} shows that $\iaccessible[ctrl_b]{\I}{\Itb}$ and $\iaccessible[plant_b]{\Itb}{\It}$.

\qed
\end{proof}

\fi
\end{document}

%% file: ruledefs.tex

\providecommand{\axkey}[1]{\textcolor{vblue}{#1}}%
\cinferenceRuleStore[diamond|$\didia{\cdot}$]{diamond axiom}
{\linferenceRule[equiv]
  {\lnot\dbox{\ausprg}{\lnot \ausfml}}
  {\axkey{\ddiamond{\ausprg}{\ausfml}}}
}
{}
\cinferenceRuleStore[diamondax|$\didia{\cdot}$]{diamond axiom}
{\linferenceRule[equiv]
  {\lnot\dbox{\ausprgax}{\lnot \ausfmlax}}
  {\axkey{\ddiamond{\ausprgax}{\ausfmlax}}}
}
{}
\cinferenceRuleStore[assignb|$\dibox{:=}$]{assignment / substitution axiom}
{\linferenceRule[equiv]
  {p(\genDJ{x})}
  {\axkey{\dbox{\pupdate{\umod{x}{\genDJ{x}}}}{p(x)}}}
}
{}
\cinferenceRuleStore[assignbax|$\dibox{:=}$]{assignment / substitution axiom}
{\linferenceRule[equiv]
  {p(\aconst)}
  {\axkey{\dbox{\pupdate{\umod{x}{\aconst}}}{p(x)}}}
}
{}
\cinferenceRuleStore[Dassignb|$\dibox{:=}$]{differential assignment}
{\linferenceRule[equiv]
{p(\astrm)}
{\axkey{\dbox{\Dupdate{\Dumod{\D{x}}{\astrm}}}{p(\D{x})}}}
}
{}
\cinferenceRuleStore[testb|$\dibox{?}$]{test}
{\linferenceRule[equiv]
  {(\ivr \limply \ausfml)}
  {\axkey{\dbox{\ptest{\ivr}}{\ausfml}}}
}{}%
\cinferenceRuleStore[testbax|$\dibox{?}$]{test}
{\linferenceRule[equiv]
  {(q \limply p)}
  {\axkey{\dbox{\ptest{q}}{p}}}
}{}%
\cinferenceRuleStore[evolveb|$\dibox{'}$]{evolve}
{\linferenceRule[equiv]
  {\lforall{t{\geq}0}{\dbox{\pupdate{\pumod{x}{\solf(t)}}}{p(x)}}}
  {\axkey{\dbox{\pevolve{\D{x}=\genDE{x}}}{p(x)}}}
}{\m{\D{\solf}(t)=\genDE{\solf}}}
\cinferenceRuleStore[choiceb|$\dibox{\cup}$]{axiom of nondeterministic choice}
{\linferenceRule[equiv]
  {\dbox{\ausprg}{\ausfml} \land \dbox{\busprg}{\ausfml}}
  {\axkey{\dbox{\pchoice{\ausprg}{\busprg}}{\ausfml}}}
}{}%
\cinferenceRuleStore[choicebax|$\dibox{\cup}$]{axiom of nondeterministic choice}
{\linferenceRule[equiv]
  {\dbox{\ausprgax}{\ausfmlax} \land \dbox{\busprgax}{\ausfmlax}}
  {\axkey{\dbox{\pchoice{\ausprgax}{\busprgax}}{\ausfmlax}}}
}{}%
\cinferenceRuleStore[evolveinb|$\dibox{'}$]{evolve}
{\linferenceRule[equiv]
  {
        \lforall{t{\geq}0}{\big(
          (\lforall{0{\leq}s{\leq}t}{q(\solf(s))})
          \limply
          \dbox{\pupdate{\pumod{x}{\solf(t)}}}{p(x)}
        \big)}
      }
  {
        \dbox{\pevolvein{\D{x}=\genDE{x}}{q(x)}}{p(x)}
  }
}{}
\cinferenceRuleStore[composeb|$\dibox{{;}}$]{composition} 
{\linferenceRule[equiv]
  {\dbox{\ausprg}{\dbox{\busprg}{\ausfml}}}
  {\axkey{\dbox{\ausprg;\busprg}{\ausfml}}}
}{}%
\cinferenceRuleStore[composebax|$\dibox{{;}}$]{composition} 
{\linferenceRule[equiv]
  {\dbox{\ausprgax}{\dbox{\busprgax}{\ausfmlax}}}
  {\axkey{\dbox{\ausprgax;\busprgax}{\ausfmlax}}}
}{}%
\cinferenceRuleStore[iterateb|$\dibox{{}^*}$]{iteration/repeat unwind} 
{\linferenceRule[equiv]
  {\ausfml \land \dbox{\ausprg}{\dbox{\prepeat{\ausprg}}{\ausfml}}}
  {\axkey{\dbox{\prepeat{\ausprg}}{\ausfml}}}
}{}%
\cinferenceRuleStore[iteratebax|$\dibox{{}^*}$]{iteration/repeat unwind} 
{\linferenceRule[equiv]
  {\ausfmlax \land \dbox{\ausprgax}{\dbox{\prepeat{\ausprgax}}{\ausfmlax}}}
  {\axkey{\dbox{\prepeat{\ausprgax}}{\ausfmlax}}}
}{}%
\cinferenceRuleStore[K|K]{K axiom / modal modus ponens}
{\linferenceRule[impl]
  {\dbox{\ausprg}{(\ausfml\limply\busfml)}}
  {(\dbox{\ausprg}{\ausfml}\limply\axkey{\dbox{\ausprg}{\busfml}})}
}{}%
\cinferenceRuleStore[Kax|K]{K axiom / modal modus ponens}
{\linferenceRule[impl]
  {\dbox{\ausprgax}{(\ausfmlax\limply\busfmlax)}}
  {(\dbox{\ausprgax}{\ausfmlax}\limply\axkey{\dbox{\ausprgax}{\busfmlax}})}
}{}%
\cinferenceRuleStore[I|II]{loop induction}
{\linferenceRule[impl]
  {\dbox{\prepeat{\ausprg}}{(\ausfml\limply\dbox{\ausprg}{\ausfml})}}
  {(\ausfml\limply\axkey{\dbox{\prepeat{\ausprg}}{\ausfml}})}
}{}%
\cinferenceRuleStore[Ieq|I]{loop induction}
{\linferenceRule[equiv]
  {\ausfml \land \dbox{\prepeat{\ausprg}}{(\ausfml\limply\dbox{\ausprg}{\ausfml})}}
  {\axkey{\dbox{\prepeat{\ausprg}}{\ausfml}}}
}{}%
\cinferenceRuleStore[Ieqax|I]{loop induction}
{\linferenceRule[equiv]
  {\ausfmlax \land \dbox{\prepeat{\ausprgax}}{(\ausfmlax\limply\dbox{\ausprgax}{\ausfmlax})}}
  {\axkey{\dbox{\prepeat{\ausprgax}}{\ausfmlax}}}
}{}%
\dinferenceRuleStore[backiterateb|\usebox{\backiterateb}]{backwards iteration/repeat unwind}
{\linferenceRule[equiv]
  {\ausfml \land \dbox{\prepeat{\ausprg}}{\dbox{\ausprg}{\ausfml}}}
  {\axkey{\dbox{\prepeat{\ausprg}}{\ausfml}}}
}{}
\dinferenceRuleStore[iterateiterateb|$\dibox{{}^*{}^*}$]{double iteration}
{\linferenceRule[equiv]
  {\dbox{\prepeat{\ausprg}}{\ausfml}}
  {\axkey{\dbox{\prepeat{\ausprg};\prepeat{\ausprg}}{\ausfml}}}
}{}
\dinferenceRuleStore[iterateiterated|$\didia{{}^*{}^*}$]{double iteration}
{\linferenceRule[equiv]
  {\ddiamond{\prepeat{\ausprg}}{\ausfml}}
  {\axkey{\ddiamond{\prepeat{\ausprg};\prepeat{\ausprg}}{\ausfml}}}
}{}
\cinferenceRuleStore[B|B]{Barcan and converse}
{\linferenceRule[equiv]
        {\ddiamond{\ausprg}{\lexists{x}{\ausfml}}}
        {\lexists{x}{\ddiamond{\ausprg}{\ausfml}}}
}{\m{x{\not\in}\ausprg}}
\cinferenceRuleStore[V|V]{vacuous $\dbox{}{}$}
{\linferenceRule[impl]
  {p}
  {\axkey{\dbox{\ausprg}{p}}}
}{\m{FV(p)\cap BV(\ausprg)=\emptyset}}
\cinferenceRuleStore[Vax|V]{vacuous $\dbox{}{}$}
{\linferenceRule[impl]
  {p}
  {\axkey{\dbox{a}{p}}}
}{}
\cinferenceRuleStore[G|G]{$\dbox{}{}$ generalization} 
{\linferenceRule[formula]
  {\ausfml}
  {\dbox{\ausprg}{\ausfml}}
}{}%
\cinferenceRuleStore[Gax|G]{$\dbox{}{}$ generalization} 
{\linferenceRule[formula]
  {\ausfmlax}
  {\dbox{\ausprgax}{\ausfmlax}}
}{}%
\cinferenceRuleStore[genaax|$\forall{}$]{$\forall{}$ generalisation}
{\linferenceRule[formula]
  {p(x)}
  {\lforall{x}{p(x)}}
}{}
\cinferenceRuleStore[MPax|MP]{modus ponens}
{\linferenceRule[formula]
  {p\limply q \quad p}
  {q}
}{}
\cinferenceRuleStore[Mb|M${\dibox{\cdot}}$]{$\dbox{}{}$ monotone}
{\linferenceRule[formula]
  {\ausfml\limply \busfml}
  {\dbox{\ausprg}{\ausfml}\limply\dbox{\ausprg}{\busfml}}
}{}%
\cinferenceRuleStore[M|M]{$\ddiamond{}{}$ monotone / $\ddiamond{}{}$-generalization}
{\linferenceRule[formula]
  {\ausfml\limply\busfml}
  {\ddiamond{\ausprg}{\ausfml}\limply\ddiamond{\ausprg}{\busfml}}
}{}%

\dinferenceRuleStore[Mbr|M\rightrule]
{$\ddiamond{}{}/\dbox{}{}$ generalization=M=G+K} 
{\linferenceRule[sequent]
  {\lsequent[L]{} {\dbox{\ausprg}{\busfml}} 
  &\lsequent[g]{\busfml} {\ausfml}}
  {\lsequent[L]{} {\dbox{\ausprg}{\ausfml}}}
}{}%

\dinferenceRuleStore[loop|loop]{inductive invariant}
{\linferenceRule[sequent]
  {\lsequent[L]{} {\inv}
  &\lsequent[g]{\inv} {\dbox{\ausprg}{\inv}}
  &\lsequent[g]{\inv} {\ausfml}}
  {\lsequent[L]{} {\dbox{\prepeat{\ausprg}}{\ausfml}}}
}{}%
\dinferenceRuleStore[invind|ind]{inductive invariant}
{\linferenceRule[sequent]
  {\lsequent[\globalrule]{\ausfml}{\dbox{\ausprg}{\ausfml}}}
  {\lsequent{\ausfml}{\dbox{\prepeat{\ausprg}}{\ausfml}}}
}{}%
\cinferenceRuleStore[con|con]{loop convergence right} 
{\linferenceRule[formula]
  {\lsequent[G]{\mapply{\var}{v}\land v>0}{\ddiamond{\ausprg}{\mapply{\var}{v-1}}}}
  {\lsequent[L]{\lexists{v}{\mapply{\var}{v}}}
      {\axkey{\ddiamond{\prepeat{\ausprg}}{\lexists{v{\leq}0}{\mapply{\var}{v}}}}}}
}{v\not\in\ausprg}
\dinferenceRuleStore[congen|con]{loop convergence}
{\linferenceRule[sequent]
  {\lsequent[L]{}{\lexists{v}{\mapply{\var}{v}}}
  &\lsequent[G]{}{\lforall{v{>}0}{({\mapply{\var}{v}}\limply{\ddiamond{\ausprg}{\mapply{\var}{v-1}})}}}
  &\lsequent[G]{\lexists{v{\leq}0}{\mapply{\var}{v}}}{\busfml}
  }
  {\lsequent[L]{}{\ddiamond{\prepeat{\ausprg}}{\busfml}}}
}{v\not\in\ausprg}

\dinferenceRuleStore[band|${[]\land}$]{$\dbox{\cdot}{\land}$}
{\linferenceRule[equiv]
  {\dbox{\ausprg}{\ausfml} \land \dbox{\ausprg}{\busfml}}
  {\axkey{\dbox{\ausprg}{(\ausfml\land\busfml)}}}
}{}%

\dinferenceRuleStore[Hoarecompose|H${;}$]{Hoare $;$}
{\linferenceRule
  {A\limply\dbox{\ausprg}{E} & E\limply\dbox{\busprg}{B}}
  {A \limply \dbox{\ausprg;\busprg}{B}}
}{}%
\dinferenceRuleStore[composebrexplicit|$\dibox{{;}}$\rightrule]{$;$}
{\linferenceRule
  {A\limply\dbox{\ausprg}{\dbox{\busprg}{B}}}
  {A \limply \dbox{\ausprg;\busprg}{B}}
}{}


\cinferenceRuleStore[notr|$\lnot$\rightrule]{$\lnot$ right}
{\linferenceRule[sequent]
  {\lsequent[L]{\asfml}{}}
  {\lsequent[L]{}{\lnot \asfml}}
}{}%
\cinferenceRuleStore[notl|$\lnot$\leftrule]{$\lnot$ left}
{\linferenceRule[sequent]
  {\lsequent[L]{}{\asfml}}
  {\lsequent[L]{\lnot \asfml}{}}
}{}%
\cinferenceRuleStore[andr|$\land$\rightrule]{$\land$ right}
{\linferenceRule[sequent]
  {\lsequent[L]{}{\asfml}
    & \lsequent[L]{}{\bsfml}}
  {\lsequent[L]{}{\asfml \land \bsfml}}
}{}%
\cinferenceRuleStore[andl|$\land$\leftrule]{$\land$ left}
{\linferenceRule[sequent]
  {\lsequent[L]{\asfml , \bsfml}{}}
  {\lsequent[L]{\asfml \land \bsfml}{}}
}{}%
\cinferenceRuleStore[orr|$\lor$\rightrule]{$\lor$ right}
{\linferenceRule[sequent]
  {\lsequent[L]{}{\asfml, \bsfml}}
  {\lsequent[L]{}{\asfml \lor \bsfml}}
}{}%
\cinferenceRuleStore[orl|$\lor$\leftrule]{$\lor$ left}
{\linferenceRule[sequent]
  {\lsequent[L]{\asfml}{}
    & \lsequent[L]{\bsfml}{}}
  {\lsequent[L]{\asfml \lor \bsfml}{}}
}{}%
\cinferenceRuleStore[implyr|$\limply$\rightrule]{$\limply$ right}
{\linferenceRule[sequent]
  {\lsequent[L]{\asfml}{\bsfml}}
  {\lsequent[L]{}{\asfml \limply \bsfml}}
}{}%
\cinferenceRuleStore[implyl|$\limply$\leftrule]{$\limply$ left}
{\linferenceRule[sequent]
  {\lsequent[L]{}{\asfml}
    & \lsequent[L]{\bsfml}{}}
  {\lsequent[L]{\asfml \limply \bsfml}{}}
}{}%
\cinferenceRuleStore[equivr|$\lbisubjunct$\rightrule]{$\lbisubjunct$ right}
{\linferenceRule[sequent]
  {\lsequent[L]{\asfml}{\bsfml}
   & \lsequent[L]{\bsfml}{\asfml}}
  {\lsequent[L]{}{\asfml \lbisubjunct \bsfml}}
}{}%
\cinferenceRuleStore[equivl|$\lbisubjunct$\leftrule]{$\lbisubjunct$ left}
{\linferenceRule[sequent]
  {\lsequent[L]{\asfml\limply\bsfml, \bsfml\limply\asfml}{}}
  {\lsequent[L]{\asfml \lbisubjunct \bsfml}{}}
}{}%
\cinferenceRuleStore[id|id]{identity}
{\linferenceRule[sequent]
  {}
  {\lsequent[L]{\asfml}{\asfml}}
}{}%
\cinferenceRuleStore[cut|cut]{cut}
{\linferenceRule[sequent]
  {\lsequent[L]{}{\cusfml}
  &\lsequent[L]{\cusfml}{}}
  {\lsequent[L]{}{}}
}{}%
\cinferenceRuleStore[weakenr|W\rightrule]{weakening right}
{\linferenceRule[sequent]
  {\lsequent[L]{}{}}
  {\lsequent[L]{}{\asfml}}
}{}%
\cinferenceRuleStore[weakenl|W\leftrule]{weakening left}
{\linferenceRule[sequent]
  {\lsequent[L]{}{}}
  {\lsequent[L]{\asfml}{}}
}{}%
\cinferenceRuleStore[exchanger|P\rightrule]{exchange right}
{\linferenceRule[sequent]
  {\lsequent[L]{}{\bsfml,\asfml}}
  {\lsequent[L]{}{\asfml,\bsfml}}
}{}%
\cinferenceRuleStore[exchangel|P\leftrule]{exchange left}
{\linferenceRule[sequent]
  {\lsequent[L]{\bsfml,\asfml}{}}
  {\lsequent[L]{\asfml,\bsfml}{}}
}{}%
\cinferenceRuleStore[contractr|c\rightrule]{contract right}
{\linferenceRule[sequent]
  {\lsequent[L]{}{\asfml,\asfml}}
  {\lsequent[L]{}{\asfml}}
}{}%
\cinferenceRuleStore[contractl|c\leftrule]{contract left}
{\linferenceRule[sequent]
  {\lsequent[L]{\asfml,\asfml}{}}
  {\lsequent[L]{\asfml}{}}
}{}
\cinferenceRuleStore[closeTrue|$\top$\rightrule]{close by true}
{\linferenceRule[sequent]
  {}
  {\lsequent[L]{}{\ltrue}}
}{}%
\cinferenceRuleStore[closeFalse|$\bot$\leftrule]{close by false}
{\linferenceRule[sequent]
  {}
  {\lsequent[L]{\lfalse}{}}
}{}%

\cinferenceRuleStore[CE|CE]{congequiv congruence of equivalences on formulas}
{\linferenceRule[formula]
  {\ausfml \lbisubjunct \busfml}
  {\contextapp{C}{\ausfml} \lbisubjunct \contextapp{C}{\busfml}}
}{}%
\dinferenceRuleStore[CEr|CE\rightrule]{congequiv congruence of equivalences on formulas}
{\linferenceRule[formula]
  {\lsequent[L]{} {\contextapp{C}{\busfml}}
  &\lsequent[g]{} {\ausfml \lbisubjunct \busfml}}
  {\lsequent[L]{} {\contextapp{C}{\ausfml}}}
}{}%
\dinferenceRuleStore[CEl|CE\leftrule]{congequiv congruence of equivalences on formulas}
{\linferenceRule[formula]
  {\lsequent[L]{\contextapp{C}{\busfml}} {}
  &\lsequent[g]{} {\ausfml \lbisubjunct \busfml}}
  {\lsequent[L]{\contextapp{C}{\ausfml}} {}}
}{}%

\cinferenceRuleStore[allr|$\forall$\rightrule]{$\lforall{}{}$ right}
{\linferenceRule[sequent]
  {\lsequent[L]{}{p(y)}}
  {\lsequent[L]{}{\lforall{x}{p(x)}}}
}{\m{y\not\in\Gamma{,}\Delta{,}\lforall{x}{p(x)}}}%
\cinferenceRuleStore[alll|$\forall$\leftrule]{$\lforall{}{}$ left instantiation}
{\linferenceRule[sequent]
  {\lsequent[L]{p(\astrm)}{}}
  {\lsequent[L]{\lforall{x}{p(x)}}{}}
}{arbitrary term $\astrm$}
\cinferenceRuleStore[existsr|$\exists$\rightrule]{$\lexists{}{}$ right}
{\linferenceRule[sequent]
  {\lsequent[L]{}{p(\astrm)}}
  {\lsequent[L]{}{\lexists{x}{p(x)}}}
}{arbitrary term $\astrm$}
\cinferenceRuleStore[existsl|$\exists$\leftrule]{$\lexists{}{}$ left}
{\linferenceRule[sequent]
  {\lsequent[L]{p(y)} {}}
  {\lsequent[L]{\lexists{x}{p(x)}} {}}
}{\m{y\not\in\Gamma{,}\Delta{,}\lexists{x}{p(x)}}}%

\cinferenceRuleStore[qear|\usebox{\Rval}]{quantifier elimination real arithmetic}
{\linferenceRule[sequent]
  {}
  {\lsequent[g]{\Gamma}{\Delta}}
}{\text{if}~\landfold_{\ausfml\in\Gamma} \ausfml \limply \lorfold_{\busfml\in\Delta} \busfml ~\text{is valid in \LOS[\reals]}}%

\dinferenceRuleStore[allGi|i$\forall$]{inverse universal generalization / universal instantiation}
{\linferenceRule[sequent]
  {\lsequent[L]{} {\lforall{x}{\ausfml}}}
  {\lsequent[L]{} {\ausfml}}
}{}

\cinferenceRuleStore[applyeqr|=\rightrule]{apply equation}
{\linferenceRule[sequent]
  {\lsequent[L]{x=\astrm}{p(\astrm)}}
  {\lsequent[L]{x=\astrm}{p(x)}}
}{}%
\cinferenceRuleStore[applyeql|=\leftrule]{apply equation}
{\linferenceRule[sequent]
  {\lsequent[L]{x=\astrm,p(\astrm)}{}}
  {\lsequent[L]{x=\astrm,p(x)}{}}
}{}%

\dinferenceRuleStore[alldupl|$\forall\forall$\leftrule]{$\lforall{}{}$ left instantiation retaining duplicates}
{\linferenceRule[sequent]
  {\lsequent[L]{\lforall{x}{p(x)},p(\astrm)}{}}
  {\lsequent[L]{\lforall{x}{p(x)}}{}}
}{}

\dinferenceRuleStore[choicebrinsist|$\dibox{\cup}\rightrule$]{}
{\linferenceRule
  {\lsequent[L]{}{\dbox{\asprg}{\ausfml}\land\dbox{\bsprg}{\ausfml}}}
  {\lsequent[L]{}{\dbox{\pchoice{\asprg}{\bsprg}}{\ausfml}}}
}{}
\dinferenceRuleStore[choiceblinsist|$\dibox{\cup}\leftrule$]{}
{\linferenceRule
  {\lsequent[L]{\dbox{\asprg}{\ausfml}\land\dbox{\bsprg}{\ausfml}}{}}
  {\lsequent[L]{\dbox{\pchoice{\asprg}{\bsprg}}{\ausfml}}{}}
}{}
\dinferenceRuleStore[choicebrinsist2|$\dibox{\cup}\rightrule2$]{}
{\linferenceRule
  {\lsequent[L]{}{\dbox{\asprg}{\ausfml}}
  &\lsequent[L]{}{\dbox{\bsprg}{\ausfml}}}
  {\lsequent[L]{}{\dbox{\pchoice{\asprg}{\bsprg}}{\ausfml}}}
}{}
\dinferenceRuleStore[choiceblinsist2|$\dibox{\cup}\leftrule2$]{}
{\linferenceRule
  {\lsequent[L]{\dbox{\asprg}{\ausfml},\dbox{\bsprg}{\ausfml}}{}}
  {\lsequent[L]{\dbox{\pchoice{\asprg}{\bsprg}}{\ausfml}}{}}
}{}
\dinferenceRuleStore[cutr|cut\rightrule]{cut right}
{\linferenceRule[sequent]
  {\lsequent[L]{}{\bsfml}
  &\lsequent[L]{}{\bsfml\limply\asfml}}
  {\lsequent[L]{}{\asfml}}
}{}
\dinferenceRuleStore[cutl|cut\leftrule]{cut left}
{\linferenceRule[sequent]
  {\lsequent[L]{\bsfml} {}
  &\lsequent[L]{}{\asfml\limply\bsfml}}
  {\lsequent[L]{\asfml} {}}
}{}


\cinferenceRuleStore[Dplus|$+'$]{derive sum}
{\linferenceRule[eq]
  {\der{\asdtrm}+\der{\bsdtrm}}
  {\axkey{\der{\asdtrm+\bsdtrm}}}
}
{}
\cinferenceRuleStore[Dplusax|$+'$]{derive sum}
{\linferenceRule[eq]
  {\der{\asdtrmax}+\der{\bsdtrmax}}
  {\axkey{\der{\asdtrmax+\bsdtrmax}}}
}
{}
\cinferenceRuleStore[Dminus|$-'$]{derive minus}
{\linferenceRule[eq]
  {\der{\asdtrm}-\der{\bsdtrm}}
  {\axkey{\der{\asdtrm-\bsdtrm}}}
}
{}
\cinferenceRuleStore[Dminusax|$-'$]{derive minus}
{\linferenceRule[eq]
  {\der{\asdtrmax}-\der{\bsdtrmax}}
  {\axkey{\der{\asdtrmax-\bsdtrmax}}}
}
{}
\cinferenceRuleStore[Dtimes|$\cdot'$]{derive product}
{\linferenceRule[eq]
  {\der{\asdtrm}\cdot \bsdtrm+\asdtrm\cdot\der{\bsdtrm}}
  {\axkey{\der{\asdtrm\cdot \bsdtrm}}}
}
{}
\cinferenceRuleStore[Dtimesax|$\cdot'$]{derive product}
{\linferenceRule[eq]
  {\der{\asdtrmax}\cdot \bsdtrmax+\asdtrmax\cdot\der{\bsdtrmax}}
  {\axkey{\der{\asdtrmax\cdot \bsdtrmax}}}
}
{}
\cinferenceRuleStore[Dquotient|$/'$]{derive quotient}
{\linferenceRule[eq]
  {\big(\der{\asdtrm}\cdot \bsdtrm-\asdtrm\cdot\der{\bsdtrm}\big) / \bsdtrm^2}
  {\axkey{\der{\asdtrm/\bsdtrm}}}
}
{}
\cinferenceRuleStore[Dquotientax|$/'$]{derive quotient}
{\linferenceRule[eq]
  {\big(\der{\asdtrmax}\cdot \bsdtrmax-\asdtrmax\cdot\der{\bsdtrmax}\big) / \bsdtrmax^2}
  {\axkey{\der{\asdtrmax/\bsdtrmax}}}
}
{}
\cinferenceRuleStore[Dconst|$c'$]{derive constant}
{\linferenceRule[eq]
  {0}
  {\axkey{\der{\aconst}}}
  \hspace{3cm}
}
{\text{for numbers or constants~$\aconst$}}
\cinferenceRuleStore[Dvar|$x'$]{derive variable}
{\linferenceRule[eq]
  {\D{x}}
  {\axkey{\der{x}}}
}
{\text{for variable~$x\in\allvars$}}

\cinferenceRuleStore[DE|DE]{differential effect} 
{\linferenceRule[viuqe]
  {\axkey{\dbox{\pevolvein{\D{x}=\genDE{x}}{\ivr}}{\ausfml}}}
  {\dbox{\pevolvein{\D{x}=f(x)}{\ivr}}{\dbox{\axeffect{\Dupdate{\Dumod{\D{x}}{\genDE{x}}}}}{\ausfml}}}
}
{}%
\cinferenceRuleStore[DEax|DE]{differential effect} 
{\linferenceRule[viuqe]
  {\axkey{\dbox{\pevolvein{\D{x}=\genDE{x}}{q(x)}}{\ausfmlax}}}
  {\dbox{\pevolvein{\D{x}=f(x)}{q(x)}}{\dbox{\axeffect{\Dupdate{\Dumod{\D{x}}{\genDE{x}}}}}{\ausfmlax}}}
}
{}%


\cinferenceRuleStore[Dand|${\land}'$]{derive and}
{\linferenceRule[equiv]
  {\der{\asfml}\land\der{\bsfml}}
  {\axkey{\der{\asfml\land\bsfml}}}
}
{}
\cinferenceRuleStore[Dor|${\lor}'$]{derive or}
{\linferenceRule[equiv]
  {\der{\asfml}\land\der{\bsfml}}
  {\axkey{\der{\asfml\lor\bsfml}}}
}
{}
\cinferenceRuleStore[diffweaken|DW]{differential evolution domain} 
{\linferenceRule[viuqe]
  {\axkey{\dbox{\pevolvein{\D{x}=\genDE{x}}{\ivr}}{\ousfml[x]}}}
  {\dbox{\pevolvein{\D{x}=\genDE{x}}{\ivr}}{(\axeffect{\ivr}\limply \ousfml[x])}}
}
{}
\cinferenceRuleStore[diffweakenax|DW]{differential evolution domain} 
{\linferenceRule[viuqe]
  {\axkey{\dbox{\pevolvein{\D{x}=\genDE{x}}{q(x)}}{p(x)}}}
  {\dbox{\pevolvein{\D{x}=\genDE{x}}{q(x)}}{(\axeffect{q(x)}\limply p(x))}}
}
{}
\cinferenceRuleStore[dW|dW]{differential weakening}
{\linferenceRule[sequent]
  {\lsequent[g]{\ivr} {\ousfml[x]}}
  {\lsequent[g]{\Gamma} {\dbox{\pevolvein{\D{x}=f(x)}{\ivr}}{\ousfml[x]},\Delta}}
}
{}
\cinferenceRuleStore[DI|DI]{differential induction}
{\linferenceRule[lpmi]
  {\big(\axkey{\dbox{\pevolvein{\D{x}=\genDE{x}}{\ivr}}{\ousfml[x]}}
  \lbisubjunct \dbox{\ptest{\ivr}}{\ousfml[x]}\big)}
  {(\ivr\limply\dbox{\pevolvein{\D{x}=\genDE{x}}{\ivr}}{\axeffect{\der{\ousfml[x]}}})}
}
{}
\cinferenceRuleStore[DIax|DI]{differential induction}
{\linferenceRule[lpmi]
  {\big(\axkey{\dbox{\pevolvein{\D{x}=\genDE{x}}{q(x)}}{p(x)}}
  \lbisubjunct \dbox{\ptest{q(x)}}{p(x)}\big)}
  {(q(x)\limply\dbox{\pevolvein{\D{x}=\genDE{x}}{q(x)}}{\axeffect{\der{p(x)}}})}
}
{}
\cinferenceRuleStore[DIlight|DI]{differential induction}
{\linferenceRule[lpmi]
  {\big(\axkey{\dbox{\pevolvein{\D{x}=\genDE{x}}{\ivr}}{\ousfml[x]}}
  \lbisubjunct \dbox{\ptest{\ivr}}{\ousfml[x]}\big)}
  {\dbox{\pevolvein{\D{x}=\genDE{x}}{\ivr})}{\axeffect{\der{\ousfml[x]}}}}
}
{}

\cinferenceRuleStore[dI|dI]{differential invariant}
{\linferenceRule[sequent]
  {\lsequent[g]{\ivr}{\Dusubst{\D{x}}{\genDE{x}}{\der{F}}}}
  {\lsequent{F}{\dbox{\pevolvein{\D{x}=\genDE{x}}{\ivr}}{F}}}
}{}
\cinferenceRuleStore[DC|DC]{differential cut}
{\linferenceRule[lpmi]
  {\big(\axkey{\dbox{\pevolvein{\D{x}=\genDE{x}}{\ivr}}{\ousfml[x]}} \lbisubjunct \dbox{\pevolvein{\D{x}=\genDE{x}}{\ivr\land \axeffect{\ousfmlc[x]}}}{\ousfml[x]}\big)}
  {\dbox{\pevolvein{\D{x}=\genDE{x}}{\ivr}}{\axeffect{\ousfmlc[x]}}}
}
{}
\cinferenceRuleStore[DCax|DC]{differential cut}
{\linferenceRule[lpmi]
  {\big(\axkey{\dbox{\pevolvein{\D{x}=\genDE{x}}{q(x)}}{p(x)}} \lbisubjunct \dbox{\pevolvein{\D{x}=\genDE{x}}{q(x)\land \axeffect{r(x)}}}{p(x)}\big)}
  {\dbox{\pevolvein{\D{x}=\genDE{x}}{q(x)}}{\axeffect{r(x)}}}
}
{}
\cinferenceRuleStore[dC|dC]{differential cut}
{\linferenceRule[sequent]
  {\lsequent[L]{}{\dbox{\pevolvein{\D{x}=\genDE{x}}{\ivr}}{\axeffect{\cusfml}}}
  &\lsequent[L]{}{\dbox{\pevolvein{\D{x}=\genDE{x}}{(\ivr\land \axeffect{\cusfml})}}{\ousfml[x]}}}
  {\lsequent[L]{}{\dbox{\pevolvein{\D{x}=\genDE{x}}{\ivr}}{\ousfml[x]}}}
}{}
\cinferenceRuleStore[DGanyode|DG]{differential ghost variables (unsound!)}
{\linferenceRule[viuqe]
  {\axkey{\dbox{\pevolvein{\D{x}=\genDE{x}}{\ivr}}{\ousfml[x]}}}
  {\lexists{y}{\dbox{\pevolvein{\D{x}=\genDE{x}\syssep\axeffect{\D{y}=g(x,y)}}{\ivr}}{\ousfml[x]}}}
}
{}
\cinferenceRuleStore[DG|DG]{differential ghost variables}
{\linferenceRule[viuqe]
  {\axkey{\dbox{\pevolvein{\D{x}=\genDE{x}}{\ivr}}{\ousfml[x]}}}
  {\lexists{y}{\dbox{\pevolvein{\D{x}=\genDE{x}\syssep\axeffect{\D{y}=a(x)\cdot y+b(x)}}{\ivr}}{\ousfml[x]}}}
}
{}
\cinferenceRuleStore[DGax|DG]{differential ghost variables}
{\linferenceRule[viuqe]
  {\axkey{\dbox{\pevolvein{\D{x}=\genDE{x}}{q(x)}}{p(x)}}}
  {\lexists{y}{\dbox{\pevolvein{\D{x}=\genDE{x}\syssep\axeffect{\D{y}=a(x)\cdot y+b(x)}}{q(x)}}{p(x)}}}
}
{}
\cinferenceRuleStore[dG|dG]{dG}
{\linferenceRule[sequent]
  {\lsequent[L]{} {\lexists{y}{\dbox{\pevolvein{\D{x}=f(x)\syssep\axeffect{\D{y}=a(x)\cdot y+b(x)}}{\oivr[x]}}{\ousfml[x]}}}
  }
  {\lsequent[L]{} {\dbox{\pevolvein{\D{x}=f(x)}{\oivr[x]}}{\ousfml[x]}}}
}
{}%

\dinferenceRuleStore[assignbeqr|$\dibox{:=}_=$]{assignb}
  {\linferenceRule[sequent]
    {\lsequent[L]{y=\austrm} {p(y)}}
    {\lsequent[L]{} {\dbox{\pupdate{\umod{x}{\austrm}}}{p(x)}}}
  }
  {\text{$y$ new}}

\cinferenceRuleStore[DSax|DS]{(constant) differential equation solution} 
{\linferenceRule[viuqe]
  {\axkey{\dbox{\pevolvein{\D{x}=\aconst}{q(x)}}{p(x)}}}
  {\lforall{t{\geq}0}{\big((\lforall{0{\leq}s{\leq}t}{q(x+\aconst\itimes s)}) \limply \dbox{\pupdate{\pumod{x}{x+\aconst\itimes t}}}{p(x)}\big)}}
}
{}

\dinferenceRuleStore[DIeq0|DI]{differential invariant axiom}
{\linferenceRule[lpmi]
  {\big(\axkey{\dbox{\pevolve{\D{x}=\genDE{x}}}{\,\astrm=0}} \lbisubjunct \astrm=0\big)}
  {\dbox{\pevolve{\D{x}=\genDE{x}}}{\,\axeffect{\der{\astrm}=0}}}
}
{}
\dinferenceRuleStore[diffindeq0|dI]{differential invariant $=0$ case}
{\linferenceRule[sequent]
  {\lsequent{~}{\Dusubst{\D{x}}{\genDE{x}}{\der{\astrm}}=0}}
  {\lsequent{\astrm=0}{\dbox{\pevolve{\D{x}=\genDE{x}}}{\astrm=0}}}
}{}
\cinferenceRuleStore[Liec|dI$_c$]{}
{\linferenceRule
  {\lsequent{\ivr}{\Dusubst{\D{x}}{\genDE{x}}{\der{\astrm}}=0}}
  {\lsequent{}{\lforall{c}{\big(\astrm=c \limply \dbox{\pevolvein{\D{x}=\genDE{x}}{\ivr}}{\astrm=c}\big)}}}
}{}


\cinferenceRuleStore[DIeq|DI$_=$]{differential induction $=$ case}
{\linferenceRule[lpmi]
  {\big(\axkey{\dbox{\pevolvein{\D{x}=\genDE{x}}{\ivr}}{\asdtrm=\bsdtrm}}
  \lbisubjunct \dbox{\ptest{\ivr}}{\asdtrm=\bsdtrm}\big)}
  {\dbox{\pevolvein{\D{x}=\genDE{x}}{\ivr})}{\axeffect{\der{\asdtrm}=\der{\bsdtrm}}}}
}
{}

\dinferenceRuleStore[diffindgen|dI']{differential invariant}
{\linferenceRule[sequent]
  {\lsequent[L]{}{\inv}
  &\lsequent[g]{\ivr}{\Dusubst{\D{x}}{\genDE{x}}{\der{\inv}}}
  &\lsequent[g]{\inv}{\psi}
  }
  {\lsequent[L]{}{\dbox{\pevolvein{\D{x}=\genDE{x}}{\ivr}}{\psi}}}
}{}
\cinferenceRuleStore[diffindunsound|dI$_{??}$]{unsound}
{\linferenceRule[sequent]
  {\lsequent{\ivr\land\inv}{\Dusubst{\D{x}}{\genDE{x}}{\der{\inv}}}}
  {\lsequent{\inv}{\dbox{\pevolvein{\D{x}=\genDE{x}}{\ivr}}{\inv}}}
}{}

\dinferenceRuleStore[introaux|iG]{introduce discrete ghost variable}
{\linferenceRule[sequent]
  {\lsequent[L]{}{\dbox{\axeffect{\pupdate{\pumod{y}{\astrm}}}}{p}}}
  {\lsequent[L]{} {p}}
}{\text{$y$ new}}
\dinferenceRuleStore[diffaux|dA]{differential auxiliary variables}
{\linferenceRule[sequent]
  {\lsequent[\globalrule]{}{\inv\lbisubjunct\lexists{y}{G}}
  &\lsequent{G} {\dbox{\pevolvein{\D{x}=\genDE{x}\syssep\axeffect{\D{y}=a(x)\cdot y+b(x)}}{\ivr}}{G}}}
  {\lsequent{\inv} {\dbox{\pevolvein{\D{x}=\genDE{x}}{\ivr}}{\inv}}}
}{}
\cinferenceRuleStore[randomd|$\didia{{:}*}$]{nondeterministic assignment}
{\linferenceRule[equiv]
  {\lexists{x}{\ousfml[x]}}
  {\axkey{\ddiamond{\prandom{x}}{\ousfml[x]}}}
}{}
\cinferenceRuleStore[randomb|$\dibox{{:}*}$]{nondeterministic assignment}
{\linferenceRule[equiv]
  {\lforall{x}{\ousfml[x]}}
  {\axkey{\dbox{\prandom{x}}{\ousfml[x]}}}
}{}

\cinferenceRuleStore[box|$\dibox{\cdot}$]{box axiom}
{\linferenceRule[equiv]
  {\lnot\ddiamond{\ausprg}{\lnot\ausfml}}
  {\axkey{\dbox{\ausprg}{\ausfml}}}
}
{}
\cinferenceRuleStore[assignd|$\didia{:=}$]{assignment / substitution axiom}
{\linferenceRule[equiv]
  {p(\genDJ{x})}
  {\axkey{\ddiamond{\pupdate{\umod{x}{\genDJ{x}}}}{p(x)}}}
}
{}
\cinferenceRuleStore[evolved|$\didia{'}$]{evolve}
{\linferenceRule[equiv]
  {\lexists{t{\geq}0}{\ddiamond{\pupdate{\pumod{x}{\solf(t)}}}{p(x)}}\hspace{1cm}}
  {\axkey{\ddiamond{\pevolve{\D{x}=\genDE{x}}}{p(x)}}}
}{\m{\D{\solf}(t)=\genDE{\solf}}}
\cinferenceRuleStore[evolveind|$\didia{'}$]{evolve}
{\linferenceRule[equiv]
  {\lexists{t{\geq}0}{\big((\lforall{0{\leq}s{\leq}t}{q(\solf(s))}) \land 
  \ddiamond{\pupdate{\pumod{x}{\solf(t)}}}{p(x)}\big)}}
  {\axkey{\ddiamond{\pevolvein{\D{x}=\genDE{x}}{q(x)}}{p(x)}}}
}{\m{\D{\solf}(t)=\genDE{\solf}}}
\cinferenceRuleStore[testd|$\didia{?}$]{test}
{\linferenceRule[equiv]
  {\ivr \land \ausfml}
  {\axkey{\ddiamond{\ptest{\ivr}}{\ausfml}}}
}{}
\cinferenceRuleStore[choiced|$\didia{\cup}$]{axiom of nondeterministic choice}
{\linferenceRule[equiv]
  {\ddiamond{\ausprg}{\ausfml} \lor \ddiamond{\busprg}{\ausfml}}
  {\axkey{\ddiamond{\pchoice{\ausprg}{\busprg}}{\ausfml}}}
}{}
\cinferenceRuleStore[composed|$\didia{{;}}$]{composition}
{\linferenceRule[equiv]
  {\ddiamond{\ausprg}{\ddiamond{\busprg}{\ausfml}}}
  {\axkey{\ddiamond{\ausprg;\busprg}{\ausfml}}}
}{}
\cinferenceRuleStore[iterated|$\didia{{}^*}$]{iteration/repeat unwind pre-fixpoint, even fixpoint}
{\linferenceRule[equiv]
  {\ausfml \lor \ddiamond{\ausprg}{\ddiamond{\prepeat{\ausprg}}{\ausfml}}}
  {\axkey{\ddiamond{\prepeat{\ausprg}}{\ausfml}}}
}{}
\cinferenceRuleStore[duald|$\didia{{^d}}$]{dual}
{\linferenceRule[equiv]
  {\lnot\ddiamond{\ausprg}{\lnot\ausfml}}
  {\axkey{\ddiamond{\pdual{\ausprg}}{\ausfml}}}
}{}
\cinferenceRuleStore[dualb|$\dibox{{^d}}$]{dual}
{\linferenceRule[equiv]
  {\lnot\dbox{\ausprg}{\lnot\ausfml}}
  {\axkey{\dbox{\pdual{\ausprg}}{\ausfml}}}
}{}
\cinferenceRuleStore[FP|FP]{iteration is least fixpoint / reflexive transitive closure RTC, equivalent to invind in the presence of R}
{\linferenceRule[formula]
  {\ausfml \lor \ddiamond{\ausprg}{\busfml} \limply \busfml}
  {\ddiamond{\prepeat{\ausprg}}{\ausfml} \limply \busfml}
}{}
\cinferenceRuleStore[invindg|ind]{inductive invariant for games}
{\linferenceRule[formula]
  {\ausfml\limply\dbox{\ausprg}{\ausfml}}
  {\ausfml\limply\dbox{\prepeat{\ausprg}}{\ausfml}}
}{}

\dinferenceRuleStore[dchoiced|$\didia{{\cap}}$]{Demon's choice}
{
\axkey{\ddiamond{\dchoice{\ausprg}{\busprg}}{\ausfml}} \lbisubjunct \ddiamond{\ausprg}{\ausfml} \land \ddiamond{\busprg}{\ausfml}
}{}
\dinferenceRuleStore[dchoiceb|$\dibox{{\cap}}$]{Demon's choice}
{
\axkey{\dbox{\dchoice{\ausprg}{\busprg}}{\ausfml}} \lbisubjunct \dbox{\ausprg}{\ausfml} \lor \dbox{\busprg}{\ausfml}
}{}
\dinferenceRuleStore[diterateb|$\dibox{\drepeat{}}$]{Demon's repetition}
{\linferenceRule[equiv]
  {\ausfml \lor \dbox{\ausprg}{\dbox{\drepeat{\ausprg}}{\ausfml}}}
  {\axkey{\dbox{\drepeat{\ausprg}}{\ausfml}}}
}{}
\dinferenceRuleStore[diterated|$\didia{\drepeat{}}$]{Demon's repetition}
{\linferenceRule[equiv]
  {\ausfml \land \ddiamond{\ausprg}{\ddiamond{\drepeat{\ausprg}}}{\ausfml}}
  {\axkey{\ddiamond{\drepeat{\ausprg}}{\ausfml}}}
}{}
\dinferenceRuleStore[dinvindg|ind$\drepeat{}$]{inductive invariant for games}
{\linferenceRule[formula]
  {\ausfml\limply\ddiamond{\ausprg}{\ausfml}}
  {\ausfml\limply\ddiamond{\drepeat{\ausprg}}{\ausfml}}
}{}
\dinferenceRuleStore[dFP|FP$\drepeat{}$]{dual iteration is least fixpoint in Demon's winning strategy}
{\linferenceRule[formula]
  {\ausfml \lor \dbox{\ausprg}{\busfml} \limply \busfml}
  {\dbox{\drepeat{\ausprg}}{\ausfml} \limply \busfml}
}{}

\cinferenceRuleStore[US|US]{uniform substitution}
{\linferenceRule[formula]
  {\phi}
  {\applyusubst{\sigma}{\phi}}
}{}

\cinferenceRuleStore[linequs|$\exists$lin]{linear equation uniform substitution}
{\linferenceRule[impl]
  {b\neq0}
  {\big(\lexists{x}{(b\cdot x+c=0 \land q(x))}
  \lbisubjunct {q(-c/b)}\big)}
}{}

\dinferenceRuleStore[FA|FA]{First arrival}
{\ddiamond{\prepeat{\ausprg}}{\ausfml} \limply \ausfml \lor \ddiamond{\prepeat{\ausprg}}{(\lnot\ausfml\land\ddiamond{\ausprg}{\ausfml})}
}{}
\dinferenceRuleStore[Mor|M]{monotonicity axiom}
{\ddiamond{\ausprg}{(\ausfml\lor\busfml)}
\lbisubjunct
\ddiamond{\ausprg}{\ausfml} \lor \ddiamond{\ausprg}{\busfml}
}{}
\dinferenceRuleStore[VK|VK]{vacuous possible $\dbox{}{}$}
{\linferenceRule[impl]
  {p}
  {(\dbox{\ausprg}{\ltrue}{\limply}\dbox{\ausprg}{p})}
  \qquad
}{\m{\freevars{p}\cap \boundvars{\ausprg}=\emptyset}}
\dinferenceRuleStore[R|R]{Regular}
{\linferenceRule[formula]
  {\ausfml_1\land\ausfml_2\limply\busfml}
  {\dbox{\ausprg}{\ausfml_1} \land \dbox{\ausprg}{\ausfml_2} \limply \dbox{\ausprg}{\busfml}}
}{}